\documentclass[usenatbib]{mn2e}

\usepackage{bm}        
\usepackage[english]{babel}
\usepackage{amsfonts,amsmath,amssymb,mathrsfs,times,graphicx,array,float,hyperref}

\newcommand{\eq}{\begin{equation}}
\newcommand{\eeq}{\end{equation}}
\newcommand{\de}{{\mathrm d}}
\newcommand{\vta}[1]{\vert \boldsymbol{a}_{#1}        \vert}

\newcommand{\vtl}   {\vert \boldsymbol{       {\ell}} \vert}

\newcommand {\kms}{\,\rm km \, s^{-1}}

\title[The evolution of massive black holes and their spins in their galactic hosts]
{The evolution of massive black holes and their spins in their galactic hosts}
\author[E.\ Barausse]{Enrico Barausse\thanks{CITA National Fellow; email: ebarauss@uoguelph.ca}\\
Department of Physics, University of Guelph, Guelph, Ontario, N1G 2W1, Canada\\ and 
Maryland Center for Fundamental
    Physics \& Joint Space-Science Institute, Department of Physics,\\ University of Maryland, College
    Park, MD 20742, USA
}
\begin{document}

\date{}

\maketitle

\label{firstpage}

\begin{abstract}
Future space-based gravitational-wave detectors, such as LISA/SGO or a similar European mission (eLISA/NGO), will
 measure the masses and spins of massive black holes up to very high redshift, and in 
principle discriminate among different models for their evolution.
Because the masses and spins change as a result of both accretion from the interstellar medium and 
the black-hole mergers that are expected to naturally occur in the hierarchical formation of galaxies,
their evolution is inextricably entangled with that of their galactic hosts.
On the one hand, the amount of gas present in galactic nuclei regulates the changes in the 
black-hole masses and spins through accretion, and affects 
the mutual orientation of the spins before mergers by exerting gravito-magnetic torques on them. 
On the other hand, massive black holes play a central role in galaxy formation because of 
the feedback exerted by AGN activity on the growth of structures.
In this paper, we study the mass and spin evolution of massive black holes within a semianalytical galaxy-formation model 
that follows the evolution of dark-matter halos along merger trees, as well as that of the baryonic
components (hot gas, stellar and gaseous bulges, and stellar and gaseous galactic disks).
This allows us to study the mass and spin evolution in a self-consistent way, by taking into account the effect of the 
gas present in galactic nuclei both during the accretion phases and during mergers. Also, 
we present predictions, as a function of redshift, 
for the fraction of gas-rich black-hole mergers -- in which the spins prior to the merger are aligned due to the gravito-magnetic torques 
exerted by the circumbinary disk -- as opposed to gas-poor mergers, in which the orientation of the spins before the merger is roughly isotropic. 
These predictions may be tested by LISA or similar spaced-based gravitational-wave detectors
such as eLISA/NGO or SGO.
\end{abstract}


\section{\label{sec:intro}Introduction}

Massive black-hole (MBH) mergers are expected to be the brightest sources of gravitational waves for future
space-based detectors such as the Laser Interferometer Space Antenna (LISA)~\citep{lisa1,lisa2,lisa3} 
or a similar mission led by ESA (eLISA/NGO, see~\citet{elisa,elisa_yellowbook})
or NASA (SGO, see~\citet{SGO}). These detectors are expected to be capable of observing 
tens or even hundreds of merger events during their lifetime, up to redshift $z\sim 10$ 
or larger~\citep{sesana_light_heavy_rates,sesana,lisa_taskforce_rates}, and measure
black-hole masses and spins with astonishing accuracy ($\sim 10^{-3}$ for the masses and $10^{-2}$ for 
the spins, see~\citet{berti_buonanno_will,harmonics,harmonics_err1,harmonics_err2,harmonics2}). Also, they should be able to tell a binary of black holes with aligned spins 
from one with misaligned (and therefore precessing) spins by looking at the higher-order harmonics of the gravitational waveforms~\citep{harmonics,harmonics_err1,harmonics_err2,harmonics2}.

Massive black holes are also expected to play a crucial role in galaxy formation: 
in fact, in Active Galactic Nuclei (AGNs) accretion onto the central MBH is
believed to power jets or disk winds capable of exerting a feedback on the growth of structures, by ejecting gas from 
the interstellar medium (ISM) and from the intergalactic medium (IGM)~\citep{reservoir,reservoir2,qso_feedback1,qso_feedback2,qso_feedback3},
therefore quenching star formation in low-redshift, massive galaxies.
Indeed, this feedback mechanism is a central ingredient of our current
understanding of galaxy formation, because it helps explaining 
why large dark-matter halos present low baryonic masses, resulting in a sharp cutoff at the high-mass end of the
stellar mass function that is not observed in the halo mass function \citep{smf_bell,benson_quenching}. 
Also, it helps making sense of the so-called ``anti-hierarchical'' evolution (or ``downsizing'')
of baryonic structures, i.e. the fact that the most massive galaxies are dominated by old stellar populations, while
low-mass galaxies generally present young stellar populations and longer-lasting star-formation activity~\citep{downsizing,downsizing2},
thus suggesting that massive galaxies assemble at higher redshift than
low-mass ones. Naively, this behavior may seem in contrast with the ``bottom-up''formation of dark-matter halos, 
which assemble hierarchically
by a series of mergers, but it can be reproduced, at least in its main features, by semianalytical galaxy-formation models
including the effect of AGN feedback~\citep{downsizing3}. 

This link between MBHs and the larger-scale galactic properties works also in 
the opposite direction, because it is the amount of cold gas
present in galactic nuclei that regulates accretion onto the MBHs, and therefore their mass and spin evolution. 
Also, the 100-pc scale circumbinary disks that are thought to form
after gas-rich mergers of galaxies~\citep{mayer07} exert torques on the spins of the MBHs, 
aligning them by the time the binary's separation has shrunk to $\sim 0.1$ pc~\citep{bardeen_petterson,tamara,perego,dotti,dotti2011},
and further spin alignment occurs from this separation to the merger as a result of Post-Newtonian resonances~\citep{resonances,kesden_berti_sperhake}.
Not only do these effects profoundly influence the spin and mass evolution -- because if the black-hole spins are aligned before the merger, the spin
of the final black hole is larger~\citep{FAUspin,RITspin,AEIspin1,AEIspin2,bkl,kesden} 
and its mass lower~\citep{FAUspin,AEImass,RITspin,kesden} -- 
but they also affect whether the MBH resulting from the
merger remains in the galaxy or gets ejected. In fact,
numerical-relativity simulations of black-hole binaries have shown that
mergers can produce final black holes with large kick
velocities relative to the center of mass of the initial binary configuration, due to anisotropic emission of gravitational
waves. In particular, for equal-mass configurations with initial black-hole spins lying on the orbital plane of the binary, the kick
velocity can be as large as 2500-4000 km/s~\citep{kickRIT,kickJena}, and velocities as high as 5000 km/s may occur
for configurations with off-equatorial spins~\citep{kickRIT2,kickRIT_massimo_marta}. Such velocities are larger than the typical galaxy escape velocities, thus leading to the
staggering conclusion that most galaxies may not host a MBH. This would be in stark contrast with observations that most galaxies do host
a MBH at low redshifts, and it may be bad news for hierarchical galaxy formation models, which as mentioned above heavily rely on the feedback from MBHs. 
However, for binaries
with aligned spins  the kick velocity is considerably lower and typically not sufficient
for ejecting the final black hole from the galaxy. Because, as mentioned above, 
circumbinary disks tend to align the black-hole spins prior to mergers, it
is clear that tracking the evolution of gas in galaxies is crucial to correctly
 predict how many galaxies host a MBH, let alone the MBH mass and spin
evolution. 

What makes galaxy formation a difficult problem is the huge range of scales that are involved, which go from the
Gpc scale of the present cosmological horizon and the Mpc scale of typical $z\sim 0$ galactic halos, through the 10-kpc scale of galactic disks and kpc
scale of galactic spheroids, down to the 100-pc scale of circumbinary disks and pc scale of MBH accretion disks, and finally to the $10^{-6}$--$10^{-7}$ pc
scale at which MBH mergers take place. This huge dynamical range, together with the complex nature of the processes, often nonlinear and dissipative,
that take place on small scales (``subgrid physics''), makes the problem very difficult to solve numerically in full generality.
In fact, while the current paradigm of cosmological structure formation (the $\Lambda$CDM model)
has enjoyed remarkable success in reproducing large-scale observations [such as the cosmic microwave
background fluctuations~\citep{wmap7,wmap_bare}; the large
scale clustering of galaxies (\citet{wmap_combo,BAO}, and references
therein); the cosmic shear field measured through weak gravitational
lensing (\citet{weak_lensing} and references therein); the small
scale power spectrum of Lyman-alpha forest sources~\citep{lyman_alpha}; the number density of galaxy clusters (\citet{clusters} and references
therein)], a global understanding of galaxy formation can presently be attempted only
by means of semi-analytical models~\citep{kauffmann93,cole94,cole_lacey,somerville99,somerville08,qso_feedback1,qso_feedback2,galform2010,morgana},
or more ambitiously with hydrodynamical simulations 
\citep{sph1,sph2,sph3,sph4,sph5,sph6,sph7,dotti,dotti2011,
2004astro.ph..3044O,2005ApJ...623..650K,2003ApJ...587...13T,2008ApJ...672..888T,2008A&A...477...79D,2011MNRAS.412.2154B,2011ApJ...729..125G,2011MNRAS.414.3656S,2009MNRAS.400..100S,2011MNRAS.414..195T}.

In this paper we use a semianalytical model for the formation of galaxies in a $\Lambda$CDM universe to study the evolution of the 
spins and masses of MBHs in a self-consistent way, taking into account both the feedback of the MBHs on
the growth of structures and the influence of the galactic nuclear gas on the MBH accretion history and on the spin-alignment prior to 
mergers (which in turns affects the spin evolution, 
as well as the kick velocities of MBHs and thus their possible ejection from galaxies).
For the purpose of this investigation, we adopt the 
widely accepted scenario in which the IGM collapses into disk structures, which give rise 
to bulges (``spheroids'') when disrupted by major (i.e. comparable-mass) galactic mergers, or when they become self-gravitating 
and develop bar-instabilities.
To describe the evolution of dark-matter halos, we use 
a full extended Press-Schechter merger tree based on \citet{parkinson08}, 
and then evolve the baryonic components along the merger-tree branches,
 employing analytical prescriptions at the 
``nodes'' of the tree to mimic the effects of the mergers, and taking into account environmental effects (tidal stripping, tidal evaporation and dynamical
friction) using the results of \citet{environment}.

The MBH are evolved within this model starting from two possible
scenarios for their seeds, namely a light-seed ($M_{\rm seed}\sim 150 M_\odot$) scenario -- in which the MBHs form 
as remnants of Pop III stars at $z\sim20$~\citep{madau_rees} --
and a heavy-seed ($M_{\rm seed}\sim 10^5 M_\odot$) scenario -- in which MBHs form  from
the collapse of massive protogalactic disks at $10\lesssim z \lesssim 15$~\citep{KBD,BVR,heavy3}.
The MBHs are then evolved along the ``branches'' of the merger tree,
using the ``reservoir'' model of~\citet{reservoir,reservoir2} to describe the gas present in the nuclear region,
and taking into account the feedback they exert on the growth of structure by means of accretion-powered jets. 
As for the accretion mechanism, 
we follow \citet{dotti_colpi,dotti2011} and assume that the nuclear cold gas accretes coherently (i.e., with a fixed angular momentum direction) 
onto the MBH in a gas-rich environment, thus resulting in a steady increase of the MBH spin~\citep{bardeen70,Thorne74}. In a
gas-poor environment, due to the absence of a rotationally supported structure, we assume that the MBH accretes chaotically 
(i.e. in lumps of material with essentially random orientations of the orbital angular momentum), which results 
on average in a decrease of its spin~\citep{chaotic}.
Similarly, prior to a  MBH merger, we assume that the MBH spins are aligned due to the gravito-magnetic torques exerted by the circumbinary disk if
the nuclear environment is gas rich (``wet merger''), whereas we assume that they are randomly oriented in a gas-poor 
nuclear environment (``dry merger'')~\citep{bardeen_petterson,tamara,perego,dotti,dotti2011}.

We stress this idealized picture will be at least partially
modified by processes such as star formation and feedback, which 
can inject energy and angular momentum into the circumnuclear disk
and potentially break the coherence of the accretion flow. However,
recent simulations by \citet{dotti2011}
show that these effects are probably not sufficient to create
strongly turbulent environments and completely randomize the orbits of the accreting gas in the case
of quasars fed by galaxy mergers,
and that in these systems the accretion flow retains, at least partially, its coherence.
The situation could be different for quasars fed by disk instabilities at high redshifts, where cold
flows are important and accretion
may take place by massive clumps (e.g. \citet{bournaud,dubois}). However, these clumps
are expected to infall from large-scale disks, and might retain at least partially the disk's
angular momentum down to the MBH.
In fact, observational hints for at least partial alignment 
come from the results of \citet{lagos2011} (see also \citet{lagos2009}), who studied
the relative orientation between AGNs and their host galaxies in the
Sloan Digital Sky Survey, finding that random orientations are ruled out at high confidence.

The effect of MBH mergers is accounted for by using analytical formulas 
reproducing the results of numerical-relativity simulations.
In fact, while the latter are the only way to study black-hole mergers rigorously, 
these simulations are very time-expensive and are not a viable option to cover the whole parameter space of black-hole binaries.
Recently, the interface between numerical and analytical relativity has produced a number of approaches which employ 
a combination of post-Newtonian theory, general-relativistic perturbation theory and/or fits to numerical data to reproduce different aspects of black-hole 
binaries, such as the
gravitational waveforms (see for instance the effective-one-body approach, 
e.g.~\citet{buonanno_damour99,DJS,BB_eob,pan_etal}, or the hybrid waveforms, e.g. \citet{hybrid}), 
the kick velocity~\citep{kick_goddard1,kick_goddard2,kick_goddard3,kickRIT,kickRITeta2,kickRIT2}, the final mass~\citep{FAUspin,RITspin,AEImass,kesden}
and the final spin~\citep{FAUspin,RITspin,AEIspin1,AEIspin2,bkl}. For this investigation, we use the formula of \citet{AEIspin2} for the spin,
the formulas of~\citet{FAUspin,AEImass} for the mass, and the formula of~\citet{kick_goddard3} for the kick velocity.

Because we track the evolution of baryonic structures, and in particular of the gas present
in galactic nuclei, along the dark-matter merger trees, we
can attempt to discriminate between chaotic or coherent accretion onto the MBHs,
and between aligned-spin and precessing-spin MBH mergers. This improves upon
previous models, e.g. \citet{berti_volonteri,volonteri05}, \citet{fanidakis1,fanidakis2} and \citet{lagos2009}, which considered 
\textit{either} chaotic \textit{or} coherent accretion,
and \textit{either} aligned \textit{or} misaligned mergers.\footnote{Indeed, \citet{berti_volonteri,volonteri05} did not describe the baryonic components and therefore
did  not try to distinguish between gas-poor and gas-rich nuclear environments.
On the contrary, \citet{fanidakis1,fanidakis2} and \citet{lagos2009} modelled the baryonic components in detail, but did not try to infer whether accretion is chaotic or coherent. Also, 
\citet{fanidakis1,fanidakis2} always considered randomly oriented spins prior to mergers.}
We stress that LISA/SGO or eLISA/NGO will be able to test
our model not only by measuring the MBH masses and spins as a function of redshift~\citep{sesana_light_heavy_rates,sesana}, but for each merger event it should also be able to 
determine whether the MBH binary has precessing or aligned spins, by looking at the higher order harmonics of 
the gravitational waveforms~\citep{harmonics,harmonics_err1,harmonics_err2,harmonics2}.

This paper is organized as follows. In Sec.~\ref{sec:model} we present our semianalytical 
galaxy-formation model in detail, focusing on dark matter in Sec.~\ref{sec:DM}, 
on the baryonic components in Sec.~\ref{sec:baryons}, and on merger and environmental 
effects in Sec.~\ref{sec:environment}. In Sec.~\ref{sec:calibration}
we calibrate the free parameters of our model to reproduce existing observations at $z=0$ and at $z>0$.
In Sec.~\ref{sec:predictions1} we present our predictions 
for the character of MBH mergers (i.e. whether they involve aligned or precessing spins), while
in Sec.~\ref{sec:predictions2} we study the evolution of the spins of MBHs as a function of redshift. In Sec.~\ref{sec:conclusion} we draw 
our conclusions and present plans for future work.

Throughout this paper, we assume a flat $\Lambda$CDM cosmology with
 $\Omega_{\rm DM}=0.227$, $\Omega_{b}=0.0456$, $H_0=70.4 {\rm\, km/(s\, Mpc)}$
and $\sigma_8=0.809$~\citep{wmap7,wmap_bare,wmap_combo}.

\section{Physical model}\label{sec:model}

\subsection{The dark-matter merger trees}
\label{sec:DM}

For the dark-matter merger and accretion evolution, we adopt
the extended Press-Schechter formalism 
of~\citet{cole_lacey}, as modified by \citet{parkinson08}. This
algorithm reproduces the statistical properties of the dark-matter
merger trees produced with cosmological numerical N-body simulations~\citep{millenium_run,cole07}.
More specifically, we start our merger trees at an initial redshift $z=20$
in the case of a light-seed MBH scenario (in which MBHs form as remnants of Pop III stars~\citep{madau_rees}),
while in the case of a heavy-seed scenario, in which MBHs form from the collapse of massive protogalactic disks~\citep{KBD,BVR,heavy3},
we start our merger trees at $z=15$. 
In the light-seed scenario, we assume that a halo of total mass $M_{\rm vir}$ forming at $15<z\leq20$ contains
a black-hole seed with mass $M_{\rm seed}=150 M_\odot$ if $M_{\rm vir}>1.1\times 10^7 h^{-1} M_\odot$.
This corresponds to populating halos 
collapsing from the large-$\sigma$ peaks of the primordial density
field~\citep{madau_rees,volonteri_haardt_madau}.\footnote{In our cosmology, a mass of $1.1\times 10^7 h^{-1} M_\odot$
corresponds to the cosmological Jeans mass collapsing
at $z=20$ from the $3.5\sigma$-peaks of the primordial density
field~\citep{volonteri_haardt_madau}.} In the heavy-seed scenario, instead,
we place  black-hole seeds with mass $M_{\rm seed}=10^5 M_\odot$
in halos forming at $10<z\leq15$ with virial mass~\citep{KBD,sesana_light_heavy_rates} 
\begin{equation}
M_{\rm vir}>10^7 
\left(\frac{1+z}{18} \right)^{-3/2} 
\left(\frac{\lambda}{0.04}\right)^{-3/2} M_\odot\,,
\end{equation}
where $\lambda$ is 
the halo spin parameter that we will introduce shortly.
In both scenarios, because little is known about the spin of the seeds,
we choose the spin parameter $a_{\rm bh}=cJ_{\rm bh}/(GM^2_{\rm bh})$ of each black-hole seed 
randomly from a uniform distribution
$-1\leq a_{\rm bh} \leq 1$. We stress, however, that the predictions of our model, and in particular
those regarding the spins of MBHs, are qualitatively independent of this assumption as long as ones looks
at MBHs of mass $M_{\rm bh}\gtrsim 3 M_{\rm seed}$. This is because a black hole loses memory
of its initial spin when it accretes a mass comparable to its own (i.e., if a black hole of mass $M_{\rm bh}$ accretes
coherently, which as we will show is the case at high redshifts, its spin becomes maximal after accreting a
mass $\lesssim 2 M_{\rm bh}$~\citep{bardeen70}).

Also, when it forms, a halo of total mass $M_{\rm vir}$ is assumed to contain
unprocessed hot gas (see Sec.~\ref{sec:baryons}) with mass 
$M_{\rm hot}=f_{\rm coll} M_{\rm vir}$, 
where the baryonic collapse fraction $f_{\rm coll}$ is given by
\begin{equation}\label{fcoll}
   f_{\rm coll}(M_{\rm vir},z) = \frac{f_b}{(1+0.26M_{f}(z)/M_{\rm vir})^{3}}\,,
\end{equation}
with $f_b=\Omega_b/\Omega_m\approx0.16$ (with $\Omega_m=\Omega_b+\Omega_{\rm DM}$).
The filtering mass as a function of redshift, $M_{f}(z)$, accounts for
 the effect of the ionizing extragalactic UV background produced by massive stars and quasars, which is able to partially
reduce the baryonic content in low-mass systems~\citep{reionization_collapse_suppression}.
In particular, we calculate $M_{f}(z)$ using the equations in Appendix B of \citet{filtering_mass},
assuming $z_{\rm overlap}=11$ and $z_{\rm reion}=10$~\citep{wmap_bare} ($z_{\rm overlap}$ and $z_{\rm reion}$ respectively correspond to the redshift at which multiple HII regions overlap, and to the redshift at which most of the medium is ionized).
After the initial formation redshift, 
additional hot gas is brought in from the IGM by the dark matter that accretes onto the halo, and we therefore assume
\begin{equation}\label{hot_gas_accr}
\dot{M}_{\rm inf}= f_{\rm coll} \dot{M}_{\rm vir}\,.
\end{equation}
where the baryonic collapse fraction $f_{\rm coll}$ is given again, as a function of redshift, by Eq.~\eqref{fcoll}.

The resolution $\Delta M$ of the merger trees (i.e., the mass scale below which 
 matter is assumed to accrete on an existing halo rather than give rise to a merger) is
chosen to keep the computational time to acceptable levels while following the
dark-matter halos (and the baryonic components within them) to very high redshifts.
In particular, 
we set 
$\Delta M=\min(10^{-3} M_0,10^{10} M_\odot)\times(1+z)^{-3.5}$, $M_0$ being the final mass of the halo at $z=0$.
The value of $\Delta M$ at $z=0$ is comparable or smaller than the resolutions typically
used by semianalytical galaxy formation models (see e.g. \citet{cole_lacey,somerville08}),
and the redshift dependence is introduced following \citet{volonteri_haardt_madau}
in order to track the merger tree to high redshifts. However,
because our simulations are computationally rather expensive (especially for large virial masses),
in order to further cap the computational time while ensuring a range of masses
that is sufficient to allow both minor and major mergers at all redshifts,
at each redshift step we also stop
following the branches that have mass smaller than
$\delta \times M_{\max}(z)$, where $\delta=0.01$ and $M_{\max}(z)$ is the mass
of the most massive halo at that redshift.

The virial radius $r_{\rm vir}$ of a halo is
related to its mass $M_{\rm vir}$ by the standard relation 
$M_{\rm vir}=4 \pi r_{\rm vir}^3 \rho_{\rm crit} \Delta_{\rm c}/3$,
where $\rho_{\rm crit}$ is the critical density, and where the 
density contrast at virialization, $\Delta_{\rm c}$, is calculated
following \citet{virial_radius}. The halo density is assumed to be
described by the fitting function of \citet{NFW} (NFW)
\begin{equation}\label{profile}
    \rho_{_{\rm NFW}}(r) = \rho_{\rm s} \left(\frac{r}{r_{\rm s}}\right)^{-1} \left(1+\frac{r}{r_{\rm s}}\right)^{-2}\,,
\end{equation}
truncated at the virial radius of the halo.
The scale radius $r_{\rm s}$ of the NFW profile is related to the virial radius by the so-called concentration parameter
$c(z) \equiv r_{\rm vir}/r_{\rm s}$. Imposing that the total halo mass equal $M_{\rm vir}$, one immediately obtains that
the scale density  $\rho_{\rm s}$ is given by
$\rho_{\rm s}=M_{\rm vir}(z) /[4\pi r_{s}^{3} f(c)]$,
with
\begin{equation}\label{fc}
    f(c) = \ln(1+c)-\frac{c}{1+c}\,.
\end{equation}

 The concentration parameter has been studied by several authors
\citep{bullock,wechsler,zhao1,zhao2,maccio},
and  found to present a large scatter for a fixed halo mass, but to scale
generally with the halo's main-progenitor history. Following  \citet{bullock} and \citet{wechsler}, 
we adopt here a scaling $c(z) \propto 1/(1+z)$ for the main-progenitor history after the halo formation:
\begin{equation}\label{c_mah}
c_{_{\rm MPH}}(z) = \max\left(\frac{c_0}{1+z},c_{\rm f}\right)\,,
\end{equation}
where the concentration at formation is $c_{\rm f}=4.1$~\citep{wechsler} and
the concentration at $z=0$ is set following \citet{maccio}:
\begin{equation}
\log_{10} c_0 = 1.071 - 0.098\, \left[\log_{10} \left({M_{\rm vir,0}\over
M_{\odot}}\right)-12\right]\,.
\end{equation}
The concentration of halos not belonging to the main-progenitor history, however, is not expected
to be given by Eq.~(\ref{c_mah}). Indeed, smaller halos are expected to be more concentrated (cf. \citet{zhao1,zhao2},
as well as the environmental effects described by \citet{bullock}). To account for this effect, we adopt the 
high-concentration limit of the expressions of \citet{zhao1}, which give $c\propto M_{\rm vir}^{(\alpha-1)/(3\alpha)}$  
(with $\alpha=0.48$) at a fixed redshift.
Combining this scaling with Eq.~(\ref{c_mah}) we obtain the expression for the concentration of
a halo of mass $M_{\rm vir}(z)$ at redshift $z$:
\begin{equation}\label{concentration}
c(z,M_{\rm vir}(z)) = \max\left[\frac{c_0}{1+z} \left(\frac{M_{\rm vir}(z)}{M_{\rm MPH}(z)}\right)^{(\alpha-1)/(3\alpha)},c_{\rm f}\right]\,.
\end{equation}
For simplicity, in this expression we assume that the main-progenitor history is given by $M_{\rm MPH}(z)=M_{\rm vir,0}\exp(-S a_f z)$, with
$S=2$ and $a_f=c_{\rm f}/c_0$~\citep{wechsler}.\footnote{Applying Eq.~(\ref{concentration}) 
using the main-progenitor history extracted from the 
merger tree under consideration would be more difficult to implement in our code,
and it is not clear that this procedure would be more accurate than the simple one 
that we use in this paper.}

The angular momentum of each halo is determined by the halo's \textit{spin parameter}, defined as
$\lambda = J_{\rm vir}E_{\rm vir}^{1/2}M_{\rm vir}^{-5/2}G^{-1}$, where $E_{\rm vir}$ and $J_{\rm vir}$ 
are the total energy and angular momentum of the halo. We assign a spin
parameter to a halo with no progenitors drawing from a log-normal distribution
with median value  $\bar{\lambda} = 0.039$ and standard deviation 
$\sigma= <\sqrt{({\ln\lambda}-\ln\bar{\lambda})^2}> =0.53$~\citep{spin_parameter,cole_lacey}. We then assume
that this spin parameter remains unchanged along the cosmic history of the halo, except if it experiences
a merger with a second halo of comparable mass (i.e. if $M_{\rm vir\,2}/M_{\rm vir\,1}>0.3$), in which
case we randomize the spin of the resulting halo by drawing from the same log-normal distribution. 
(We will discuss this in more detail in Sec.~\ref{sec:environment}).

\subsection{The baryonic matter}
\label{sec:baryons}
\subsubsection{The hot gas phase}
\label{sec:hot}
We assume that the hot unprocessed gas phase is isothermal at the halo's virial temperature $T_{\rm vir}$,  and 
in hydrostatic equilibrium within the NFW profile, such that
\begin{equation}
    \rho_{\rm hot}(r) = \rho_{0} \exp \left[ -\frac{27}{2} \beta \left\{ 1-\frac{\ln(1+r/r_{s})}{r/r_{s}}   \right\}   \right],
\end{equation}
with
\begin{equation}
   \beta = \frac{ 8 \pi \mu m_{p} G \rho_{s} r_{s}^{2}}{27 k_{B} T_{\rm vir}}\,,
\end{equation}
where $m_p$ is the proton mass, $\mu$ is the mean molecular mass and $\rho_{0}$ is calculated by normalizing 
to the total hot-gas mass at the redshift under consideration. 
The hot gas cools on a timescale
$t_{\rm cool}$ given, at each point of its density distribution, by the standard expression
\begin{equation}\label{coolingtime}
    t_{\rm cool}(r)=\frac{3 \rho_{\rm hot}(r)k_{B}T_{\rm vir}}{2 \mu m_{p} n_{e}(r)^{2}\Lambda(T_{\rm vir},Z)},
\end{equation}
where $n_e(r)$ is the electron number density (which is proportional to $\rho_{\rm hot}(r)$), $\Lambda(T,Z)$ is 
the cooling function given by the tabulated results of \citet{cooling_function}, 
and where we assume that the hot gas is unprocessed and therefore has primordial metallicity $Z=10^{-3}Z_{\odot}$.

If the cooling time of the hot gas is shorter than its free-fall time, 
given approximately, at each radius, by
\begin{equation}\label{ff_time}
t_{\rm dyn}(r)=\sqrt{\frac{3 \pi}{32 G \bar{\rho}_{_{\rm NFW}}(r)}}
\end{equation}
(where $\bar{\rho}_{_{\rm NFW}}(r)$ is the NFW average density within a radius $r$),
then the hot-gas phase cools ``fast'' and undergoes gravitational collapse. If instead the cooling time
is longer than the free-fall time, then the cooling is ``slow'' and the hot gas cools down through a sequence of
quasi-hydrostatic equilibrium states. We therefore assume that the hot-gas phase
is transferred into a cold-gas phase with rate
\begin{equation}\label{hot_gas_coll}
    \dot{M}_{\rm coll}(z) = 4 \pi \int_{0}^{r_{\rm vir}(z)} \frac{r^{2} \rho_{\rm hot}(r,z)}{t_{\rm coll}(r,z)} {\rm d} r\,,
\end{equation}
where $t_{\rm coll}(r,z)=\max(t_{\rm cool}(r,z),t_{\rm dyn}(r,z))$. 

However, on top of this ``classical'' picture employed already by early 
semianalytical galaxy-formation 
models (see e.g.~\citet{kauffmann93,cole94,cole_lacey,somerville99}),
we also include the effects of cold accretion flows~\citep{cold_accr1,cold_accr2,cold_accr3}, which have been
shown to be the predominant mechanism leading to the formation of low-mass systems. In halos with mass lower than the
critical mass
\begin{equation}
   M_{\rm c} = M_{\rm shock} \max[1, 10^{1.3(z-z_{\rm c})}]\,,
\end{equation}
where $M_{\rm shock} =2\times10^{12} M_{\odot}$ and $z_{\rm c}=3.2$, 
we assume that all the gas accreted from the IGM is \textit{not} shock heated 
to the halo's virial temperature, but streams in on a dynamical time~\citep{cold_accr1,cold_accr2,cold_accr3}, thus
enhancing star 
formation at high redshifts relative to the scenario where the accreting gas is shock heated.
From the point of view
of our model, this is equivalent to assuming
directly $t_{\rm coll} = t_{\rm dyn}$ in (\ref{hot_gas_coll}), 
for halos with $M_{\rm vir}< M_{\rm c}$.
Finally, in order to mimic the effect of ram pressure~\citep{rampressure} and clumpy accretion~\citep{clumpy,clumpy2}, 
which are expected to quench the cooling
of the hot gas at low redshifts in large halos, we set  $t_{\rm coll}$ to the Hubble time 
in halos with $M_{\rm vir}(z)>10^{13} M_\odot$ and $z<2$.

\subsubsection{Density profile of the baryonic structures}

For the growth of the baryonic structures, we
here adopt the widely accepted scenario in which the hot-gas collapse gives rise
to a disk of cold gas with mass $M_{d,{\rm gas}}$, where star formation may occur and result
in the formation of a stellar disk with mass $M_{d,{\rm stars}}$. These disks may 
be disrupted by major galactic mergers and bar instabilities, thus
forming a gaseous bulge with mass $M_{b,{\rm gas}}$
and a stellar bulge with mass $M_{b,{\rm stars}}$.
Stellar formation takes place in the gaseous bulge as well,
further contributing to $M_{b,{\rm stars}}$. 
Also, as we will explain below, we allow
part of the gaseous bulge to flow into a disk-like reservoir with mass $M_{\rm res}$, which feeds the 
central MBHs during the accretion phase,
and which forms the circumbinary disk expected to surround black-hole binaries after galactic mergers.

Assuming a dissipationless collapse of the hot gas into the gaseous disk, 
the angular momentum conservation allows one to relate the halo's virial radius and spin parameter 
to the radius of the disk.
In particular, adopting an exponential surface-density profile for the gaseous disk,
\begin{equation}
    \Sigma_{\rm d}(r,z) = \Sigma_{0}(z) {\rm e}^{-r/r^{\rm gas}_{d}(z)}\,,
\end{equation}
the scale radius is given by \citep{mo_mao_white}
\begin{equation}
r^{\rm gas}_{d}(z) = \frac{\lambda}{\sqrt{2}} \frac{j_{d}}{m_{d}}  \frac{r_{\rm vir}(z)}{\sqrt{f_c}} f_r(\lambda, c, m_{d}, j_{d})\,,
\end{equation}
with
\begin{equation}
f_c={c \over 2}~{1-1/(1+c)^2-2\ln(1+c)/(1+c) \over [c/(1+c)-\ln(1+c)]^2}\,
\end{equation}
and
\begin{equation}\label{f_func}
    f_r(\lambda, c, m_{d}, j_{d}) = 2 \left[ \int_{0}^{ \infty } e^{-u}u^{2}\frac{V_{c}(r_{d}u)}{V_{c}(r_{\rm vir})}\de u \right]^{-1}\,,
\end{equation}
where $V_c(r)$ is the velocity profile of the composite system 
(bulges, reservoir, disks, hot gas and dark matter) and where $m_d$ and $j_d$ are the ratios between 
the total mass and angular momentum of the disk and those of the halo. 
More specifically, we take $m_d = (M_{\rm d}^{\rm stars}+M_{\rm d}^{\rm gas})/M_{\rm vir}$ 
and assume $j_d=m_d$~\citep{mo_mao_white}.

In the calculation of the disk's scale radius, we also account for the adiabatic halo response, which affects the
velocity profile $V_c$ entering Eq.~(\ref{f_func}), using the standard prescription of \citet{blumenthal}.
In particular, the angular momentum conservation implies that the  halo contraction following
 the baryonic collapse is described by
\begin{equation}\label{contraction1}
   M_{i}(r_{i})r_{i} = M_{f}(r_{f})r_{f}\,,
\end{equation}
where $r_i$ and $r_f$ are the initial and final radius of the shell under consideration;
$M_{i}(r_{i})$ is the mass of the dark matter and hot gas contained in a radius $r_i$, calculated with
the initial mass distribution (which is assumed to be given by the NFW density profile for both the 
dark matter and the hot gas);
and $M_{f}(r_{f})$  is the mass of the composite system (dark matter, hot gas, disks, bulges and reservoir) contained in
the radius $r_f$.
Moreover, the mass conservation ensures 
\begin{align}\label{contraction2}
  & M_{f}(r_{f}) = \nonumber \\&M_{d}(r_{f})+M_{b}(r_f) +M_{\rm res}(r_{f})+M_{\rm hot}(r_f)+M_{\rm DM}(r_{f})=\nonumber \\
  & M_{d}(r_{f})+M_{b}(r_f) +M_{\rm res}(r_{f}) + (1-f_{\rm gal})M_{i}(r_{i})\,,
\end{align}
where $M_X(r)$ is the mass of component $X$ within a radius $r$, and where 
$f_{\rm gal}=M_{\rm gal}/M_{\rm vir}$ (with $M_{\rm gal}=M_d+M_b+M_{\rm res}$, $M_d$ and $M_b$ being the total, i.e. gaseous and stellar, disk
and bulge masses). 
Assuming spherical collapse without shell crossing, we can adopt the ansatz 
$r_{f}=\Gamma r_{i}$, where $\Gamma=\mbox{const}$ is the contraction factor~\citep{blumenthal}.
Eqs.~(\ref{contraction1}) and~(\ref{contraction2}) can then be solved numerically for $\Gamma$, which
then allows one to include the effect of the adiabatic contraction following the baryon collapse in the calculation
of the velocity profile $V_c$ of the composite system.

For the stellar disk, we assume an exponential surface-density profile with scale radius $r^{\rm stars}_d=r_d^{\rm gas}/2$ 
following \citet{somerville08} and \citet{avila_reese} (see also \citet{dutton_sfr}, who show that
the gaseous disk is theoretically expected to have a larger scale radius than the stellar disk).
As for the gaseous and stellar bulges,  
we assume that they are described by the Hernquist profile~\citep{hernquist}
\begin{equation}\label{bulge_density}
    \rho_{b}^{*}(r)= \frac{M^{*}_{b}}{2 \pi}\frac{r_{b}}{r(r+r_{b})^{3}}\,,\quad *=\mbox{stars, gas}\,,
\end{equation}
where the scale radius $r_b$ (which we assume to be the same for the stellar and gaseous components) 
is related to the half-light radius $R_{\rm eff}$  by $r_{b}=R_{\rm eff}/1.8153$~\citep{hernquist}. 
Using the fit of \citet{shen_fit} for $R_{\rm eff}$ , 
\[  {\log_{10}(R_{\rm eff}/\mbox{kpc})= }  \left\{
\begin{array}{ll}
    & -5.54 + 0.56\log_{10}\left(\frac{M_{b}}{M_\odot}\right)       \\
& \qquad\mbox{   for $\log_{10}\left(\frac{M_{b}}{M_\odot}\right) > 10.3$} \\
&\\
  &  -1.21 + 0.14 \log_{10}\left(\frac{M_{b}}{M_\odot}\right)    \\& \qquad\mbox{   for $\log_{10}\left(\frac{M_{b}}{M_\odot}\right) \leq 10.3$}
\end{array}\right. \]
we can express the scale radius in terms of the total bulge mass
$M_b=M_{b,{\rm stars}}+M_{b,{\rm gas}}$.

For simplicity\footnote{The reservoir's geometry 
is needed for instance to calculate the velocity  $V_c$ of the composite system, entering e.g. in
Eq.~(\ref{f_func}), in the calculation of the adiabatic halo contraction factor $\Gamma$ [Eqs.~(\ref{contraction1}) and~(\ref{contraction2})],
and in that
the total gravitational potential $\phi$
appearing in Eqs.~(\ref{snfb2}) and~(\ref{snfb1}). However, the specific choice of the reservoir's
geometry  does not impact our final results significantly, because of its small size relative 
to the other components.}, 
we also assume that the reservoir can be described by an exponential surface-density profile
with scale radius $r_{\rm res}$ proportional to the influence radius of the central MBH, i.e.
$r_{\rm res}=\alpha G M_{_{\rm bh}}/V_{\rm vir}^2$, with $\alpha\approx100$.

\subsubsection{Star formation, supernova feedback and disk instabilities}
\label{star_formation_section}

Assuming that star formation in galactic disks only happens 
inside dense molecular clouds, which are well traced by the HCN luminosity~\citep{hcn1,hcn2}, we 
follow \citet{blitz,dutton_sfr} and
take the disk star-formation rate (SFR) to be proportional to the molecular-cloud surface density as traced by HCN, $\Sigma_{\rm mol, HCN}$:
\begin{equation}\label{disk_sfr}
   \dot{\Sigma}_{\rm sfr} = \tilde{\epsilon}_{\rm sf} \Sigma_{\rm mol, HCN}\,,
\end{equation}
where $\tilde{\epsilon}_{\rm sf}=13 \mbox{Gyr}^{-1}$. From this expression one easily obtains the total SFR 
by integrating over the surface of the gaseous disk.

More specifically, we write $\Sigma_{\rm mol,HCN}$ as the product of 
the HCN fraction $R_{\rm HCN}=\Sigma_{\rm mol,HCN}/\Sigma_{\rm mol}$ 
with the total molecular surface density of the disk, $\Sigma_{\rm mol}$, which in turn
we write as $\Sigma_{\rm mol}=f_{\rm mol} \Sigma_{\rm d,gas}$, $f_{\rm mol}$
being the molecular fraction of the disk's gas.
For $R_{\rm HCN}$ we use the fitting relation of \citet{blitz},
\begin{equation}
R_{\rm HCN} = 0.1\times(1+\Sigma_{\rm mol}/(200 M_{\odot}{\rm pc}^{-2}))^{0.4}\,,
\end{equation}
and we relate $f_{\rm mol}=R_{\rm mol}/(R_{\rm mol}+1)$ (with $R_{\rm mol}=\Sigma_{\rm mol}/\Sigma_{\rm atom}$) 
to the mid-plane pressure $P_{\rm mp}$ of the disk, following again
\citet{blitz}:
\begin{equation}
R_{\rm mol} = \left(\frac{P_{\rm mp}/k_B}{4.3\times10^{4}} \right)^{0.92}\,,
\end{equation}
where the pressure and the Boltzmann constant are in CGS units.
For the mid-plane pressure of the disk, we assume~\citep{midplane_pressure,blitz,dutton_sfr}
\begin{equation}
P_{\rm mp} = \frac{\pi}{2}G \Sigma_{g} \left(   \Sigma_{g}  + ( \sigma_{g}/\sigma_{s} ) \Sigma_{s} \right)\,,
\end{equation}
where we use $\sigma_{g}/\sigma_{s} = 0.1$~\citep{dutton_sfr}.
 
We stress that in high-mass (and thus high-density) galaxies, 
where the molecular fraction $f_{\rm mol} \approx 1$, this star-formation prescription reduces to the 
standard Schmidt-Kennicutt star formation power law~\citep{kennicutt}, 
$\dot{\Sigma}_{\rm sfr}= {\epsilon}_{\rm sf} [\Sigma_{d,{\rm gas}}/(M_\odot {\rm pc}^{-2})]^{n}$ with $n=1.4$ and $\epsilon_{\rm sf}=1.6\times10^{-4} M_{\odot} {\rm kpc}^{-2} {\rm yr}^{-1}$, 
whereas in low-density systems the exponent $n$ approaches $2.84$. As a result, the star formation law that we adopt
suppresses star formation in low-mass galaxies, in accordance with observations 
(see \citet{dutton_sfr,blitz} for a detailed discussion).

The feedback from supernova events is expected to eject cold gas from the disk. To do so, 
the energy released by supernova explosions in the disk must be sufficient to unbind the cold gas. We therefore 
compare, at each radius, the amount of energy 
released by  these explosions with the
binding energy, and write the total amount of cold gas ejected from the system as
\begin{equation}
     \dot{M}^d_{\rm SN}(z) = 2 \pi \int_{0}^{r_{\rm vir}} r \dot{\Sigma}_{\rm SN}(r,z) \de r\,,
\end{equation}
where 
\begin{equation}\label{snfb2}
        \dot{\Sigma}_{\rm SN}(r,z) = -\frac{\epsilon_{\rm SN} E_{\rm SN} \eta_{\rm SN} \dot{\Sigma}_{\rm sfr}(r,z)}{\phi(r,z)}\,.
\end{equation}
Here, $\phi(r,z)$ is the binding energy per unit mass of the composite system (bulges, disks, reservoir, hot gas and dark matter),
$\eta_{\rm SN} $ is the number of Type II supernovae expected per solar mass of stars 
formed\footnote{Following \citet{romano}, we adopt a \citet{chabrier} initial mass function (IMF) between $0.001 M_\odot$ and 1 $M_\odot$,
and a power law ${\rm d} N/{\rm d} m_{\star}\propto m_{\star}^{-2.7}$, therefore steeper than the standard Salpeter IMF, between  1 $M_\odot$ and
 100 $M_\odot$. This IMF gives $\eta_{\rm SN} =5\times10^{-3} M_\odot^{-1}$.},
$E_{\rm SN} =10^{44}$ J is the kinetic energy released per supernova event, and $\epsilon_{\rm SN} $ is a parameter ranging from 0 to 1 and regulating the efficiency with which 
the supernova energy is transferred to the cold gas. 
We stress that the supernova feedback is most effective in low-mass systems, which present shallower potential 
wells from which the cold gas can easily escape due to supernova explosions.

Also, disks are known to develop bar instabilities when they become self-gravitating, thus
getting disrupted and transferring  material to the bulge components~\citep{bar_instability,bar_instability2}.
We assume that a stellar or gaseous disk is stable if, respectively,
\begin{equation}\label{stability}
\frac{V_{\rm c}(2.2 r_d)}{(GM^*_{d}/r^*_d)^{1/2}} > \alpha_{\rm crit}^{*}\,\quad *={\rm stars\,, gas}\,,
\end{equation}
where $\alpha^{\rm stars}_{\rm crit}=1.1$ and $\alpha^{\rm gas}_{\rm crit}=0.9$~\citep{bar_instability,bar_instability2}. 
If the disk becomes unstable, we assume that it gets disrupted 
in a dynamical time and transfers its material (either stars or gas) to the corresponding bulge component.

As for the gaseous bulge, we assume that star formation takes place on a dynamical timescale. More specifically, we assume that 
the SFR per spherical shell in the gaseous bulge can be calculated as
\begin{equation}\label{bulge_sfr}
   \frac{\de{\psi}_{b}}{\de r}(r,t)= 4\pi r^2 \frac{\rho_{b,{\rm gas}}(r)}{t_{\rm gas}(r)}\,,
\end{equation}
where $t_{\rm gas}(r)=\sqrt{3\pi/(32 G \rho_{b,{\rm gas}})}$ is the local dynamical time for the bulge gas. 
Eq.~(\ref{bulge_sfr}) can then be integrated over all radii to give the total SFR in the bulge.
As in the disk case, we assume that the star formation ejects cold gas as a result of supernova explosions, at a rate
\begin{equation}\label{snfb1}
    \dot{M}_{\rm b,gas}^{\rm SN} (t) = -\int\frac{\epsilon_{\rm SN} E_{\rm SN} \eta_{\rm SN} \de\psi_{b}(r,t)/\de r}{\phi(r,t)}\mbox{d}r\,,
\end{equation}
where the efficiency $\epsilon_{\rm SN} $ is assumed to be the same as for the disk. Again, this feedback mechanism is 
most effective in low-mass systems.

Finally we note that the SFR rates (\ref{disk_sfr}) and  (\ref{bulge_sfr}) do not account for the
mass returned to the cold-gas phase by short-lived stars in the form of processed material. 
To include this effect, we use the instant-recycling approximation
and assume that a fraction $R=0.29$ of the mass is instantly returned into the cold-gas phase.\footnote{To calculate this return rate, we 
follow again  \citet{romano} and adopt a \citet{chabrier} IMF between $0.001 M_\odot$ and 1 $M_\odot$,
and a power law ${\rm d} N/{\rm d} m_{\star}\propto m_{\star}^{-2.7}$  between  1 $M_\odot$ and
 100 $M_\odot$.}
In particular, this implies that the
effective SFRs regulating the evolution of the disks and bulges are
\begin{gather}
\dot{M}_b^{\rm SFR}(t)= (1-R) \int\frac{\de\psi_b}{\de r} (r,t) \de r\,,\\
\dot{M}_d^{\rm SFR}(t)= (1-R) \int 2 \dot{\Sigma}_{\rm sfr} (r,t) \pi r  \de r\,.
\end{gather}

\subsubsection{The evolution of MBHs}
\label{smbh_evolution}

Star formation in the bulge is believed to force, e.g. by radiation drag~\citep{rd1,rd2,rd3}, 
part of the bulge's cold gas onto a low angular momentum circumnuclear reservoir, which feeds the 
central MBHs during the accretion phases~\citep{haiman},
and which may be identified with the circumbinary disks expected to surround MBH binaries after galactic mergers. 
Because star formation in the bulge happens in violent bursts triggered by disk instabilities (see previous section) or
by galaxy mergers (as we will explain in the next section), we identify this accretion mode with the quasar mode of MBHs.

In particular, we assume that the growth rate of the reservoir is proportional to the bulge SFR~\citep{reservoir,reservoir2} and is given by
\begin{equation}\label{mdot_res}
\dot{M}_{\rm res}= A_{\rm res}\psi_{b}(t)\,,
\end{equation}
where $A_{\rm res}$ is a free parameter of our model.
The cold gas in this reservoir then becomes available to accrete onto the central MBH at a rate
\begin{equation}\label{tacrr}
    \dot{M}_{\rm QSO} =  \frac{M_{\rm res}}{t_{\rm accr}}\,,
\end{equation}
where the timescale $t_{\rm accr}$ is a free parameter regulating the infall of the reservoir gas into the
(pc-scale) MBH accretion disk. The accretion of this gas 
then changes the MBH mass according to
\begin{equation}\label{tacrr_bis}
    \dot{M}_{\rm bh,QSO} =  \dot{M}_{\rm QSO} (1-\eta(a_{\rm bh}))\,,
\end{equation}
where the
spin-dependent efficiency $\eta(a_{\rm bh})$
measures the energy emitted  in electromagnetic radiation by the accretion disk.
More specifically, 
we follow \citet{dotti_colpi,dotti2011} and assume that the accretion onto the MBH takes place 
coherently (i.e. with a fixed angular momentum direction, cf. \citet{bardeen70,Thorne74}) 
 in a gas-rich environment, 
where gravito-magnetic torques rapidly align the 
the disk's angular momentum with the spin of the MBH.~\footnote{This is
the so-called Bardeen-Petterson effect~\citep{bardeen_petterson}, which also plays a fundamental role during gas-rich MBH mergers,
as we will explain shortly.}
(As mentioned in the introduction, this idealized coherent accretion flow may be at least partially randomized
by star formation and feedback effects in the circumnuclear disk, or by the formation of clumps in 
high-redshift disk galaxies.)
In a gas-poor environment, 
due to the absence of a rotationally supported structure, we assume that the MBH accretes chaotically~\citep{dotti_colpi} 
(i.e. in lumps of material with essentially random orientations of the orbital angular momentum, 
cf.~\citet{chaotic}).
Identifying a
gas-rich environment with
one where $M_{\rm res}> M_{\rm bh}$, we assume that the efficiency is
\begin{equation}
\eta(a_{\rm bh})= 1-E_{_{\rm ISCO}}(a_{\rm bh}) 
\end{equation}
for $M_{\rm res}> M_{\rm bh}$, with
${E}_{\rm_{\rm ISCO}}$ the specific energy at the anti-clockwise innermost stable circular orbit (ISCO) 
around
a Kerr black hole with spin parameter $a_{\rm bh}$ ranging between $-1$ (extremal spin
pointing downwards) and $1$ (extremal spin
pointing upwards)~\citep{bardeen70}.
Identifying instead a
gas-poor environment with
one where $M_{\rm res}< M_{\rm bh}$, we assume
\begin{equation}
\eta(a_{\rm bh})= 1-\frac{E_{_{\rm ISCO}}(a_{\rm bh})+E_{_{\rm ISCO}}(-a_{\rm bh})}{2}
\end{equation}
for $M_{\rm res}< M_{\rm bh}$, where
we have assumed that accretion has equal probability of happening in the
clockwise or anti-clockwise directions. (This is a simplified
version of the prescription derived by~\citet{king_alignment}.)

Because of the energy and angular momentum carried by the gas accreting onto the MBH, 
in a gas-rich environment ($M_{\rm res}> M_{\rm bh}$)
the spin parameter  $a_{\rm bh}$ increases steadily under coherent accretion:
\begin{equation}
 \dot{a}^{\rm coherent}_{\rm bh,QSO} = [{L}_{\rm_{\rm ISCO}}(a_{\rm bh})-2 a_{\rm bh} 
{E}_{\rm_{\rm ISCO}}(a_{\rm bh})]
\frac{ \dot{M}_{\rm QSO}}{{M}_{\rm bh}}\,,
\end{equation}
where again ${E}_{\rm_{\rm ISCO}}$ and ${L}_{\rm_{\rm ISCO}}$ 
are the specific energy and angular momentum at the anti-clockwise ISCO around
a Kerr black hole with spin parameter $a_{\rm bh}$ ranging between $-1$ (extremal spin
pointing downwards) and $1$ (extremal spin
pointing upwards). In a gas-poor environment ($M_{\rm res}< M_{\rm bh}$), we again assume that
accretion can happen clockwise or anticlockwise with equal probabilities, and the spin parameter decreases
(on average) under chaotic accretion, following
\begin{multline}
 \dot{a}^{\rm chaotic}_{\rm bh,QSO} = \Bigg\{
\frac{{L}_{\rm_{\rm ISCO}}(a_{\rm bh})+{L}_{\rm_{\rm ISCO}}(-a_{\rm bh})}{2}\\- a_{\rm bh} 
[{E}_{\rm_{\rm ISCO}}(a_{\rm bh})+{E}_{\rm_{\rm ISCO}}(-a_{\rm bh})]\Bigg\}
\frac{ \dot{M}_{\rm QSO}}{{M}_{\rm bh}}\,.\label{adot_chaotic}
\end{multline}

We stress that we do \textit{not} restrict the MBH accretion rate to values lower than the
Eddington rate, i.e. we allow super-Eddington \textit{mass} accretion. However, following the theory of 
thin accretion disks (\citet{shakura_sunyaev}; see also \citet{super_eddington}), we assume the MBH's bolometric luminosity to be
\begin{multline}
    {L}_{\rm bh,QSO} = \min\left\{\eta(a_{\rm bh})\dot{M}_{\rm QSO} c^2,\right.\\\left.L_{\rm Edd} \left[1+\ln\left(\frac{\eta(a_{\rm bh})\dot{M}_{\rm QSO} c^2}{L_{\rm Edd}}\right)\right]\right\}\,.
\end{multline}
Also, thin-disk accretion is believed to produce jet outflows due to the Blandford-Znajeck effect~\citep{BZ}, and the jet power is parameterized
by~\citep{meier}
\begin{eqnarray} \label{eq:ljet_sda}
L_{\rm jet,QSO} &\approx &f_{\rm jet} \times 10^{42.7} {\rm erg}\ {\rm s}^{-1} 
\left(\frac{\alpha}{0.01} \right)^{-0.1} m_9^{0.9}
\left(\frac{\dot{m}}{0.1}\right)^{6/5} \nonumber \\
    &\times& (1 + 1.1 a_{\rm bh} + 0.29 a_{\rm bh}^2), 
  \label{SD}
\end{eqnarray}
where $\alpha$ is the disk's viscosity parameter (for which we assume $\alpha=0.1$), 
$m_9=M_{\rm bh}/(10^9 M_{\odot})$, 
$\dot{m} = \dot{M}_{\rm QSO}/ (22 \,  m_9 \,  M_\odot \,  {\rm yr}^{-1})$, and where $f_{\rm jet}$ is a ``fudge'' factor
parameterizing the uncertainties affecting Eq.~\eqref{eq:ljet_sda} (\textit{e.g.,} the Blandford-Znajeck jet-outflow power 
scales quadratically with the magnetic field, which is poorly known). These jets are expected to exert a feedback on the hot-gas phase
and on the bulge gaseous component. More specifically, we assume that they eject hot gas and bulge cold gas 
from the system with rates~\citep{reservoir}
\begin{eqnarray}
            \dot{M}_{\rm b,gas}^{QSO}&=&\frac{2}{3} \frac{L_{\rm jet,QSO}}{\sigma^{2}} \frac{M_{\rm b,gas}}{M_{\rm hot} + M_{\rm b,gas}}\,,\\
            \dot{M}_{\rm hot}^{QSO}&=&\frac{2}{3} \frac{L_{\rm jet,QSO}}{\sigma^{2}} \frac{M_{\rm hot}}{M_{\rm hot} + M_{\rm b,gas}}\,.
\end{eqnarray}
with $\sigma=\,0.65 V_{\rm vir}$. 

In addition to the thin-disk quasar accretion mode, MBHs are expected to quiescently accrete matter 
directly from the hot-gas phase, when that is in quasi-hydrostatic equilibrium, through a thick advection-dominated
accretion flow (ADAF). This
is usually dubbed ``radio accretion mode'' 
and does not contribute significantly to the mass evolution of MBHs, because the accretion rate is much smaller than for the quasar mode.
However, the radio mode is believed to play an important role in galaxy formation
because it produces jet outflows that are much more powerful than those that 
would be produced by a thin disk with the same accretion rate~\citep{qso_feedback1,qso_feedback2,meier}. This
compensates the smaller mass accretion rate of the radio mode relative to the quasar mode, and enhances its feedback. 
Moreover, while the quasar feedback is triggered by starbursts in the bulge, and therefore
 intermittent and important mostly at relatively high redshifts,
the radio-mode accretion and feedback are continuous and remain efficient up to $z=0$.

We therefore assume that 
when $t_{\rm cool}>t_{\rm ff}$ (i.e., when the hot gas undergoes quasi-hydrostatic cooling, cf. Sec.~\ref{sec:hot})
the MBHs accrete directly from the hot-gas phase at the Bondi accretion rate~\citep{bondi}
\begin{equation}
 \dot{M}_{\rm bh,radio} = 4\pi\lambda_{B}\rho_{\rm hot}(G M_{\rm bh})^2/v_{\rm s}^3\,,
\end{equation}
where $\rho_{\rm hot}$ is the density of the hot gas in the center of the galaxy; $v_{\rm s}$ is the sound
velocity in the hot gas, which is of the order of the virial velocity $V_{\rm vir}$; and $\lambda_{B}$ is a constant
that depends on the adiabatic index of the gas, with $\lambda_{B}\approx 1.12$ for an isothermal gas.
The bolometric luminosity is then given~\citep{adaf_luminosity} by the ADAF luminosity
\begin{equation}
L_{\rm bol,radio}=1.3\times 10^{38}\left(\frac{M_{\rm bh}}{M_{\odot}}\right)
\left(\frac{\dot{m}^2}{\alpha^2}\right)\left(\frac{\beta}{0.5}\right)\;\mbox{erg~s$^{-1}$},
\label{disc_bol_adaf}
\end{equation}
where 
$\dot{m} = \dot{M}_{\rm bh,radio}/ (22 \,  m_9 \,  M_\odot \,  {\rm yr}^{-1})$
and $\beta$ is related to viscosity parameter
$\alpha=0.1$ by $\alpha\approx0.55(1-\beta)$~\citep{fanidakis2}. 
Because the radio-mode accretion happens through an ADAF and not through a thin disk, the rate of change of the
spin parameter is different from the quasar-mode case. More specifically, if we assume that the accretion is geometrically spherical, no angular 
momentum is transferred to the black hole and the spin parameter decreases due to the mass increase:
\begin{equation}
 \dot{a}_{\rm bh,radio} = -2 a_{\rm bh} \frac{ \dot{M}_{\rm bh,radio}}{{M}_{\rm bh}}\,.
\end{equation}
Finally, ADAF accretion is expected to produce much more powerful jets than the quasar mode, and the jet power is parameterized by~\citep{meier}
\begin{eqnarray} \label{eq:ljet_riafa}
L_{\rm jet}^{\rm radio} &\approx& f_{\rm jet}\times 10^{45.1} {\rm erg}\ {\rm s}^{-1}
\left(\frac{\alpha}{0.3}\right)^{-1} m_9 
\left(\frac{\dot{m}}{0.1}\right) g^2 \nonumber \\
&\times& (0.55 f^2 + 1.5 f a_{\rm bh} + a_{\rm bh}^2), 
  \label{RIAF}
\end{eqnarray}
where again $m_9=M_{\rm bh}/(10^9 M_{\odot})$ and
$\dot{m} = \dot{M}_{\rm bh, radio}/ (22 \,  m_9 \,  M_\odot \,  {\rm yr}^{-1})$,
and where $f$ and $g$ are dimensionless quantities, defined precisely in \citet{meier}, regulating the
actual angular velocity and azimuthal magnetic field of the system.
Following \citet{meier} we set $f = 1$ and $g = 2.3$, but we include a ``fudge'' factor $f_{\rm jet}$ 
[assumed to be the same as in Eq.~(\ref{eq:ljet_sda})] to account 
for the uncertainties in Eq.~(\ref{eq:ljet_riafa}) (e.g. the uncertainties in the magnetic field, on which
the jet power depends quadratically, and the higher-order terms in the black-hole spin, which are neglected in the
standard Blandford-Znajeck calculation, cf. \citet{BZ_quartic1,BZ_quartic2}). Like in the case of the quasar mode, 
we assume that the jets remove hot gas and bulge cold gas 
from the system with rates~\citep{reservoir}
\begin{eqnarray}
            \dot{M}_{b,{\rm gas}}^{\rm radio}&=&\frac{2}{3} \frac{L_{\rm jet,radio}}{\sigma^{2}} \frac{M_{b,{\rm gas}}}{M_{\rm hot} + M_{b,{\rm gas}}}\,,\\
            \dot{M}_{\rm hot}^{\rm radio}&=&\frac{2}{3} \frac{L_{\rm jet,radio}}{\sigma^{2}} \frac{M_{\rm hot}}{M_{\rm hot} + M_{b,{\rm gas}}}\,.
\end{eqnarray}

Finally, we stress that while the supernova feedback is most effective in low-mass systems, which presents shallow potential 
wells, the quasar and radio-mode feedbacks are most important in massive galaxies, where the quasar and radio activity is most pronounced.

\subsection{Mergers and environmental effects}
\label{sec:environment}
The prescriptions outlined in the previous sections allow us to describe the evolution of the baryonic structures
along the branches of the dark-matter merger trees. Therefore, in order to produce ``baryonic merger trees'' describing
the complete evolution of the baryons along the cosmic history, we only need a prescription to describe what happens at
the ``nodes'' of the dark-matter merger trees, i.e. when halos merge.

First of all, we should note that the ``nodes'' of the dark-matter merger trees correspond to the instant at which
the smaller halo (the ``satellite'') enters the bigger one (the ``host''). After entering the host, the
satellite halo survives as a bound substructure (a ``subhalo'') within the host. However, because of the gravitational interaction
with the particles of the host halo, known as dynamical friction, the satellite gradually loses energy and angular momentum
and slowly sinks to the center of the host. It is only when the satellite halo reaches the center of the host that the subhalo finally
dissolves and the baryonic structures of the two halos merge.

Because our dark-matter merger trees are not necessarily binary (i.e., their redshift step is 
such that a ``node'' may correspond to the merger
of more than two halos), at each node we consider the biggest halo as the host, 
and calculate the dynamical-friction timescales for the remaining
halos (the satellites) using the fitting formula proposed by \citet{df_formula}, which accurately reproduces the
merger timescales of extended halos predicted by N-body simulations:
\begin{equation}
   t_{\rm df} = \frac{R_{\rm vir}}{V_{\rm vir}} A \, {(M_{\rm host}/M_{\rm sat})^b \over \ln(1+M_{\rm host}/M_{\rm sat})}
  \exp\left[c \, \epsilon \right] \, \left[{r_c(E) \over R_{\rm vir}} \right]^d
  \label{eq:tdf_fit} \;,
\end{equation}
with $A=0.216$, $b=1.3$, $c=1.9$ and $d=1$.
Here, $\epsilon = L/L_c(E)$ and $r_c(E)$ are respectively the ``orbital circularity'' (the ratio between the satellite's angular 
momentum $L$ and the angular momentum $L_c$ of a circular orbit with the same energy $E$ as
the satellite) and 
the ``circular radius'' (the radius of a circular orbit with the same energy $E$ as
the satellite). These quantities describe the initial conditions of the infall (i.e. on what orbit the satellite enters the host). 
In particular, $\epsilon$ may vary from 0 to 1 and is drawn from a normal distribution centered on $\bar{\epsilon}\approx0.5$ and standard
deviation $\sigma \approx 0.23$~\citep{tormen,Khochfar}, while $r_c$ is derived from the periastron radius $r_p$, which N-body simulations suggest should be
 correlated 
with the circularity $\epsilon$ (more specifically, we assume $r_p=R_{\rm vir}\epsilon^{2.17}$: cf. \citet{tormen,Khochfar}).

We notice that Eq.~(\ref{eq:tdf_fit}) already includes  
environmental effects such as the tidal stripping and tidal heating, which cause the 
mass of the satellite to decrease while sinking to the host's center. This is 
because  Eq.~(\ref{eq:tdf_fit}) fits the results of N-body simulations, where these effects are automatically included.
However, Eq.~(\ref{eq:tdf_fit}) does not include the effect of the continuous accretion of dark matter and hot gas onto the host halo along cosmic time, and we
therefore correct
for this effect at each redshift step of our merger tree by rescaling the remaining dynamical friction time to account for 
the host having grown in the meantime.

Even though environmental effects are included in the dynamical friction timescale (\ref{eq:tdf_fit}), as we have mentioned they also
have the effect of reducing the satellite's mass while it sinks in the host. In particular, following \citet{environment}, 
we account for the tidal stripping (i.e. the tidal truncation of the satellite's density profile due to the average tidal force 
exerted by the host halo)
by cutting the dark-matter and hot-gas density profiles of the satellite at the tidal radius $R_{\rm tidal}$, corresponding to the
distance, from the satellite's center, at which the mean density of the satellite is comparable to the host's mean density at
the satellite's position $r$:
\begin{equation}
\bar{\rho}^{\rm NFW}_{\rm sat}(R_{\rm tidal})\approx \bar{\rho}^{\rm NFW}_{\rm host}(r)\,.
\end{equation}
More specifically, because the tidal stripping is most effective when the satellite is at the periastron of its orbit, 
we assume $r=r_p$ in this equation and perform the cut when the satellite first reaches the periastron.
Besides the tidal stripping, the satellite also loses mass due to the tidal heating, i.e. the evaporation induced by the rapidly
varying tidal forces near the periastron. We assume that this effect causes both the dark matter as well as all the baryon components to lose
mass with characteristic timescale calculated as in the Appendix B of \citet{environment}, and we assume that this mass loss 
starts when the satellites first
reaches the periastron.

When the satellite has sunk to the center of the host, a dynamical friction time $t_{\rm df}$ after it first entered the host,
the satellite halo finally loses its identity, and the baryonic structures of the host and satellite merge as well.
When such a merger happens, if the ratio $M^{\rm sat}_{\rm vir}/M^{\rm host}_{\rm vir}$  between the host and satellite halo masses
is sufficiently large, we assume that the merger between the dark-matter structures perturbs
the spin parameter of the resulting composite halo. 
More specifically, as already mentioned in Sec.~\ref{sec:DM}, 
if $M^{\rm sat}_{\rm vir}/M^{\rm host}_{\rm vir}>0.3$ we assign the final composite halo a new spin parameter $\lambda$
drawn from the same distribution used for newly formed halos, i.e.
a log-normal distribution with median value  $\bar{\lambda} = 0.039$ and standard deviation 
$\sigma= <\sqrt{({\ln\lambda}-\ln\bar{\lambda})^2}> =0.53$~\citep{spin_parameter,cole_lacey},
while if $M^{\rm sat}_{\rm vir}/M^{\rm host}_{\rm vir}<0.3$ we leave the host's spin parameter unchanged.

Also, further complications to this picture arise at the nodes where two or more halos, some or all of which containing their own subhalos, meet. 
In this case, if $M^{\rm sat}_{\rm vir}/M^{\rm host}_{\rm vir}>0.3$ for any one of the satellite halos, 
we recalculate all the dynamical friction timescales using Eq.~(\ref{eq:tdf_fit}), where we assume new values for the circularity
$\epsilon$, drawn from the same distribution used when the satellites first entered the host,
i.e. a normal distribution centered on $\bar{\epsilon}\approx0.5$ and standard
deviation $\sigma \approx 0.23$~\citep{tormen,Khochfar}. If instead $M^{\rm sat}_{\rm vir}/M^{\rm host}_{\rm vir}<0.3$ 
for \textit{all} the satellite halos, we calculate the dynamical friction times of the satellites and their subhalos 
in the host using the procedure that we just described, but leave the dynamical friction times of the subhalos of the host unchanged.
This scenario corresponds to the intuitive picture in which the incoming satellites manage to perturb and randomize 
the orbits of the host's subhalos only if they are sufficiently massive compared to the host.

We also recall that numerical simulations~\citep{1996ApJ...460..121W,2003ApJ...597..893N,2005A&A...437...69B} 
suggest that mergers where the mass ratio between the
total baryonic masses is larger than $\sim0.25-0.3$ (``major mergers'')
disrupt the galactic disks and give rise to a spheroidal component, while ``minor mergers'' 
(i.e., with mass ratio between the
total baryonic masses smaller than $\sim0.25-0.3$) do not destroy the galactic disks, although they may  
 perhaps drive the growth of the bulge by disk instabilities (see Sec. \ref{star_formation_section}).
We implement this scenario in our model by defining a merger as ``major'' if the ratio of the baryonic masses
is larger than 0.25 ($M_{\rm baryon,\,sat}/M_{\rm baryon,\,host}>0.25$ with $M_{\rm baryon}=M_{\rm d,stars}+M_{\rm d,gas}
+M_{\rm b,stars}+M_{\rm b,gas}+M_{\rm res}$), 
otherwise we define the merger as minor.
We then assume that in major mergers the disks are destroyed and their masses are added to the corresponding bulge components
\begin{align}
   &M_{d,{\rm gas}} = 0  \nonumber \\
   &M_{d,{\rm stars}} = 0  \label{merger1}  \\
   &M_{b,{\rm gas}} = M_{b,{\rm gas}}^{\rm host}  +   M_{d,{\rm gas}}^{\rm sat} \nonumber \\
  &M_{b,{\rm stars}} = M_{b,{\rm stars}}^{\rm host}  +   M_{d,{\rm stars}}^{\rm sat} \nonumber\\
  &M_{\rm res} = M_{\rm res}^{\rm host}  +   M_{\rm res}^{\rm sat} \nonumber \,,
\end{align}
while we assume that minor mergers do not affect the galactic morphology,
and therefore simply add the satellite's disk and 
bulge material to the disk and bulge component of the host galaxy:
\begin{align}
 &M_{d,{\rm gas}} = M_{d,{\rm gas}}^{\rm host} + M_{d,{\rm gas}}^{\rm sat}  \nonumber  \\
 &M_{d,{\rm stars}} = M_{d,{\rm stars}}^{\rm host} + M_{d,{\rm stars}}^{\rm sat}  \label{merger2} \\
  &M_{b,{\rm stars}} = M_{b,{\rm stars}}^{\rm host}+ M_{b,{\rm stars}}^{\rm sat} \nonumber \\
 &M_{b,{\rm gas}} = M_{b,{\rm gas}}^{\rm host}+ M_{b,{\rm gas}}^{\rm sat}   \nonumber\\
&M_{\rm res} = M_{\rm res}^{\rm host}  +   M_{\rm res}^{\rm sat} \nonumber \,.
\end{align}
Another effect of mergers is to cause significant starburst events in 
the merging galaxies. This is automatically accounted for in our model, because the disruption of the galactic disks
following a major merger channels gas into the spheroidal component, where star formation is very efficient,
because it happens on a dynamical timescale [cf. Eq.~(\ref{bulge_sfr})].

\subsubsection{Black-hole mergers}
\label{sec:BHmergers}

When the baryonic structures of the host and satellite merge, a dynamical friction
time $t_{\rm df}$ after the satellite first entered the host, 
the central MBHs do not coalesce right away, but form a binary system. 
The binary then continues to harden through ``slingshot'' interactions~\citep{slingshot}, in which stars intersecting the binary
are ejected at velocities comparable to the binary's orbital velocity, thus increasing the binding energy of the binary system.
However, the binary will soon eject all the intersecting stars, thus making the slingshot hardening inefficient. This
will cause the binary to stall at a separation of $\sim 1$ pc, unless other mechanisms intervene to make it decay
to a separation of $\sim 0.01$ pc, where gravitational-wave emission becomes 
strong enough to drive the binary's evolution
to the merger in a timescale shorter than the Hubble time. Since there is not, at present, 
a generally accepted scenario to overcome the stalling of the binary's evolution at
$\sim 1$ pc, this bottleneck is generally referred to as ``the final-parsec problem''~\citep{finalpc}.
It is generally thought, however, that the final-parsec problem is somehow solved in nature, because
 uncoalesced binaries
would result in slingshot ejection of MBHs when additional MBHs are brought in 
by successive mergers, thus resulting in off-center MBHs that seem rare or non-existent
and in too much scatter in the $M-\sigma$ relation~\citep{Msigma_scatter}.
Also, some possible mechanisms that would harden the binary until gravitational-wave emission becomes important, possibly solving
the final-parsec problem, have been proposed. For instance, the presence of a gaseous accretion disk would harden the binary on the viscosity
timescale~\citep{2002ApJ...567L...9A}; or the supply of stars available for slingshot interaction may be replenished by star-star 
encounters~\citep{2003ApJ...596..860M,2002MNRAS.331..935Y},
or as a result of the triaxial gravitational potential
that one naturally expects in galaxies forming from major mergers~\citep{2004ApJ...606..788M,finalpc1,finalpc2,finalpc3}.

Because of all these uncertainties, and because the MBH coalescence timescale is in any case expected to be small compared 
to the dynamical friction time  $t_{\rm df}$ and its 
uncertainties\footnote{For instance, for a satellite in a Milky Way type halo the dynamical friction
time is typically of a few Gyr, while the coalescence time for black-hole binaries with masses $\sim 10^6 M_\odot$ 
(roughly the mass of the Milky Way MBH)
is expected to be $\lesssim 10^7$ yr~\citep{alberto_mergertime}.}, we make the simplifying assumption
that the MBHs merge at the same time as the baryonic structures, i.e.
a dynamical friction
time $t_{\rm df}$ after the satellite first entered the host.
This approximation is adopted, to our knowledge, by virtually all semianalytical galaxy-formation models proposed so far,
and has the advantage of greatly simplifying the implementation of our model.\footnote{See however \citet{volonteri_haardt_madau} for a model
that does not make this assumption. That model, however, only includes dark-matter halos and MBHs, and does not attempt to describe the 
formation of galaxies, which makes the implementation of a realistic coalescence timescale for MBH binaries much simpler than in our case.}

When the two MBHs merge, we can determine the mass and spin of the resulting MBH remnant if we make some assumptions on
the relative orientation of the spins at large separations. As mentioned in the introduction, 
in the last few years numerical relativity has reached
a level of maturity sufficient for simulating black-hole binaries with non-aligned spins in a 
vast region of the space of parameters. Phenomenological 
formulas~\citep{FAUspin,RITspin,AEIspin1,AEIspin2,AEImass,bkl,kesden} based on Post-Newtonian theory, general-relativistic perturbation theory, 
symmetry arguments, as well as fits to the numerical-relativity
results, are not only capable of reproducing to high accuracy the simulation 
results for observables like the final mass, the final-spin magnitude and orientation, and the recoil velocity, but
also allow one to make sensible predictions for these quantities in regions of the parameter space where
simulations are still too time-expensive to ensure complete coverage (see \citet{luciano_review} for a review).

Here, we use the formula of \citet{AEIspin2}, which predicts the MBH remnant's spin magnitude
and orientation to very high accuracy (see \citet{AEIspin2,kesden_berti_sperhake}). In particular, the final-spin magnitude is
\begin{eqnarray}
\label{eq:general}
&\vert \boldsymbol{a}_{\rm fin}\vert=
\frac{1}{(1+q)^2}\Big[ \vta{1}^2 + \vta{2}^2 q^4+
 2 {\vert \boldsymbol{a}_2\vert}{\vert 
\boldsymbol{a}_1\vert} q^2 \cos \alpha\,+
\nonumber\\ 
& \hskip 0.2cm
2\left(
     {\vert \boldsymbol{a}_1\vert}\cos \beta +
     {\vert \boldsymbol{a}_2\vert} q^2  \cos \gamma
\right) {\vert \boldsymbol{{\ell}} \vert}{q}+\vert \boldsymbol{{\ell}}\vert^2 q^2
\Big]^{1/2},
\end{eqnarray}
with 
\begin{eqnarray}
\label{eq:L2}
&&\vtl = 2 \sqrt{3}+ t_2 \nu + t_3 \nu^2 +\nonumber \\
&&
 \frac{s_4}{(1+q^2)^2} \left(\vta{1}^2 + \vta{2}^2 q^4 
        + 2 \vta{1} \vta{2} q^2 \cos\alpha\right) + 
\nonumber \\
&&
\left(\frac{s_5 \nu + t_0 + 2}{1+q^2}\right)
        \left(\vta{1}\cos{\beta} + 
        \vta{2} q^2 \cos{\gamma}\right)\,.
\end{eqnarray}
Here, $q=M_{\rm bh,2}/M_{\rm bh,1}$ is the mass ratio; $\nu=q/(1+q)^2$ is the symmetric mass ratio;
$\vta{1}$ and $\vta{2}$ are the initial spin magnitudes; $\alpha$, $\beta$ ($\gamma$) are 
the angles (at large separation) respectively  
between the two spins and between spin 1 (spin 2) and the direction of the orbital angular momentum, ${\hat{\boldsymbol{L}}}$;
and $s_4 = -0.1229\pm0.0075$, $s_5 = 0.4537\pm0.1463$, $t_0=-2.8904\pm0.0359$,
$t_3 = 2.5763\pm0.4833$ and $t_2=-3.5171 \pm 0.1208$.\footnote{We note
that while \citet{AEIspin2} also present a formula predicting the orientation of
the MBH remnant's spin, that information is not necessary in our model. As we will
mention later on, in fact, our model
assumes that the spins of the two MBHs are aligned with the angular momentum of the
circumbinary disk in a gas-rich environment, in which
case the merger always produces a final spin in the same direction~\citep{aei_spin_flip}. In a gas-poor
environment, instead, we assume that the spins of the two MBHs are randomly oriented,
in which case we may apply the formula of \citet{AEIspin2} for the final-spin direction.
However, because accretion is chaotic in a gas-poor environment (cf. sec. \ref{smbh_evolution}), the information 
about the final spin direction
is 
not necessary in our model.}

Because during MBH mergers a fraction of the total mass is radiated in gravitational waves, the mass of the MBH remnant
is smaller than the initial mass of the binary. To account for this effect, 
we use the formula of \citet{AEImass}, which accurately predicts the 
final mass for equal-mass black-hole binaries with spins aligned or anti-aligned with the orbital angular momentum:
\begin{equation}
M_{\rm bh,fin}=[1 - (p_0+2 p_1 a+4 p_2 a^2)](M_{\rm bh,1}+M_{\rm bh,2})
\end{equation}
where $p_0 = 4.826\times 10^{-2}$, 
$p_1 = 1.559\times 10^{-2}$, $p_2 = 0.485\times 10^{-2}$ and $a=(a_1+a_2)/2$ 
($a_{1}$ and $a_2$ being the projections
of the spin parameters on the direction $\hat{\boldsymbol{L}}$ of the orbital angular momentum). 
For binaries with non-aligned spins or non-equal
masses, we use instead the formula of \citet{FAUspin}:
\begin{multline}
M_{\rm bh,fin}=(M_{\rm bh,1}+M_{\rm bh,2}) \times \\ [1 + 4\nu(m_0-1)+ 16 m_1  \nu^2 (a_1\cos\beta+a_2\cos\gamma)]\,
\end{multline}
where $m_0=0.9515\pm 0.001$ and $m_1=-0.013\pm 0.007$.

Also,  because the anisotropic emission of gravitational waves during the 
merger of two generic black holes  carries linear momentum away from the system, 
the MBH remnant is imparted a recoil velocity (``kick''). There has been much controversy
on the dependence of this kick velocity on the mass ratio, with \citet{kickRITeta2} advocating 
a  scaling with $\nu^2$, and  \citet{kick_goddard2}
finding a   scaling with $\nu^3$. It seems, however, that this discrepancy originated because
 \citet{kickRITeta2} and  \citet{kick_goddard2}
considered two different regions of the parameter space. A comprehensive formula valid in both regions 
was recently proposed by \citet{kick_goddard3}:
\begin{align}
& \boldsymbol{V}_{\rm recoil} = v_m \, \hat{\boldsymbol{e}}_1 + v_{\perp} (\cos\xi \, \hat{\boldsymbol{e}}_1 + \sin\xi \, \hat{\boldsymbol{e}}_2) 
+ v_{\parallel} \, \hat{\boldsymbol{e}}_3, \label{kick_v}\\
&      v_m     = A \nu^2 \sqrt{1 - 4 \nu} (1 + B \nu), \nonumber\\
& v_{\perp}     = H \frac{\nu^2}{1+q} \left( a_1^{\parallel} - q a_2^{\parallel} \right), \nonumber\\
& v_{\parallel} = \frac{K_2 \nu^2+K_3 \nu^3}{1+q}\left[qa_2^{\perp} \cos(\phi_2-\Phi_2) - a_1^{\perp} \cos(\phi_1-\Phi_1)\right]\nonumber\\
& +\frac{K_S (q-1) \nu^2}{(1+q)^3} 
\left[q^2a_2^{\perp} \cos(\phi_2-\Phi_2) + a_1^{\perp} \cos(\phi_1-\Phi_1)\right]
\nonumber
\,. 
\end{align}
Here $\hat{\boldsymbol{e}}_1$ and $\hat{\boldsymbol{e}}_3$ are orthogonal unit 
 vectors respectively in the direction of separation
and along the orbital axis \textit{just before merger}, and $\hat{\boldsymbol{e}}_2=\hat{\boldsymbol{e}}_1 \times \hat{\boldsymbol{e}}_3$
is a third unit vector orthogonal to them; $a_i^{\parallel}$ is the projection of the spin parameter $\boldsymbol{a}_i$ of black hole $i$ along
the orbital angular momentum, while $a_i^{\perp}$ is the magnitude 
of its projection  $\boldsymbol{a}_i^{\perp}$ onto the orbital plane;
$\phi_i$ is the angle of  $\boldsymbol{a}_i^{\perp}$ 
with respect to a reference angle
representing the separation of the black holes, as measured at
some point before the merger, while $\Phi_i$  represents the amount by which this angle precesses
before the merger. The angles $\Phi_i$ depend on the mass ratio and on the initial separation, and
must be determined with a numerical-relativity simulation~\citep{kick_goddard3}, which seriously
undermines the predictive power of Eq.~\eqref{kick_v}.
 The contribution $v_m$ to the kick is dubbed the ``mass-asymmetry contribution'',
because it does not depend on the spins and it disappears for equal mass binaries, 
while $v_{\perp}$ and $v_{\parallel}$ are the ``spin contributions'', which produce kicks
perpendicular and parallel to the orbital angular momentum. The angle between the mass asymmetry and spin contributions
is measured by $\xi = 215^\circ \pm 5^\circ$, while the other fitting parameters take the values  $A = 1.35 \times 10^4 \kms$, $B = -1.48$, $H = 7540
\pm 160 \kms$, $K_2 = 3.21 \pm 0.16
\times 10^4 \kms$, $K_3 =1.09  \pm 0.05
\times 10^5 \kms$ and $K_S =1.54  \pm 0.18
\times 10^4 \kms$.

Because it depends on quantities measured just before the black-hole merger, rather than defined
at large separations (as was e.g. the case for the formula~\eqref{eq:general}  for the final spin, 
cf. the discussion in \citet{AEIspin2}), 
Eq.~\eqref{kick_v} cannot be applied unambiguously in our model. In fact, the orbital axis and orbital
separation directions just before merger entering Eq.~\eqref{kick_v}, as well as the angles $\Phi_i$, 
can in general be determined
only with a full numerical-relativity simulation. 
An exception is given by binaries with aligned spins, in which case the magnitude of the recoil velocity is
independent of $\phi_i$ and $\Phi_i$, as can be seen from Eq.~\eqref{kick_v},
and the orbital axis remains unchanged during the binary's evolution.
In the general case, we assume (somewhat arbitrarily, cf. \citet{AEIspin2})
 that the orbital axis just before the merger is parallel to
the orbital axis at large separations, and 
because $\phi_2-\phi_1=\chi$ ($\chi$ being the angle between $\boldsymbol{a}_1^{\perp}$
and $\boldsymbol{a}_2^{\perp}$), 
defining the phases $\Delta_i\equiv \phi_1-\Phi_i$ we can write
$\phi_1-\Phi_1=\Delta_1$ and $\phi_2-\Phi_2=\chi+\Delta_2$ in Eq.~\eqref{kick_v}.
Using (again, a bit arbitrarily, cf. \citet{resonances,kesden_berti_sperhake}) the value of $\chi$ at large separation, we only need the
phases $\Delta_i$ to evaluate Eq.~\eqref{kick_v}. In this paper,
we choose these phases randomly from a uniform distribution.\footnote{Another choice 
may be to set $\Delta_1=\Delta_2=0$, and we have verified
that our results do not change significantly if we make this assumption, 
thus confirming the intuitive expectation that the
astrophysical effects of the gravitational recoil depend more on its overall scaling with the mass ratio $\nu$ than on the phases
$\Delta_i$.}

The recoil velocity can be as large as $4000 \kms$ for an equal-mass binary 
with antialigned equal spins lying on the equatorial plane 
(``superkick configuration'', see~\citet{kickRIT,kickJena}). Such a large velocity would be sufficient
to eject the MBH remnant from the galactic nucleus. When the spins are aligned with one another and
with the orbital angular momentum, instead, the recoil velocity is much smaller and typically insufficient 
to allow the remnant MBH to escape from the galaxy. At each black-hole merger, we therefore 
check whether the MBH remnant is retained by the spheroidal component
of the galaxy, thereby remaining available to accrete the cold gas brought in by
radiation drag when star formation occurs in the bulge [cf. Eq.~(\ref{mdot_res})], or whether it escapes.
Therefore, after each black-hole merger we compare the
recoil velocity ${V}_{\rm recoil}$ with the escape velocity $V_{\rm esc} = \sqrt{2 \phi}$, where $\phi$ is the potential due 
to the bulge (including both stars and gas) and the reservoir, evaluated at the galactic center. 
Because in the standard galaxy-formation model that we adopt here, the galactic disks form first 
and give rise to the bulges only later as a result of instabilities and major mergers, black-hole mergers
at high redshifts will happen in disk-dominated galaxies, resulting in very large fractions of MBHs being ejected.
We will discuss this more in detail in Sec.~\ref{sec:predictions1}, and hint at how this prediction may change in 
alternative scenarios of galaxy formation such as the ``two-phase model'' put forward in \citet{twophases,cook1,cook2}.
As we will see in the next sections, however, even in our current ``standard'' model the gravitational recoil does not eject
all the MBHs from their hosts. This is because the occupation fraction of black-hole seeds
at high-redshifts is smaller than 1 (cf., in Sec.~\ref{sec:DM},  the prescriptions that we adopt to 
populate the halos with black-hole seeds at high redshifts), which is enough to ensure that MBHs 
survive to low redshifts~\citep{lippai,zoltan_occ_frac,volonteri_kick,volonteri_kick2}. 
Moreover, as emphasized by \citet{schnittman_kick}, even if we populated
all halos with black-hole seeds at high redshift, 
the gravitational recoil in the first generation of
mergers would automatically decrease the MBH occupation fraction, 
and even in the case of very high ejection probabilities the occupation
fraction would settle to $\sim 50$ \% in a few more merger generations.

 Also, we stress that our recipe to determine whether a MBH is expelled from the galaxy may be overly pessimistic. In fact, as mentioned above, here we are interested in checking whether the MBH is retained in the spheroid, because it seems reasonable that a MBH
 wandering in the halo would neither accrete gas from the
circumnuclear reservoir, nor form MBH binaries after galaxy mergers. While this viewpoint may be sufficient for our purposes (since we want to study the mass and spin evolution of the nuclear MBHs and not of these halo MBH population), our ejection rates are clearly too large (especially at high $z$, when bulges are small) if one is interested in the number of MBHs that are expelled from the whole composite system (dark-matter halo and baryonic structures).

\begin{table}
\begin{center}
\begin{tabular}{ccc}
\hline 
 & {\rm light seeds} & {\rm heavy seeds} \\
\hline
 $\epsilon_{\rm SN}$  &  0.7 & 0.7  \\
 $f_{\rm jet}$    & 10   & 10\\
 $A_{\rm res}$    &  $1.1\times 10^{-2}$  &  $1.1\times 10^{-2}$  \\
 $t_{\rm accr}$   &  $4.8\times 10^{8}$ yr &   $4.8\times 10^{8}$ yr\\
\hline
\end{tabular}
\caption{The calibrated values of the free parameters of the model. These values are used to produce the figures.
\label{table:parameters}}
\end{center}
\end{table}

In order to apply the formulas for the MBH remnant's mass, spin and kick velocity that we have 
just reviewed, we need to specify not only the masses and spins of the two progenitor MBHs, but also
their relative orientation at large separations. It is well known~\citep{bardeen_petterson,tamara,perego,dotti,dotti2011} 
that if the binary's inspiral preceding the coalescence
happens in a circumbinary disk (which is expected to be present in gas-rich environments), 
the gravito-magnetic torque exerted by the disk aligns
the MBH spins with the disk's orbital angular momentum. 
The details of this alignment are weakly dependent on the equation of state of
the circumbinary disk, with a ``cold'' disk (i.e. one with polytropic index $\Gamma=7/5$, approximating
a gas with solar metallicity heated by a starburst) resulting in a residual 
angle $\lesssim 10^\circ$ between the spins and the disk's angular momentum,
and with a warmer disk (polytropic index $\Gamma=5/3$, corresponding to an adiabatic monatomic gas
and thus to a scenario where
radiative cooling is suppressed during the merger) leading to residual angles $\lesssim 30^\circ$~\citep{dotti}. 
As a result, when the dynamical interaction between the binary and the
gas creates a low-density region at a separation of about $0.1$ pc 
(the so called ``gap'', see \citet{gap}), the two spins are aligned
with one another and with the orbital angular momentum of the binary to within $10-30^\circ$. 
Because the gas density in the gap is very low,
the subsequent evolution is driven by purely gravitational effects which tend to further align 
the spins~\citep{resonances,kesden_berti_sperhake}. 
If the MBH merger happens instead in a gas-poor environment, 
the two spins are expected to be randomly distributed.

\begin{figure*}
\begin{center}
\includegraphics[width=6.cm,angle=-90]{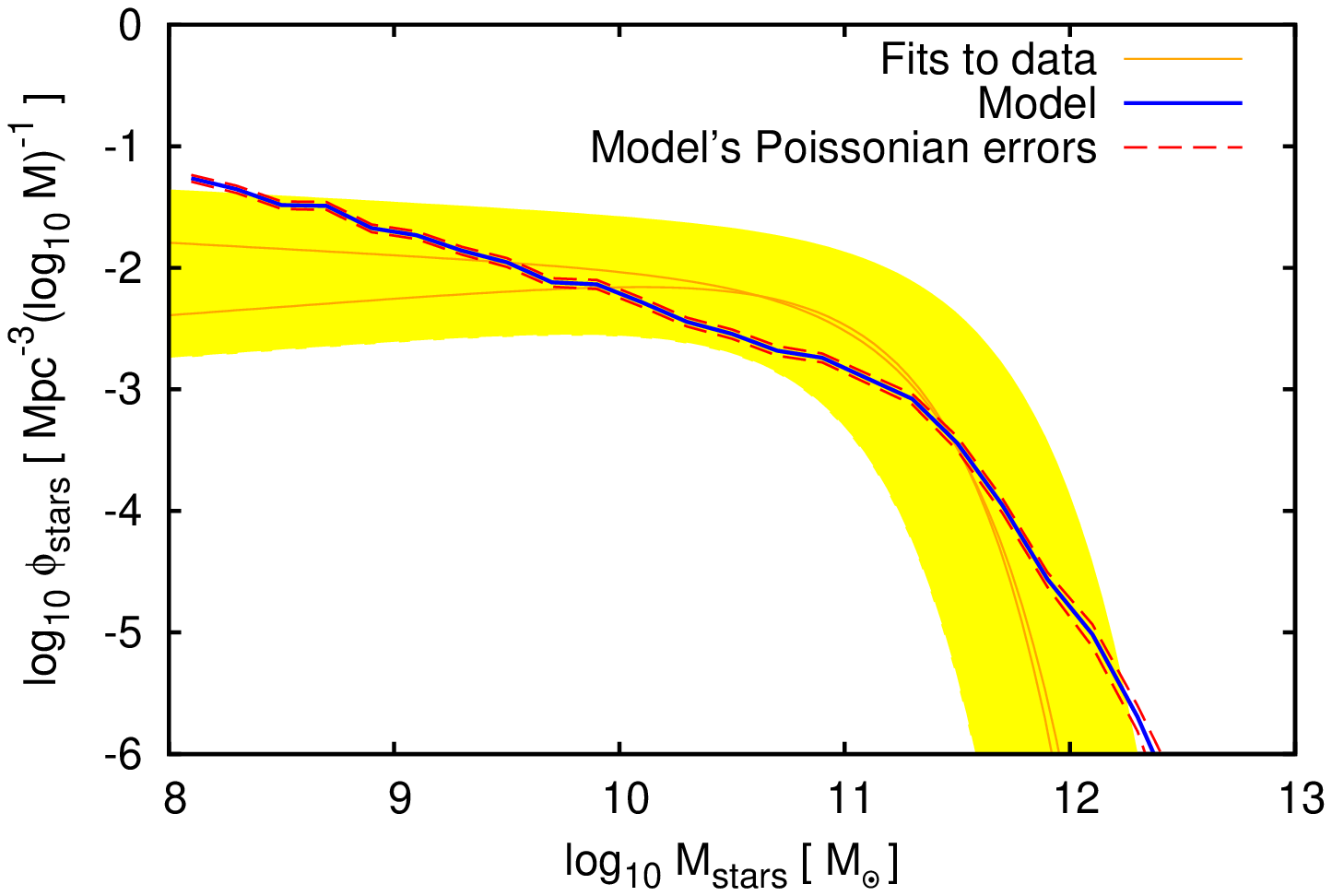}
\includegraphics[width=6.cm,angle=-90]{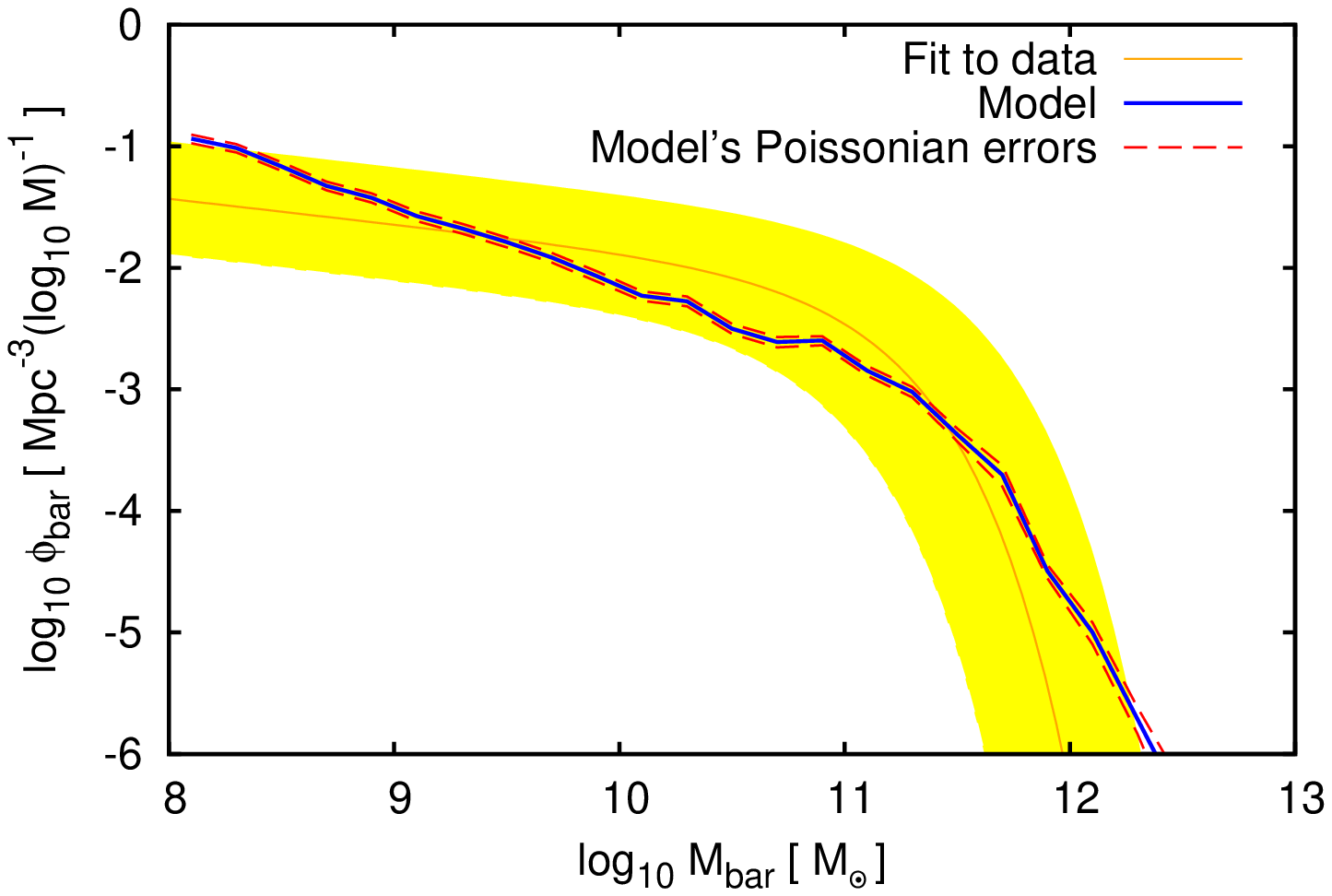}
\end{center}
\caption{
\label{fig_SMF_and_BMF_light}  
The local stellar mass function (left) and baryon mass function (right) of our model compared to the observational fits of \citet{smf_bell} and
 \citet{bmf_bell}. The observational uncertainties are shown with a shaded yellow area. These results are for the light-seed scenario, but the
heavy-seed one gives similar results.
}
\end{figure*}

Existing models for the spin evolution of MBHs do not attempt to 
distinguish between gas-rich (``wet'') and gas-poor (``dry'') mergers.
For instance, \citet{berti_volonteri} have to consider the two possibilities separately 
(i.e., either all mergers are assumed to be wet or all mergers are assumed to be
dry) because their model only includes dark-matter halos and MBHs and does not 
describe the baryonic components; \citet{fanidakis1,fanidakis2} model in great
detail the gravito-magnetic interaction of a single MBH with its own accretion disk, 
but do not include the disk's effect on the spin alignment prior to a black-hole
merger, and consider only dry mergers. Because it keeps consistently track of the evolution of the baryonic matter, 
our model offers a natural way to distinguish
between wet and dry MBH mergers. This is important for the spin evolution of MBHs, because the spin distribution
of the MBH population coming from a model with only wet mergers  differs drastically 
from that coming from a model with only dry mergers
(see Fig. 4 in \citet{berti_volonteri}). Also, dry mergers tend to give large kick velocities, in principle 
sufficient to eject the remnant MBH from the galactic nucleus, because the spins prior to merger are 
randomly oriented. Wet mergers, where
the spins are aligned, give much smaller recoil velocities~\citep{dotti}. Therefore, distinguishing between wet and dry mergers 
is vital to predict the MBH occupation number and the number of event rates for gravitational-wave detectors.

More specifically, in our model we can discriminate between dry and wet mergers by comparing the MBH binary's mass to
that of the circumbinary disk. This comparison needs to be performed carefully, however, because as explained above 
the MBH merger happens \textit{after} the galactic merger, although for simplicity this delay is not implemented in our model.
Major galaxy mergers are typically accompanied by starbursts, which in our 
model occur because the galactic disks are disrupted and feed the gaseous
bulges, where star formation is very efficient. As a result, large quantities of cold gas are forced 
into the circumbinary disk (i.e. the
``reservoir'' described in the previous sections) by radiation drag [cf. Eq.~(\ref{mdot_res})]. 
The circumbinary disk is therefore more massive during the MBH inspiral than at
the time of the galactic merger. We can therefore approximate its mass 
as $M_{\rm res}+A_{\rm res} (M_{b,{\rm gas}}/t_{\rm sf}) t_{\rm ff} $, where $M_{\rm res}$ and $M_{b,{\rm gas}}$ 
are the masses of the reservoir and bulge right after the
galactic merger [cf. Eqs.~(\ref{merger1}) and (\ref{merger2})], $M_{b,{\rm gas}}/t_{\rm sf}$ is approximately
the SFR in the bulge ($t_{\rm sf}$ being the dynamical time of the gaseous bulge) and $t_{\rm ff}$ is the characteristic
time of the MBH inspiral (i.e. the free fall time of the bulge, including both stars and gas). During the time $t_{\rm ff}$,
the two MBHs continue accreting at the expense of the circumbinary disk, and thus at a time $t_{\rm ff}$ after the galactic merger 
the mass of the binary is roughly
\begin{equation}
M_{\rm binary}\approx M_{\rm bh,1} 
\exp\left(\frac{t_{\rm ff}}{t_{\rm Edd,1}}\right)+ M_{\rm bh,2} 
\exp\left(\frac{t_{\rm ff}}{t_{\rm Edd,2}}\right)\,
\end{equation}
($t_{\rm Edd,1}$ and $t_{\rm Edd,2}$ being the Eddington-accretion timescales of the two MBHs),
 to be compared with
a mass 
\begin{multline}
M_{\rm circ\,disk}\approx M_{\rm res}+A_{\rm res}\left( \frac{M_{b,{\rm gas}}}{t_{\rm sf}}\right) t_{\rm ff}\\- M_{\rm bh,1} 
\left[\exp\left(\frac{t_{\rm ff}}{t_{\rm Edd,1}}\right)-1\right]
- M_{\rm bh,2} 
\left[\exp\left(\frac{t_{\rm ff}}{t_{\rm Edd,2}}\right)-1\right]
\end{multline}
for the circumbinary disk.
We therefore dub a merger ``dry'' if $M_{\rm binary}>M_{\rm circ\,disk}$, and in this case we assume that the spins of the two MBHs are randomly oriented at large separations. We instead consider a merger as ``wet'' if $M_{\rm binary}<M_{\rm circ\,disk}$, and in this case we assume that the spins
are exactly aligned with the orbital angular momentum.

\section{Testing the model against observations}

\label{sec:calibration}

\begin{figure}
\includegraphics[width=6.cm,angle=-90]{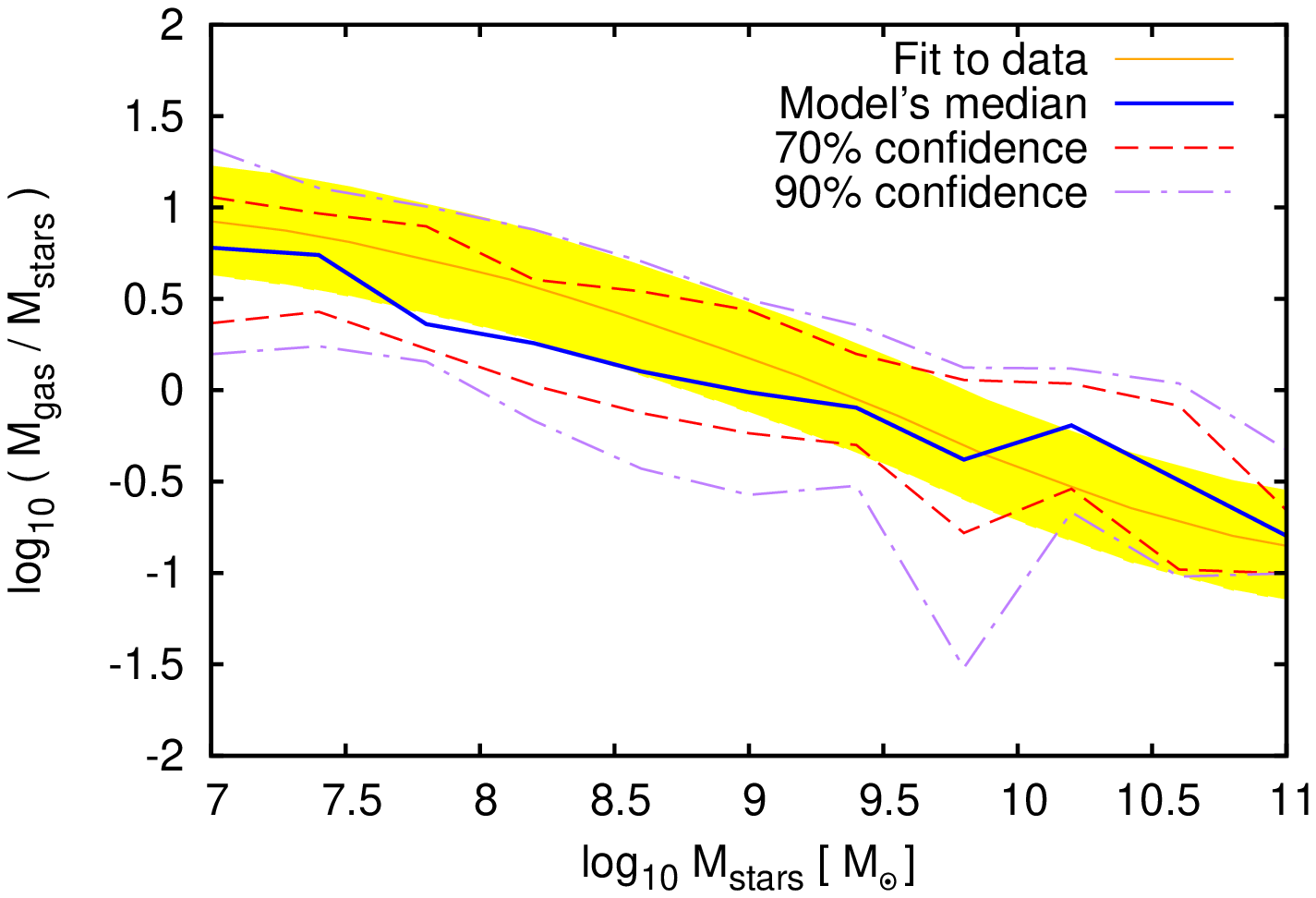}
\caption{
\label{gas_fraction_fig_light}  
The local gas fraction of our model compared with the observational parameterization of \citet{baldry_gas_fraction}.
 The observational uncertainties are shown with a shaded yellow area. These results are for the light-seed scenario, but the
heavy-seed one gives similar results.
}
\end{figure}

Semianalytical galaxy-formation models require in general several free parameters to reproduce accurately 
a large range of observational data (cf. for instance \citet{galform2010}), which makes their calibration
rather involved. Our focus here, however, is not on the galactic properties \textit{per se}, but rather on the 
spin evolution of the MBH population. We therefore allow a limited number of free parameters in the galaxy formation model that
we described in the previous sections, and calibrate them against a limited but representative
number of observations, both at $z=0$ and at higher redshifts. 

In particular, we allow the following four free parameters to vary: \textit{(i)} the supernova feedback efficiency $\epsilon_{\rm SN} $ 
[cf. Eqs.~(\ref{snfb2}) and (\ref{snfb1})], which ranges from 0 to 1 and which we assume to be 
the same for the feedback in the disks and in the bulges;
\textit{(ii)} the ``fudge'' factor $f_{\rm jet}$ [cf. Eqs.~(\ref{eq:ljet_sda}) and (\ref{eq:ljet_riafa})]
parameterizing the uncertainties in the Blandford-Znajeck effect (i.e. the strength of the magnetic field and the higher-order
terms in the black-hole spin); this factor is
assumed to vary between $0.1$ and $10$, and we assume it to be the same for the radio and quasar-mode feedback; 
\textit{(iii)} the normalization
factor $A_{\rm res}$ that describes the strength of the radiation drag regulating the growth of the circumnuclear reservoir 
[cf. Eq.~(\ref{mdot_res})]; 
in order to reproduce the $M_{\rm bh}-\sigma$ relation and the ``Magorrian'' 
$M_{\rm bh}-M_{b}$ relation, it should be $A_{\rm res}\gtrsim M_{\rm bh}/M_{b}$,
and because $M_{\rm bh}/M_{b}\sim 10^{-3}$~\citep{magorrian,haring_rix}, we
expect this parameter to be on the order of $10^{-2}$--$10^{-3}$ (cf. also 
\citet{reservoir,reservoir2,cook1,cook2,haiman}); \textit{(iv)} the timescale $t_{\rm accr}$
regulating the inflow of the circumnuclear reservoir gas into the MBH pc-scale 
accretion disk [cf. Eq.~(\ref{tacrr})]; because 
the quasar bolometric luminosity peaks at $z\approx2$ (see e.g.~\citet{Lqso}), 
this timescale is expected to be on the order of the age of the universe at $z\approx2$, i.e.
$t_{\rm accr} \sim 10^9$ yr. As we will show in the next sections, these four parameters are non-degenerate, because
each of them (roughly) regulates the predictions for different observables, 
and this makes their calibration rather straightforward.

\begin{figure*}
\begin{center}
\includegraphics[width=6.cm,angle=-90]{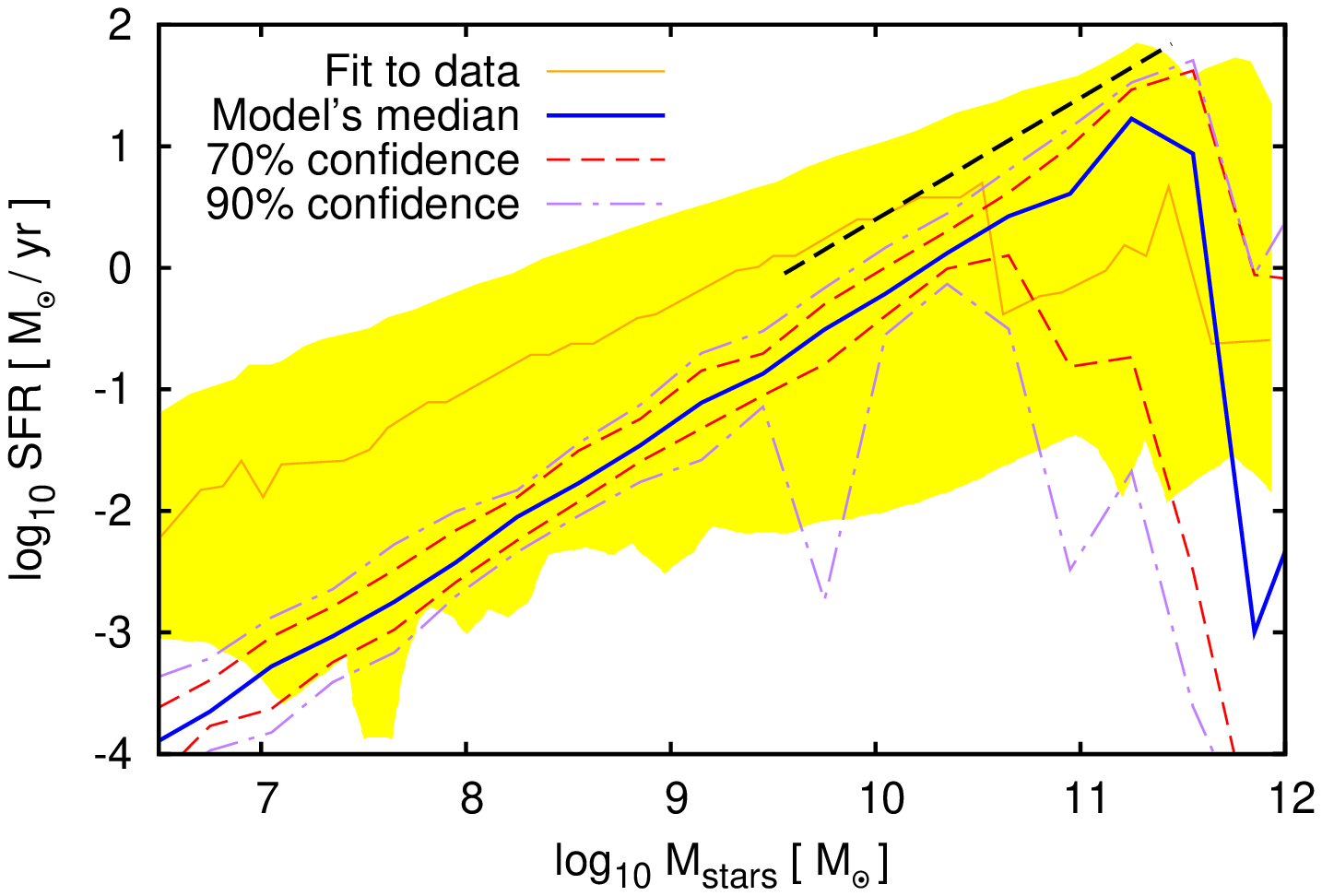}
\includegraphics[width=6.cm,angle=-90]{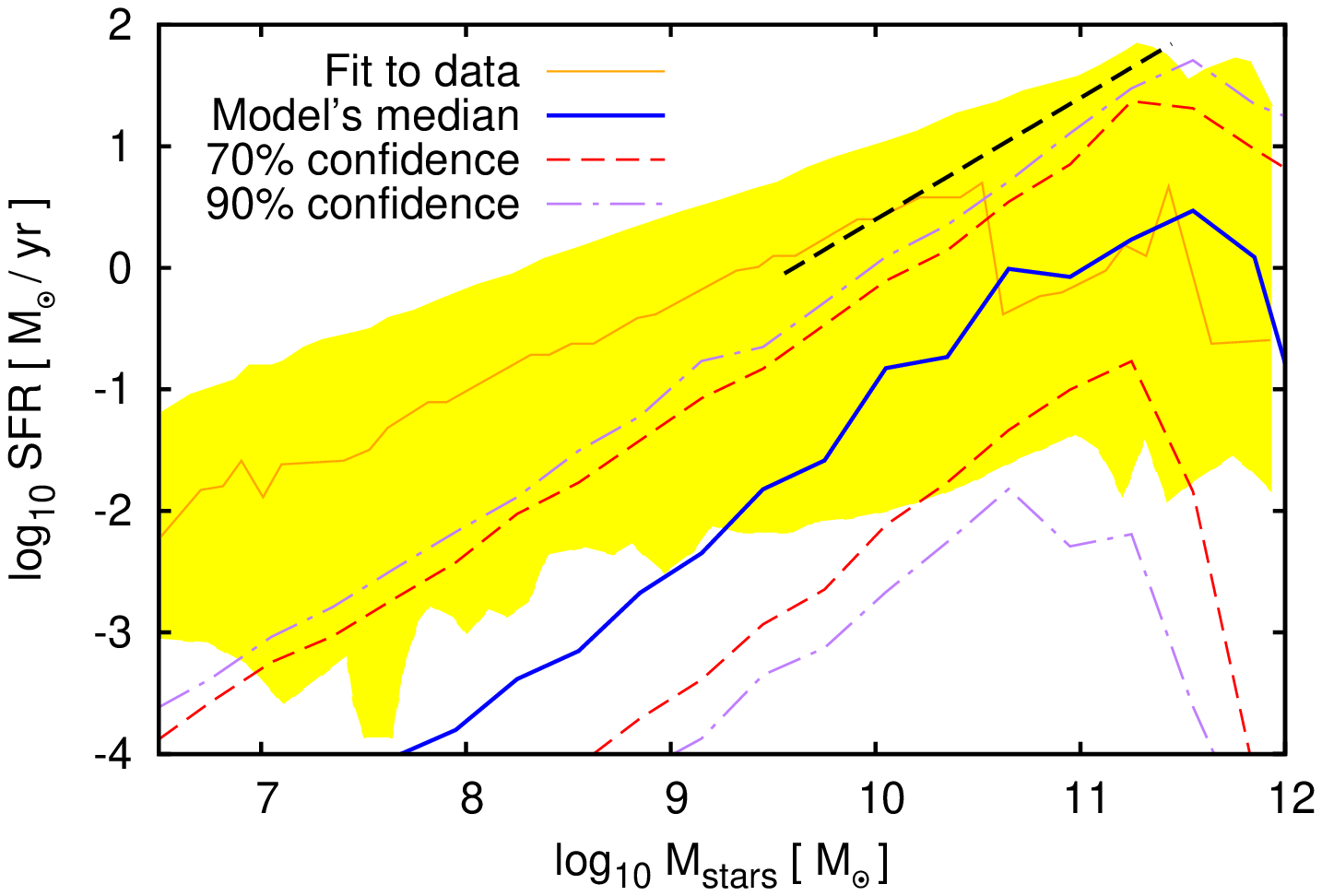}
\end{center}
\caption{
\label{sfr_fig}  
Our model's SFR (at $z=0$) compared to the observational results of \citet{sfr_brinchmann}.
 The observational 95\% confidence region is shown with a shaded yellow area.
 In order to highlight 
possible selection effects, we show the output of our model including only the central galaxies with 
SFR$>10^{-6} M_\odot/{\rm yr}$ (left panel) and both the central and satellite galaxies with 
SFR$>10^{-6} M_\odot/{\rm yr}$ (right panel). These results are for the light-seed scenario, but the
heavy-seed one gives similar results.
}
\end{figure*}

The observables that we will consider are, at $z=0$:
the stellar and baryon mass functions, the MBH mass function, the gas to stellar mass ratio,
the SFR, the $M_{\rm bh}-\sigma$ relation, the ``Magorrian'' $M_{\rm bh}-M_{b}$ relation, 
and the fraction of elliptical, spiral and irregular galaxies.
Also, we will test our model at $z>0$ by looking at the SFR density (SFRD) and at the 
quasar bolometric luminosity density as a function of $z$.
The four free parameters of the model that we mentioned above are chosen to optimize the agreement with
these observables, and their values are given in Table \ref{table:parameters}.

To obtain the results shown in the next sections,  for each of our two models (light or heavy black-hole seeds)
we have produced $\sim 1100$ halos with masses at $z=0$ between $10^{10}$ and $10^{15} M_\odot$, and we have 
weighed the contribution of each one of them 
using the Sheth-Tormen halo mass function~\citep{ST1999,ST2002} at $z=0$.
We stress that each of these halos corresponds, at $z=0$, either to a single galaxy (if there are no satellites) 
or to a group or cluster of galaxies (if satellites are present).

\begin{figure}
\includegraphics[width=6.cm,angle=-90]{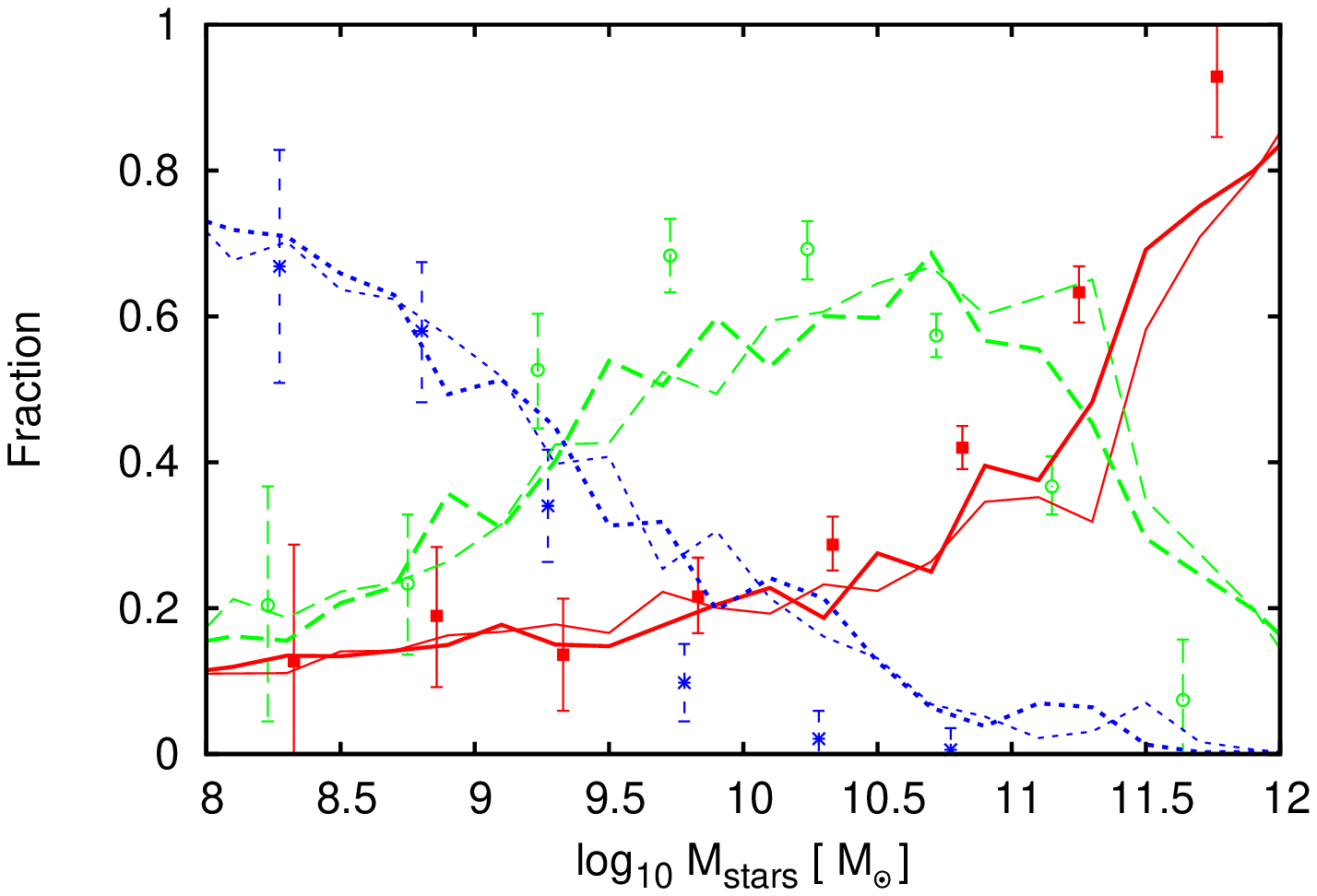}
\caption{
\label{morph_fig}  
The fraction of elliptical, spiral and irregular galaxies, at $z=0$, as a function of stellar mass.
The symbols are data from~\citet{conselice} (squares: ellipticals, circles: spirals, stars: irregulars),
 while the predictions of our model are shown with thick
lines (heavy seeds) and thin lines (light seeds).
}
\end{figure}

\subsection{Observables at $z=0$}

For the local stellar and baryon mass functions, we compare our results with the Schechter function fits by \citet{smf_bell} and 
 \citet{bmf_bell}, which were obtained with a large sample of
galaxies from the \emph{Two Micron All Sky Survey} (2MASS) and the \emph{Sloan Digital Sky Survey} (SDSS),
using simple models to convert the optical and near infrared galaxy luminosities into stellar masses
and assuming a universally-applicable stellar IMF. 
These fits are shown in Fig.~\ref{fig_SMF_and_BMF_light} (left: stellar mass function; right: baryon mass function)
with orange thin solid lines, together with
shaded yellow areas representing the observational errors. In particular, we have assumed 
0.3 dex errors in the stellar and baryon mass determinations, and 0.3 dex
in the value of the mass functions to account for the statistical errors as well as the systematic
uncertainties (mainly due to the mass-to-light ratios: see discussions in \citet{smf_bell}, 
\citet{bmf_bell} and also \citet{stellar_mass_systematics}). 
Also presented in  Fig.~\ref{fig_SMF_and_BMF_light}   is the output of our model in the light-seed scenario (the heavy-seed scenario
yields similar results), together 
with the Poissonian errors due to to finite sample of galaxies that we simulate. 
As can be seen, the model's results are
generally within the observational errors, although 
the slope is significantly different from the observations both at the low and high-mass ends.
We stress that the agreement of the model's mass functions with the observational 
estimates is the product of different and competing
processes. More specifically, the ionizing radiation background and the supernova feedback 
are effective at quenching star formation and reducing
the baryonic content in small-mass systems, while the quasar and radio-mode feedback, 
as well as the ram pressure and clumpy accretion,
are responsible for the sharp decrease of the mass function at large masses, which is significantly 
faster than what would be expected from the 
behavior of the Sheth-Tormen halo mass function alone. 
In particular, to reproduce the high-mass end of the stellar and baryon mass functions it is crucial
to calibrate the parameter  $f_{\rm jet}$, which regulates the radio and quasar mode feedback in our model.

\begin{figure*}
\begin{center}
\includegraphics[width=6.cm,angle=-90]{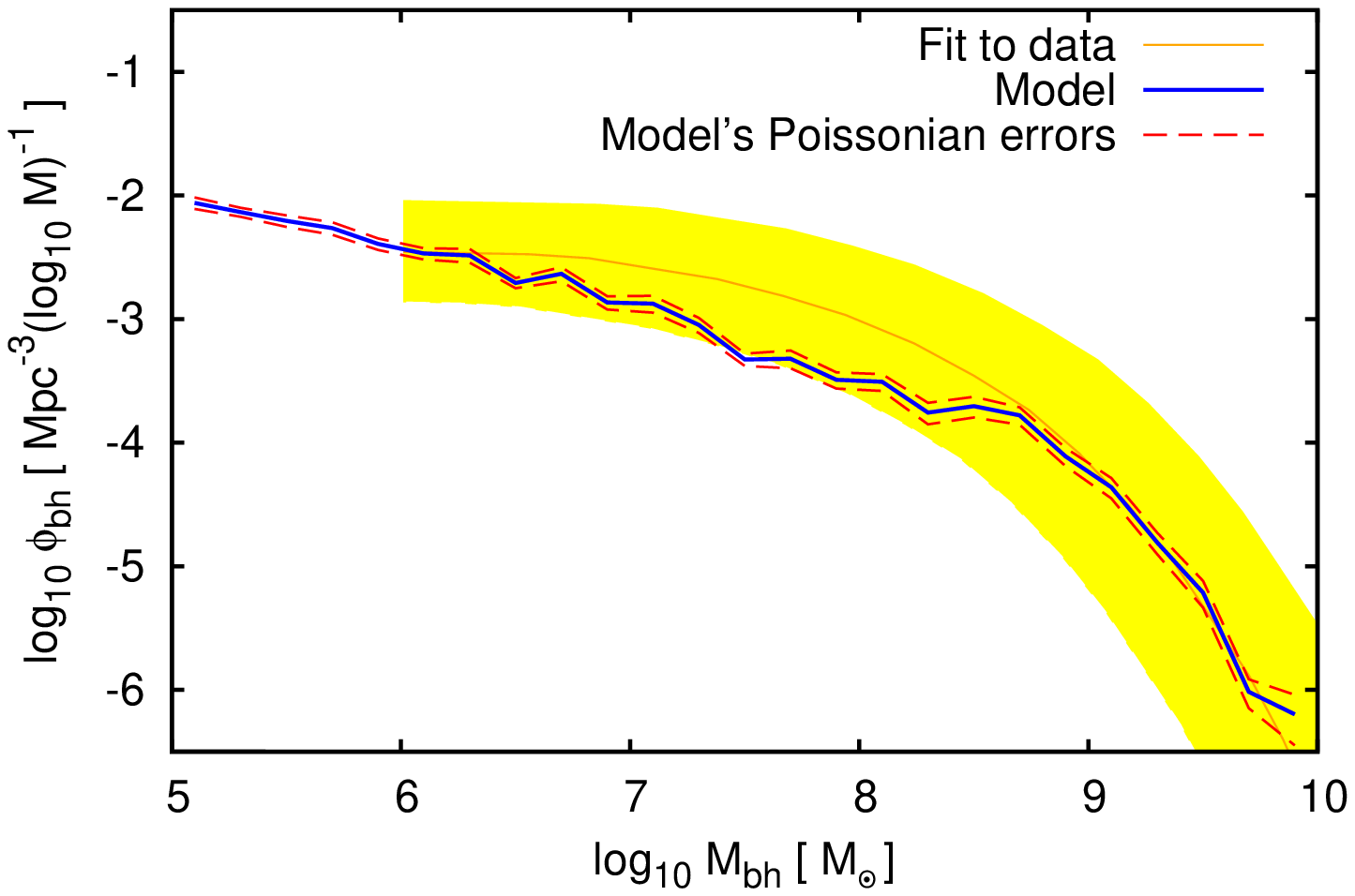}
\includegraphics[width=6.cm,angle=-90]{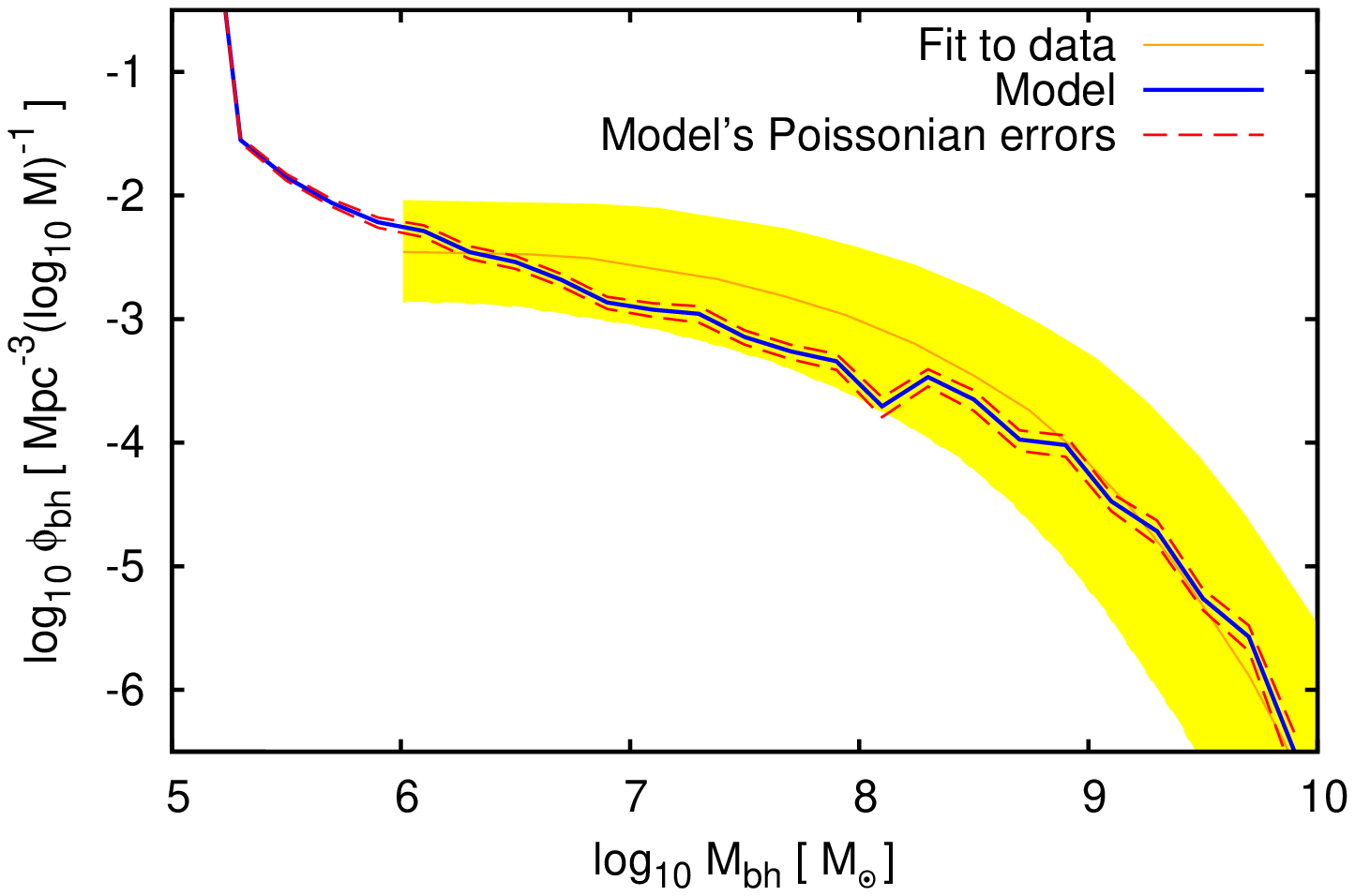}
\end{center}
\caption{
\label{BHMF_fig}  
The local MBH mass function predicted by our model in the light-seed (left) and heavy-seed (right) scenario, compared to
the observational estimate by \citet{marconi}
The observational uncertainties are shown with a shaded yellow area. These results are for the light-seed scenario, but the
heavy-seed one gives similar results.
}
\end{figure*}

In Fig.~\ref{gas_fraction_fig_light} 
we test our model against observational results for the 
ratio between gaseous and stellar masses
(``gas fraction'') at $z=0$. We show the comparison for the light-seed scenario, but the heavy-seed scenario
yields similar results.
In particular, we use the gas-fraction parameterization of \citet{baldry_gas_fraction} (orange thin solid line), and
we represent its observational uncertainties (see Fig. 11 of \citet{baldry_gas_fraction}) with a shaded yellow area.
Because the data used by \citet{baldry_gas_fraction} include only field galaxies, 
when calculating the gas fraction in our model we only include central galaxies
whose stellar mass is at least 90\% of the total stellar mass of the system (central galaxy 
plus all the satellites). Also, the data used  by \citet{baldry_gas_fraction} do not include 
gas-poor ellipticals, but those galaxies
are only a significant field population at $M_{\rm stars}\gtrsim 10^{11} M_\odot$, 
which is outside the range of Fig.~\ref{gas_fraction_fig_light}. 
Because of the scatter of our model's predictions, we present its median 
output at a given stellar mass, as well as its
70\% and 90\% confidence-level regions, i.e.
the regions containing respectively 70\% and 90\% of the galaxies produced by our model at a given stellar mass. 
As can be seen, the model is in good agreement with the observational data. 
In particular, in order to obtain a good agreement with the observed gas fraction at 
$M_{\rm stars}\sim 10^7-10^8 M_\odot$, 
it is crucial to tune the parameter $\epsilon_{\rm SN}$, which regulates the efficiency of the supernova feedback.

The SFR at $z=0$ as a function of stellar mass, as obtained from a survey $\sim 10^5$ galaxies 
with measurable  SFR in the SDSS~\citep{sfr_brinchmann}, is shown in Fig.~\ref{sfr_fig}. 
The orange thin solid line represents
the median SFR observed at a given stellar mass, while the shaded yellow area
represents the observational 95\% confidence level (i.e.
the region containing 95\% of the galaxies at a given stellar mass.). The median SFR predicted by our model
at a given stellar mass (in the light-seed scenario: the heavy-seed one gives again similar results) is represented by
the thick solid blue line, while the thin red dashed and thin purple dot-dashed lines represent the model's 
70\% and 90\% confidence regions.
In particular, because the median and confidence region of~\citet{sfr_brinchmann} may depend on the
composition of their sample of galaxies, we have plotted the output of our model when considering
only the central galaxies, where star formation is more intense (left panel),
 and when considering both the central galaxies
and the satellites (right panel). In both cases, we have neglected galaxies 
with SFR less than $10^{-6} M_\odot/{\rm yr}$,
since the sample of \citet{sfr_brinchmann} only includes galaxies with measurable SFR.

\begin{figure*}
\begin{center}
\includegraphics[width=6.cm,angle=-90]{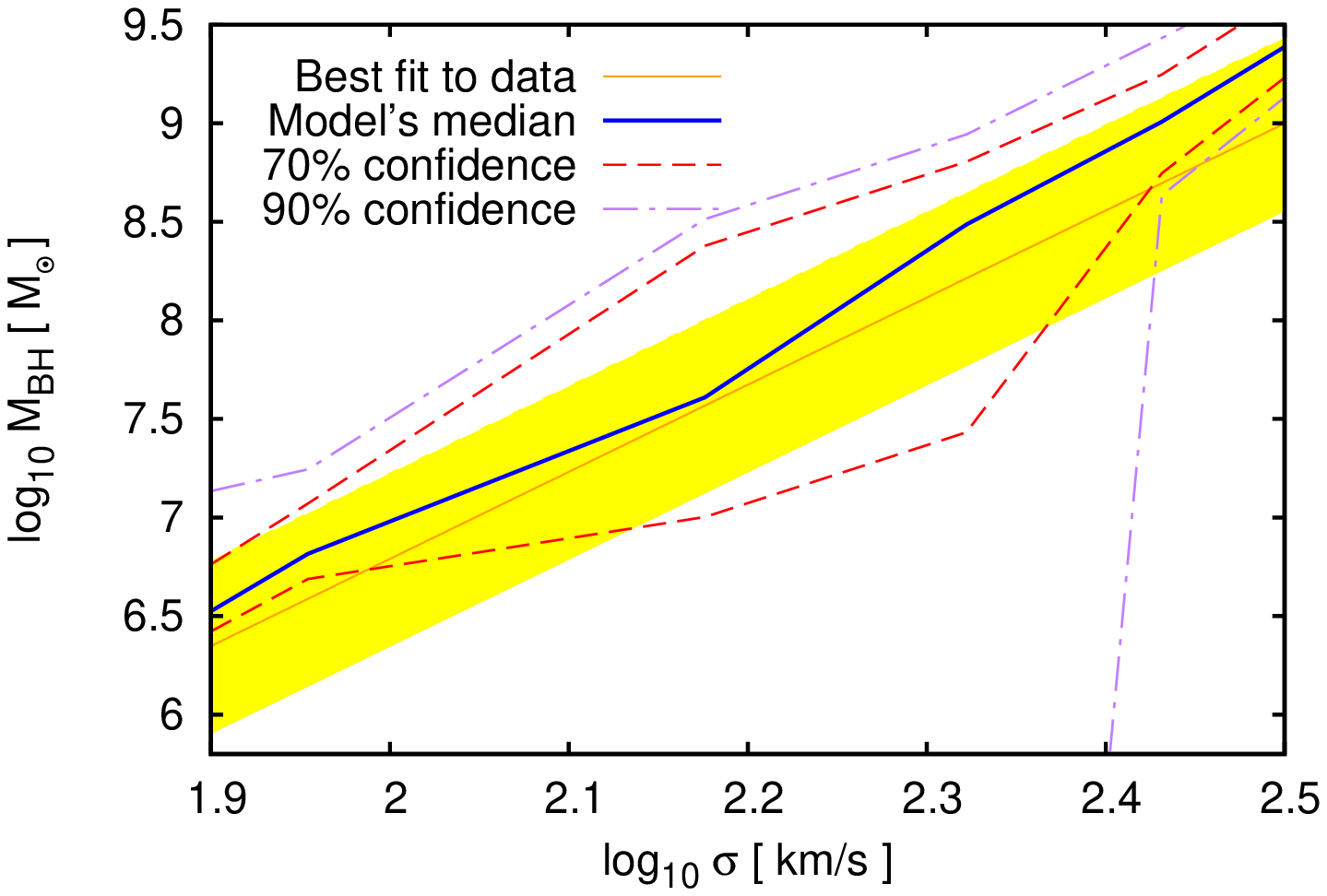}
\includegraphics[width=6.cm,angle=-90]{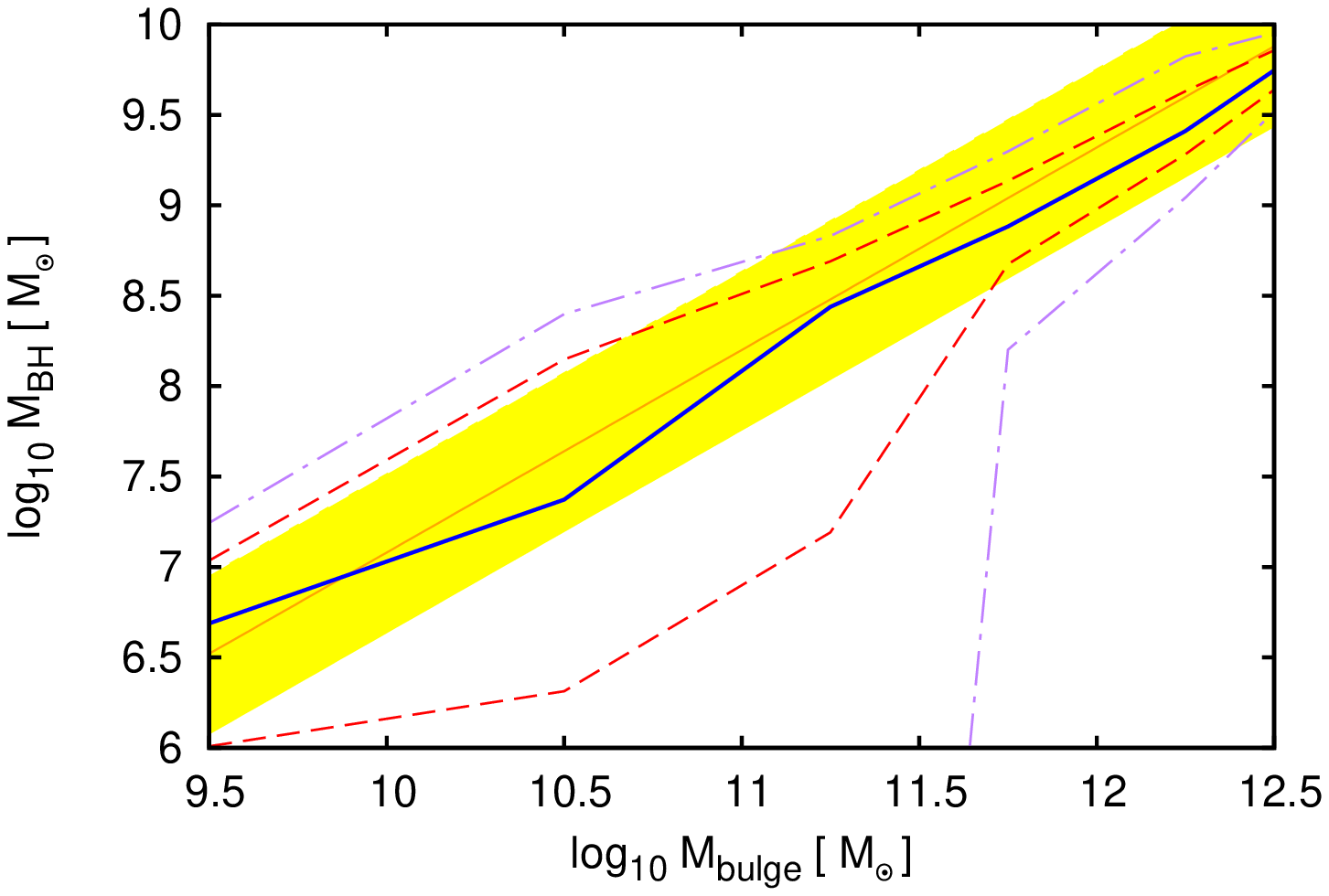}
\end{center}
\caption{
\label{msigma_fig_light}  
The predictions of our model, in the light-seed scenario, for the $M_{\rm bh}-\sigma$ and $M_{\rm bh}-M_{b}$ relations at $z=0$,
compared to the observational fits of \citet{msigma_latest} and \citet{haring_rix}, respectively. The observational uncertainties are shown with a shaded yellow area.
}
\end{figure*}

\begin{figure*}
\begin{center}
\includegraphics[width=6.cm,angle=-90]{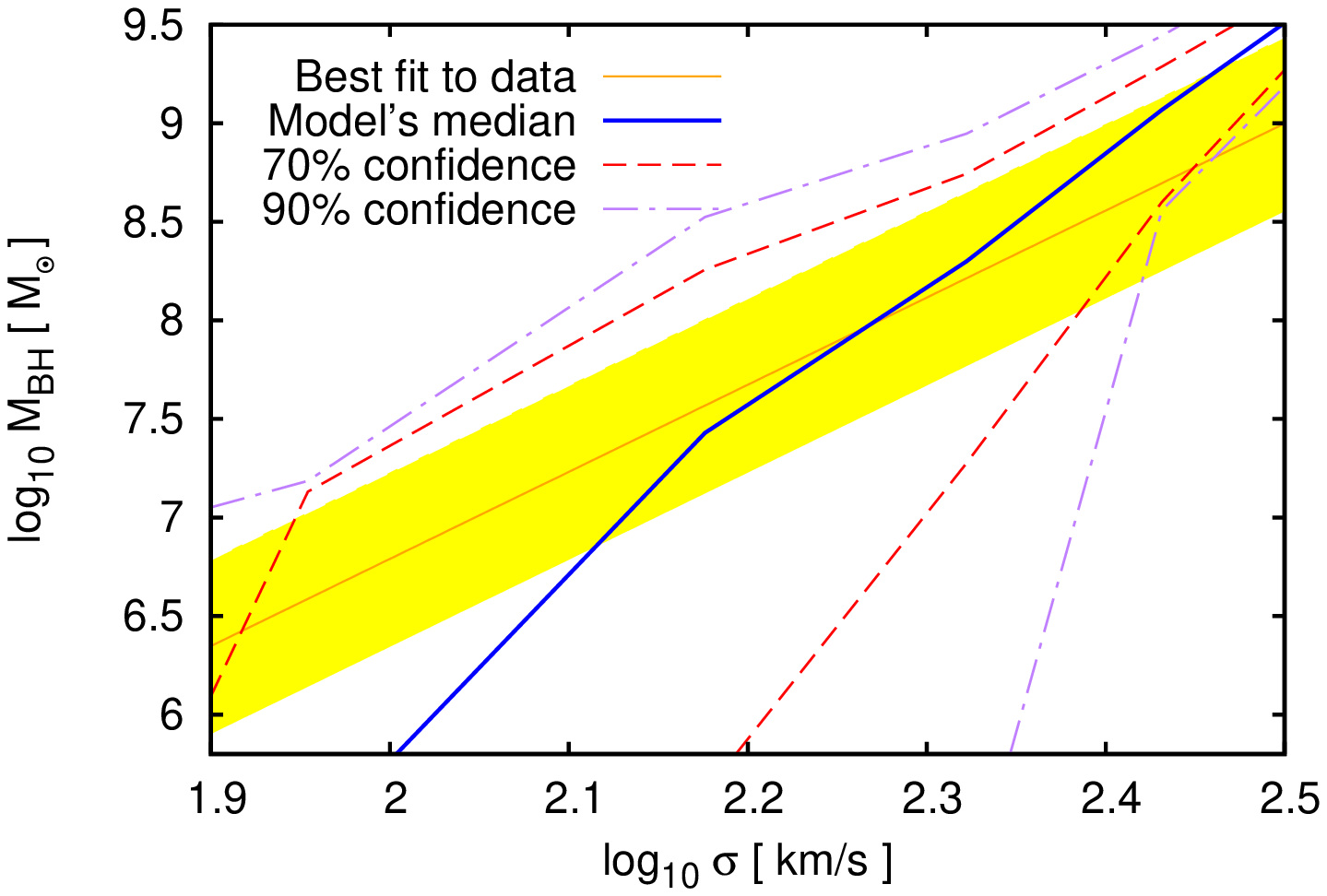}
\includegraphics[width=6.cm,angle=-90]{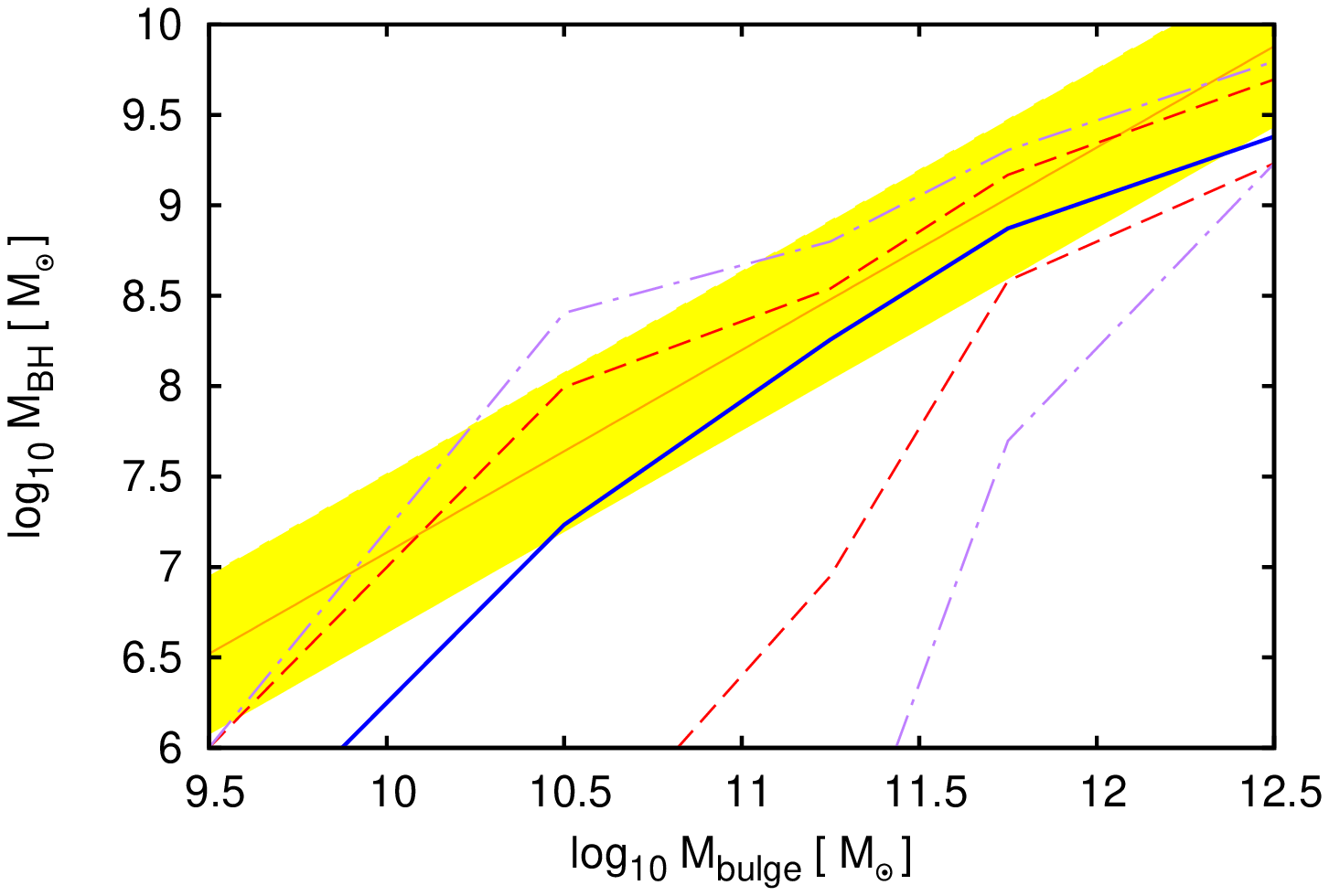}
\end{center}
\caption{
\label{msigma_fig_heavy}  
The same as in Fig.~\ref{msigma_fig_light}, but for the heavy-seed scenario.
}
\end{figure*}

As can be seen, irrespective of which of the two samples we consider, 
the median predictions of our model lie within the 95\%-confidence region of the observational
data, and reproduce the flattening of the SFR at high masses 
(due to the combined effect of quasar and radio-mode feedback, clumpy accretion
and ram pressure). However, our predictions are significantly lower than the 
median of the observations, especially at small stellar masses, essentially because they
present a steeper slope. 
More specifically, our model predicts SFR $\propto M_{\rm stars}$, while the data
by \citet{sfr_brinchmann} suggest SFR $\propto M_{\rm stars}^n$, with $n\approx 0.7$. While other observational data 
point at a slope $n$ slightly less than 1 (for instance, \citet{elbaz07} find $n\approx 0.77$ and 
\citet{salim07} find $n\approx0.65$), \citet{elbaz11} find
that at $z=0$ the SFR-$M_{\rm stars}$ relation is well fitted by
SFR$=M_{\rm stars}/(4\times10^9 {\rm yr})$
for $9.5\lesssim\log_{10} M_{\rm stars}\lesssim11.5$. 
This parameterization is represented in Fig.~\ref{sfr_fig} with a dashed thick
black line.  Also, there is some evidence that the slope $n$
is rather sensitive to selection effects: for instance, \citet{karim} find that 
for highly active star-forming galaxies
the exponent $n$ approaches 1, although it is not clear whether these galaxies are representative of the 
entire star-forming population.
In general, the slope $n\approx1$ of our model's predictions 
is due to our star formation prescriptions of Sec.~\ref{star_formation_section},
and is therefore common to most semianalytical galaxy-formation models 
(see for instance \citet{toosteep}). While it is in principle possible to change our star formation
prescription to obtain a milder slope~\citep{new_sfr}, at this point it is not clear whether this is needed, 
because it seems that the issue is still not settled from the observational point of view.

In Fig.~\ref{morph_fig}, we plot the fraction of elliptical, spiral and irregular galaxies as a function of stellar mass.
The symbols are actual data from~\citet{conselice} (squares: ellipticals, circles: spirals, stars: irregulars),
 while the predictions of our model are shown with thick
lines (heavy seeds) and thin lines (light seeds). More specifically,
within our model we follow~\citet{white_morph} and use the ratio $M_{b}/M_{\rm tot}$ (with $M_{\rm tot}=M_b+M_d$) to discriminate different morphologies.
In particular, we identify galaxies with $M_{b}/M_{\rm tot}>0.7$ with ellipticals
and represent them with solid lines; galaxies with $0.03<M_{b}/M_{\rm tot}<0.7$
with spirals and represent them with  dashed lines; 
galaxies with $M_{b}/M_{\rm tot}<0.03$ with extreme late-type or pure-disk galaxies
(and therefore with irregulars~\citep{white_morph}), and represent them with dotted lines. 
Remarkably, in spite of this simplistic classification, our model seems to reproduce
the observed morphological fractions, at least qualitatively.

To test our predictions for the MBH population at $z=0$, we look at the local MBH mass function,
at the $M_{\rm bh}-\sigma$ relation between the black-hole mass and the line-of-sight 
velocity dispersion $\sigma$ of the bulge~\citep{msigma1,msigma2,msigma_latest,msigma_extra1,msigma_extra2,msigma_extra3},
and at the relation between the MBH mass and the bulge mass, 
initially proposed by \citet{magorrian} and later updated by \citet{haring_rix}.
The predictions of our model for these observables 
mainly depend on the normalization
factor $A_{\rm res}$ regulating the growth of the circumnuclear reservoir and therefore of the MBHs, and we 
calibrate this parameter to obtain agreement with the observations.
In particular, in  Fig.~\ref{BHMF_fig} we show  the estimate 
by \citet{marconi} for the MBH mass function at $z=0$ with a thin solid orange line,
and we assume observational uncertainties, represented by shaded yellow areas, of 0.3 dex for the MBH mass
and 0.4 dex for the mass function (cf. \citet{shankar}, Fig. 5). 
The predictions of our model are instead shown in the left panel for the light-seed scenario,
and in the right-panel for the heavy-seed scenario, with error bars
representing the Poissonian errors due to to finite sample of galaxies that we simulate.
As can be seen, both scenarios produce MBH mass functions that are compatible with the observational estimate,
but they yield different predictions for $M_{\rm bh}\lesssim10^6 M_\odot$, where the mass function
is unconstrained by observations. In particular, the heavy-seed scenario obviously predicts a large number
of MBH with   $M_{\rm bh}\gtrsim M_{\rm seed}=10^5 M_\odot$, while the light-seed scenario predicts a flatter mass function.

In the left panel of Figs.~\ref{msigma_fig_light} (light-seed scenario) and~\ref{msigma_fig_heavy} (heavy-seed scenario), 
we show the parameterization of \citet{msigma_latest} for the $M_{\rm bh}-\sigma$ relation
originally discovered by \citet{msigma1,msigma2}, while in the right panel we show the
parameterization of \citet{haring_rix} for the $M_{\rm bh}-M_{b}$ relation, originally suggested by~\citet{magorrian}.
In particular,
both parameterizations are represented by thin solid orange lines, while 
the $1\sigma$ observational scatter is denoted by a yellow shaded area. We also show the median 
of our model's predictions for $M_{\rm bh}$
at any given $\sigma$ ($M_{b}$), as well as their 70\% and 90\%  
confidence regions.

More specifically, in order to calculate the median and confidence regions in our model we have
neglected the MBHs residing in satellite galaxies. Also, even for the central galaxies, we have only considered the MBHs residing in elliptical
galaxies (which we identify again with ones having $M_{b}/M_{\rm tot}>0.7$~\citep{white_morph}). This is
because the $M_{\rm bh}-\sigma$ relation is known to present a large scatter for late-type galaxies (cf. discussion in \citet{msigma_latest}
and their Fig. 1; see also \citet{msigma_extra1,msigma_extra2,msigma_extra3}), and the $M_{\rm bh}-M_{b}$ relation has been established in \citet{haring_rix} using a sample of 30 mostly elliptical galaxies.
The predictions of our model are obtained using the virial theorem to relate
the bulge velocity dispersion $V^2_{\rm bulge}$ to the bulge gravitational potential 
(which we calculate from the bulge density \eqref{bulge_density}),
and then calculating the line-of-site velocity dispersion as 
$\sigma^2=V^2_{\rm bulge}/K$, where the correction factor $K$ depends on the type of orbits of the stars in the
bulge. For instance, if the orbits were mainly along the line of sight, one would have
$K\approx1$, but $K\approx3$ for
nearly circular and isotropic orbits. Here, following \citet{correction_sigma}, we assume $K=2.1$. 
Overall, as can be seen, both the
light-seed and heavy-seed scenario reproduce the observational relations only qualitatively, because they predict a significant
number of ``outliers'' below the observed correlations, with masses $M_{\rm bh}\lesssim 10^8$. 
These outliers become even more numerous if
one includes the spiral and irregulars galaxies in the analysis, or if one also accounts for the MBHs 
in satellite galaxies. Physically, they therefore represent MBHs that will settle on the $M_{\rm bh}-\sigma$ 
and $M_{\rm bh}-M_{b}$ relations in the future, after merging with a MBH residing in a central galaxy already 
on the $M_{\rm bh}-\sigma$ and $M_{\rm bh}-M_{b}$ relations, or after a burst of star formation 
in the bulge following a disk disruption or a major merger. This population of outliers is also found
for instance in the model of \citet{volonteri_natarajan}, but remarkably not in the ``two-phase'' galaxy-formation
model proposed by \citet{twophases,cook1,cook2} (see  in particular Fig. 7 of \citet{cook2}), where bulges form at
high redshifts and MBHs settle on the $M_{\rm bh}-\sigma$ and $M_{\rm bh}-M_{b}$ relations earlier.
We will discuss this ``two-phase'' model and its possible consequences on the MBH redshift evolution
more in detail in Sec.~\ref{sec:predictions1}.

\subsection{Observables at $z>0$}

In this section we will examine the output of our model for two observables at  $z>0$, namely: 
the SFR density (SFRD) and the bolometric quasar luminosity density. 

There are at least two classes of methods to determine the redshift evolution of the SFRD, also known as the star-formation 
``cosmic history''~\citep{madau,lilly}. On the one hand,
the SFRD can be measured directly at various redshifts using instantaneous indicators. More specifically, emission from 
massive stars, which are short-lived compared to the typical star-formation timescales, can be used to extrapolate
 the IMF and obtain the SFR (see~\citet{hopkins,hopkins_err,hopkins_beacom,hopkins_beacom_err} for compilations
of recent data). A similar procedure can be followed 
with core-collapse supernova rates (see for instance~\citet{sfh_sn}),
for these stars too are short-lived compared to star formation. Alternatively, one can measure
 the distribution of stellar ages in nearby galaxies and use stellar population synthesis models to reconstruct the
star-formation cosmic history (see~\citet{fossil_sfh} for
a recent analysis).

\begin{figure}
\includegraphics[width=6.cm,angle=-90]{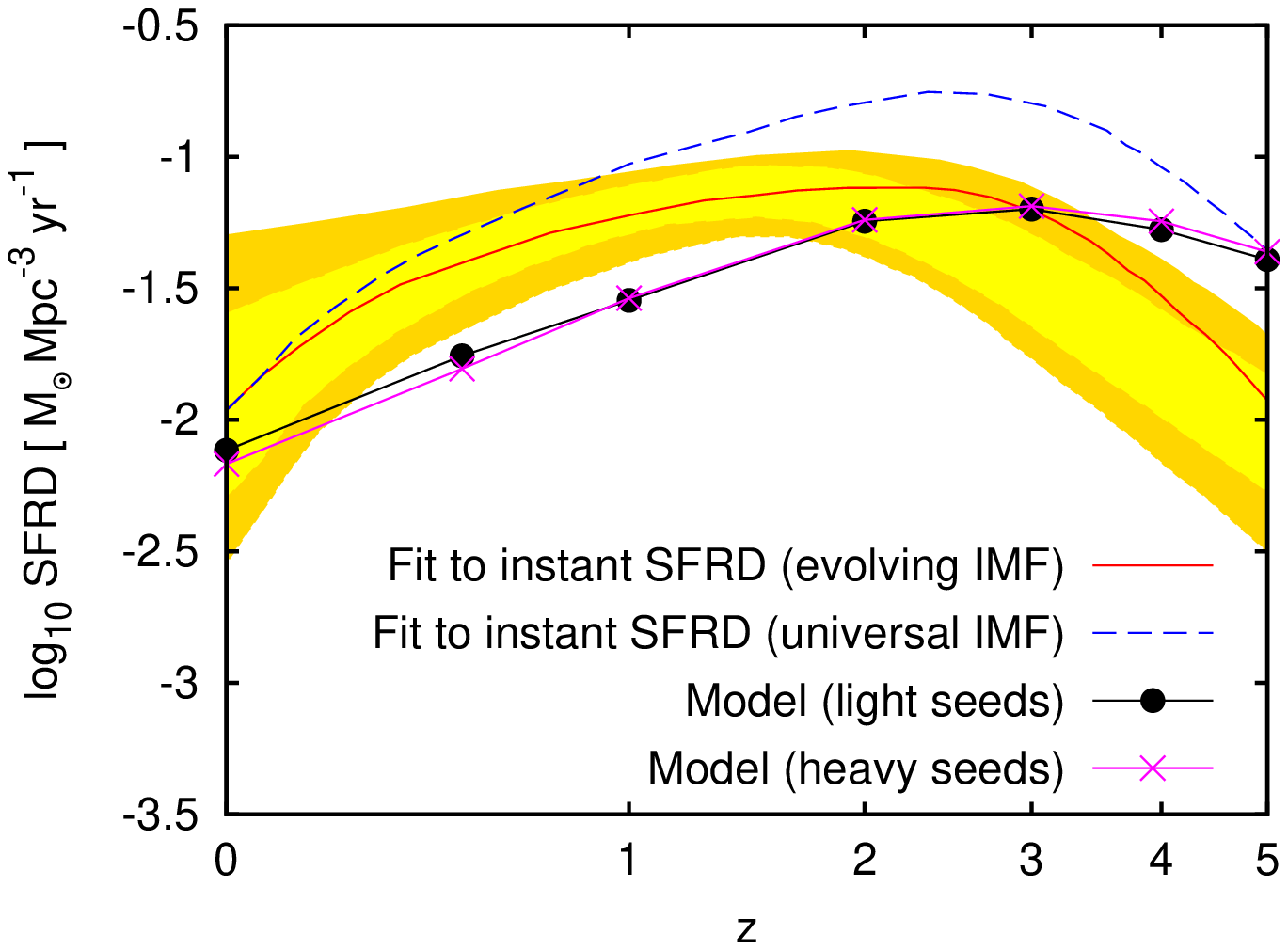}
\caption{
\label{madau_fig}
The cosmic star-formation history predicted by our model, compared to that
derived by \citet{wilkins} from observations of the stellar mass-density and assuming an
evolving IMF (yellow and orange shaded areas, representing the $1\sigma$ and $3\sigma$ uncertainty regions
of the observations). Also shown are fits to instantaneous SFRD indicators~\citep{wilkins},
assuming either a universal or an evolving IMF.
}
\end{figure}

\begin{figure}
\includegraphics[width=6.cm,angle=-90]{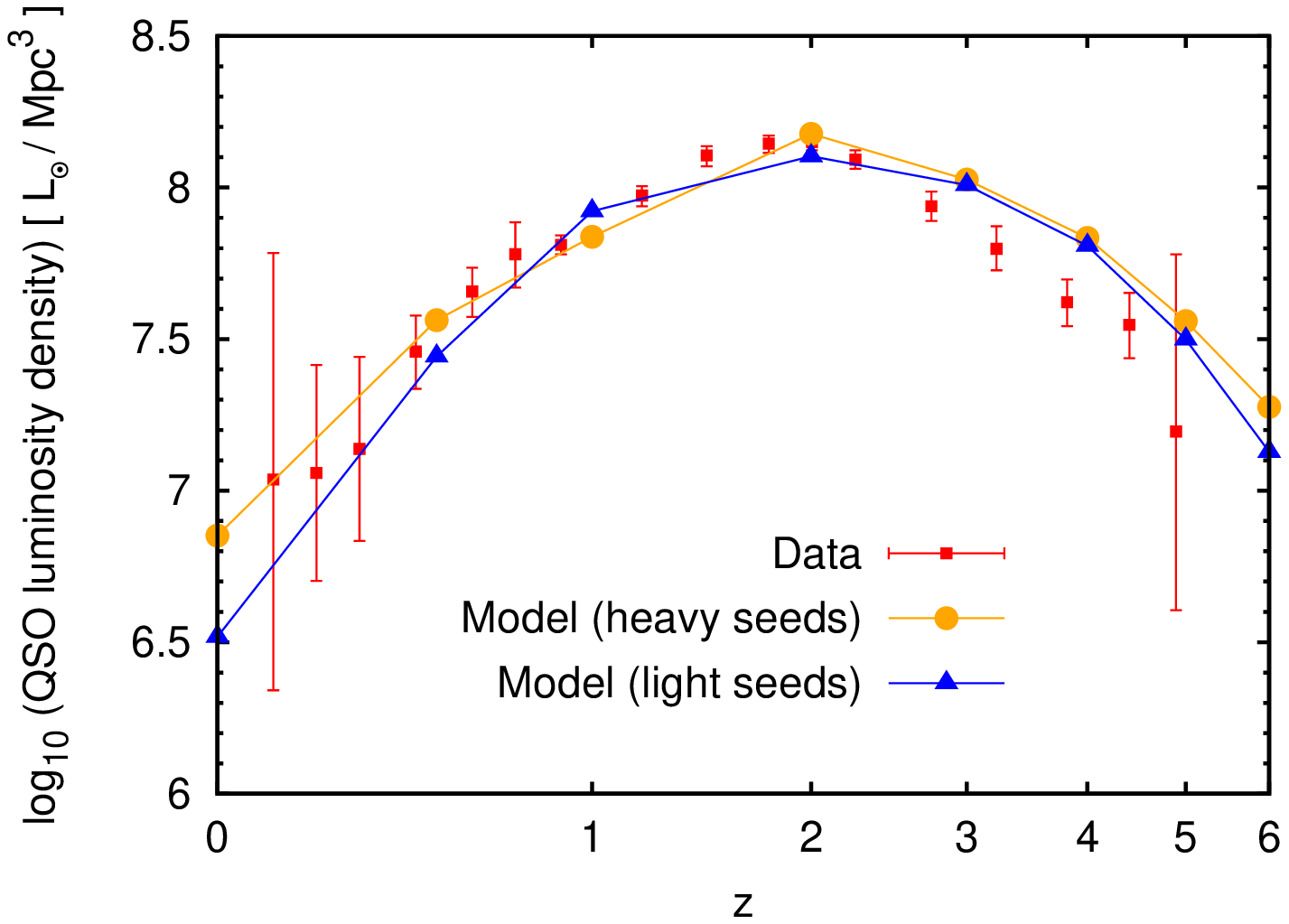}
\caption{
\label{Lqso_fig}
The prediction of our model for the bolometric quasar luminosity density as a function of redshift,
compared to data from \citet{Lqso}.
}
\end{figure}

On the other hand, integrating the cosmic star-formation history over redshift and correcting for the mass loss through supernovae and
stellar winds yields the stellar-mass density as a function of redshift. Conversely, the cosmic star-formation history can be extracted
from the stellar-mass density evolution, which can be measured independently with galaxy surveys and is sensitive to 
a larger range of masses 
than instantaneous indicators, which only probe the most massive stars. Moreover, instantaneous indicators are subject to greater uncertainties
due to the effects of dust obscuration. 
A compilation of recent measurements of the stellar-mass density as a function of
redshift is presented by \citet{wilkins}, who showed that the inferred star-formation history
agrees  with the one derived from instantaneous indicators for $z<0.7$, at least for suitable
choices of the IMF (see also \citet{hopkins_beacom_err}). At higher redshift, however,
the instantaneous indicators give larger SFRDs than the evolution of the
stellar-mass density, and this discrepancy peaks at $z\approx3$, where the difference is $\sim 0.6$ dex~\citep{wilkins,wilkins2}.

There are various possible explanations for this discrepancy, namely uncertainties in the effect of dust
on the stellar-mass and SFR estimates at high-redshifts~\citep{dust}, or incompleteness in the measured stellar-mass
density~\citep{incomplete_stellar_mass}. Another possibility is that the IMF  might evolve with redshift~\citep{wilkins,wilkins2,dave},
and there are in fact theoretical arguments and indirect observational evidence for that~\citep{van_dokkum,larson,swift_imf}.
In Fig.~\ref{madau_fig}, the yellow and orange shaded areas represent the $1\sigma$ and $3\sigma$ uncertainty regions
of the cosmic star-formation history derived by \citet{wilkins} from
the evolution of the stellar mass-density, assuming an
evolving IMF. The dashed and solid lines are \citet{wilkins}'s fits to instantaneous SFRD indicators, 
assuming respectively a universal and an evolving IMF. In this figure, we also show the output of our model. As can be seen the light-seed and 
heavy-seed scenarios predict essentially the same SFRD, but these predictions agree only qualitatively with the observations. While it should
be possible to amend our model to reproduce the observational data more closely, that would probably require including 
more free parameters than the four that we consider in this paper (as of now, we have no parameters regulating the cosmic star-formation history directly, besides those that we fixed by comparing to the $z=0$ observations of the previous section). 
We deem such a refinement premature due the discrepancies between different indicators and the problems in interpreting the SFRD observations
 that we mentioned above.

Finally, in Fig.~\ref{Lqso_fig} we show data from \citet{Lqso} for
 the bolometric quasar luminosity density as a function of redshift, and the
corresponding output of our model. As can be seen both the light-seed and heavy-seed scenarios do
a good job at reproducing the observations, and in particular the peak of the quasar luminosity
at $z\approx2$. We stress that while the normalization of the quasar luminosity density depends on
the parameter $A_{\rm res}$, which also regulates the normalization of the MBH mass function
and that of the $M_{\rm bh}-\sigma$ and $M_{\rm bh}-M_{b}$ relations (see previous section), 
its shape crucially depends on our free parameter $t_{\rm accr}$,
which we calibrate to reproduce the observations. Also, the predictions of our model as shown in Fig.~\ref{Lqso_fig}
only consider MBHs with bolometric luminosities $L>10^{10} L_\odot$. This is because
lower luminosities cannot be observationally resolved as 
confusion with normal star-forming and
starburst galaxies becomes an issue, and therefore they do not
enter in the analysis of \citet{Lqso}.

\section{The character of MBH mergers}

\label{sec:predictions1}

\begin{figure*}
 \begin{tabular}{m{0.33\textwidth}m{0.33\textwidth}m{0.33\textwidth}} 
  \includegraphics[height=0.34\textwidth,angle=-90]{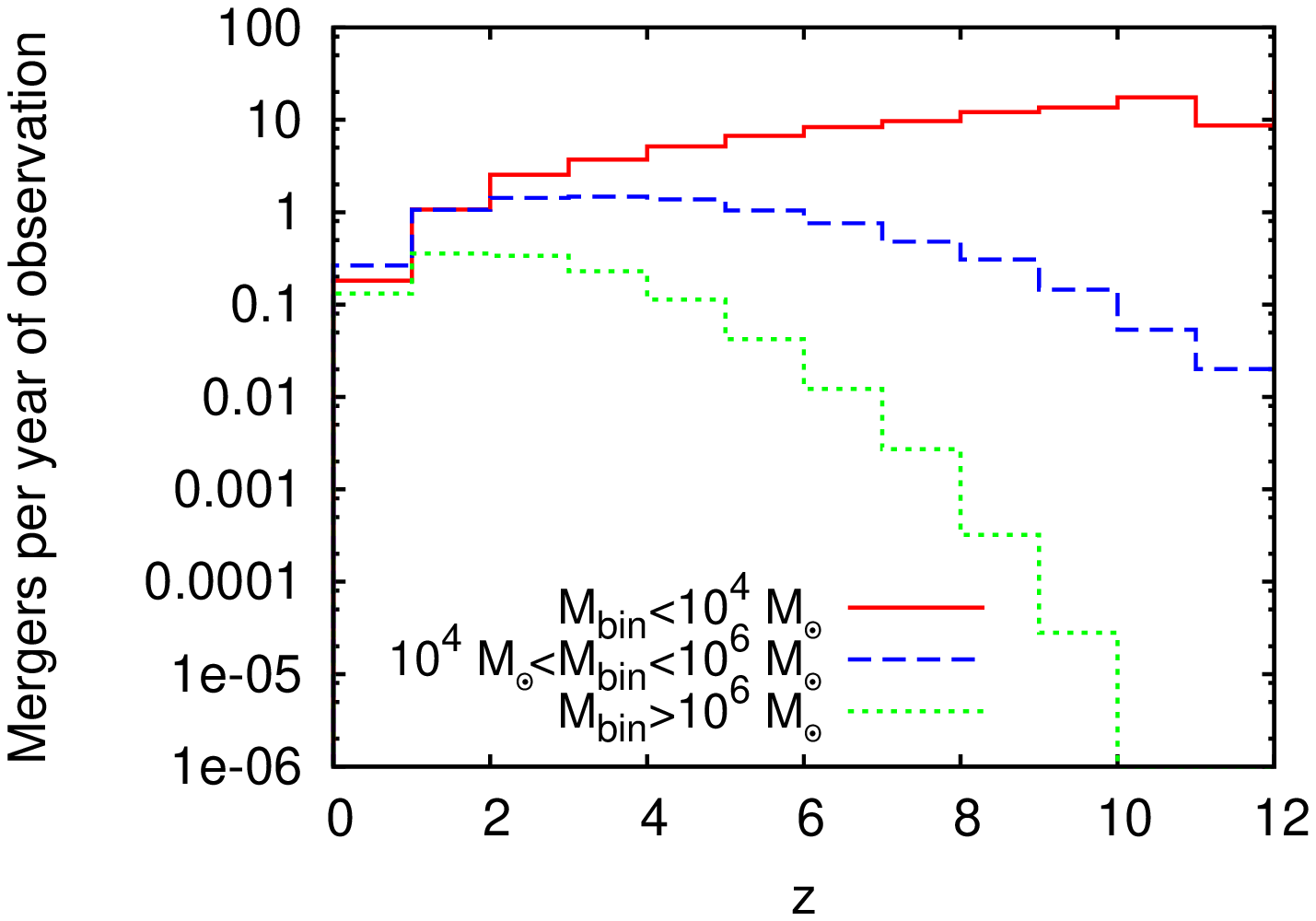}&
             \includegraphics[height=0.33\textwidth,angle=-90]{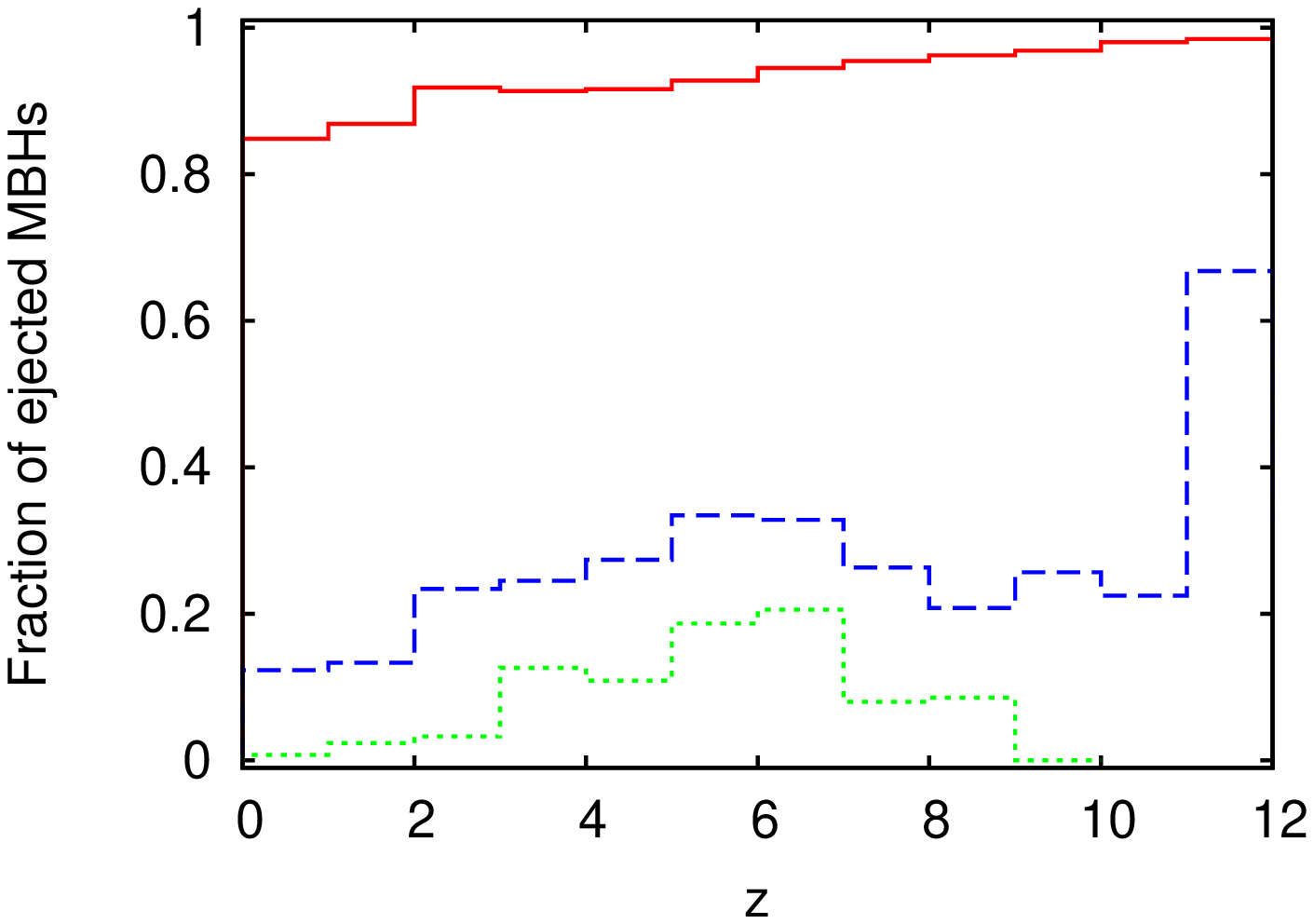}&
                \includegraphics[height=0.33\textwidth,angle=-90]{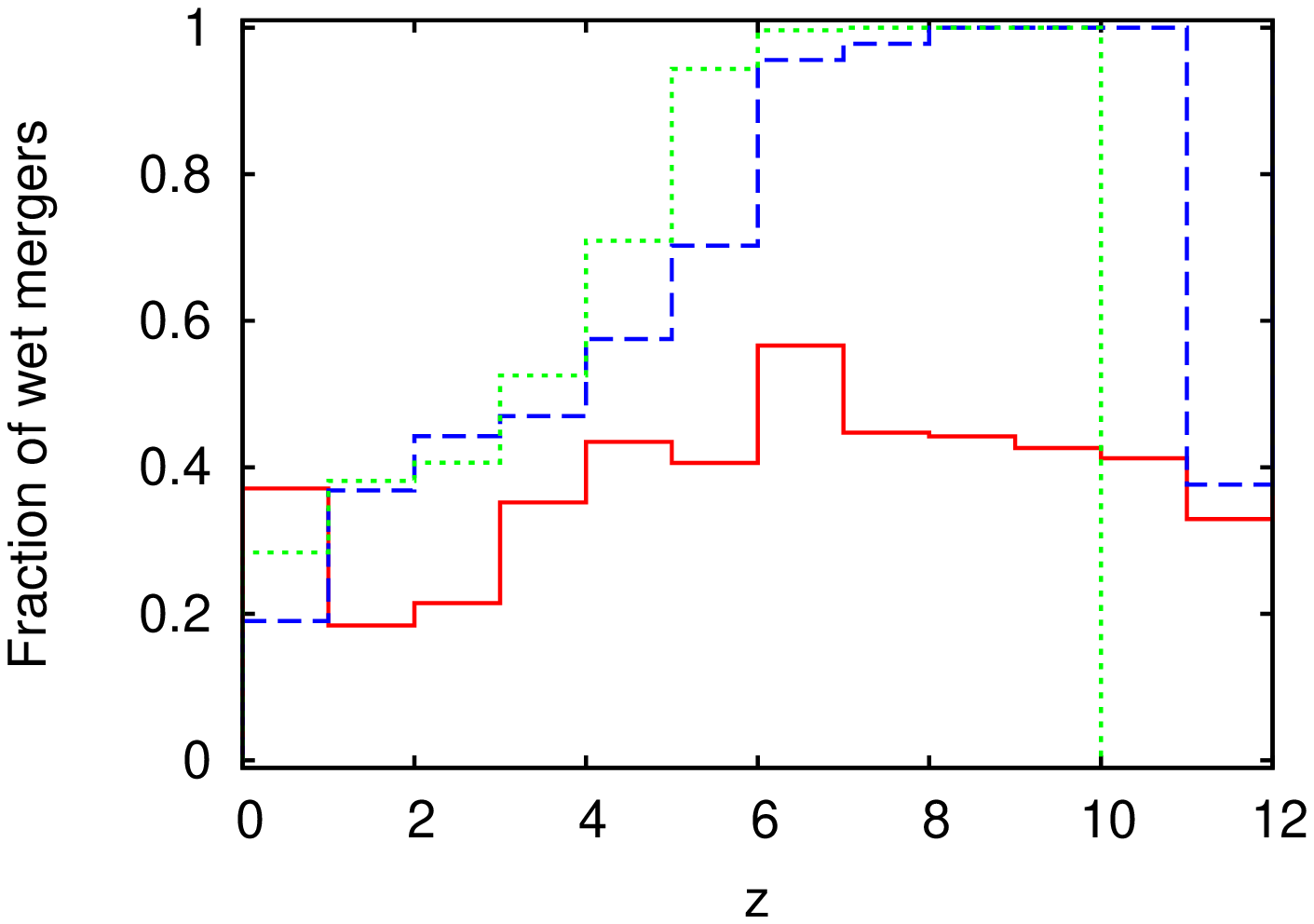} 
  \\
  \end{tabular}
\caption{The predictions of our model (in the light-seed scenario) for
the number of MBH mergers per year of observation (and unit redshift),
for the fraction of mergers producing a MBH that is ejected from the galaxy as a result of the 
gravitational recoil, and for the fraction of gas-rich (``wet'') MBH mergers,
 as a function of redshift and
in different mass ranges. We notice that the number of mergers with $M_{\rm bin}>10^6 M_\odot$ 
drops to zero for $z>10$, hence the fraction of wet mergers and that of ejected MBHs are not defined for
 $z>10$ in that mass range.
\label{mergers_light}}
\end{figure*}
\begin{figure*}
 \begin{tabular}{m{0.33\textwidth}m{0.33\textwidth}m{0.33\textwidth}} 
  \includegraphics[height=0.34\textwidth,angle=-90]{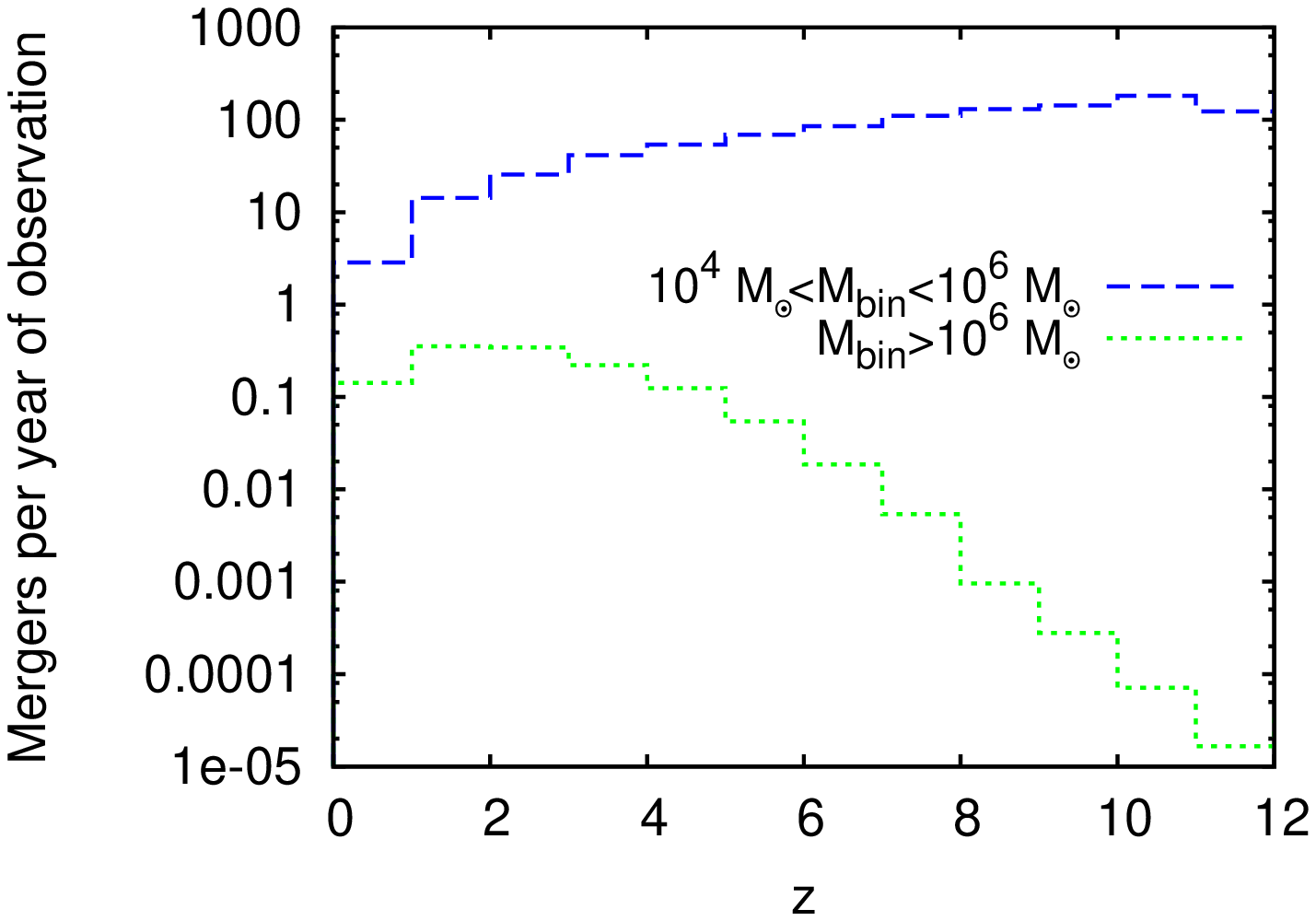}&
             \includegraphics[height=0.33\textwidth,angle=-90]{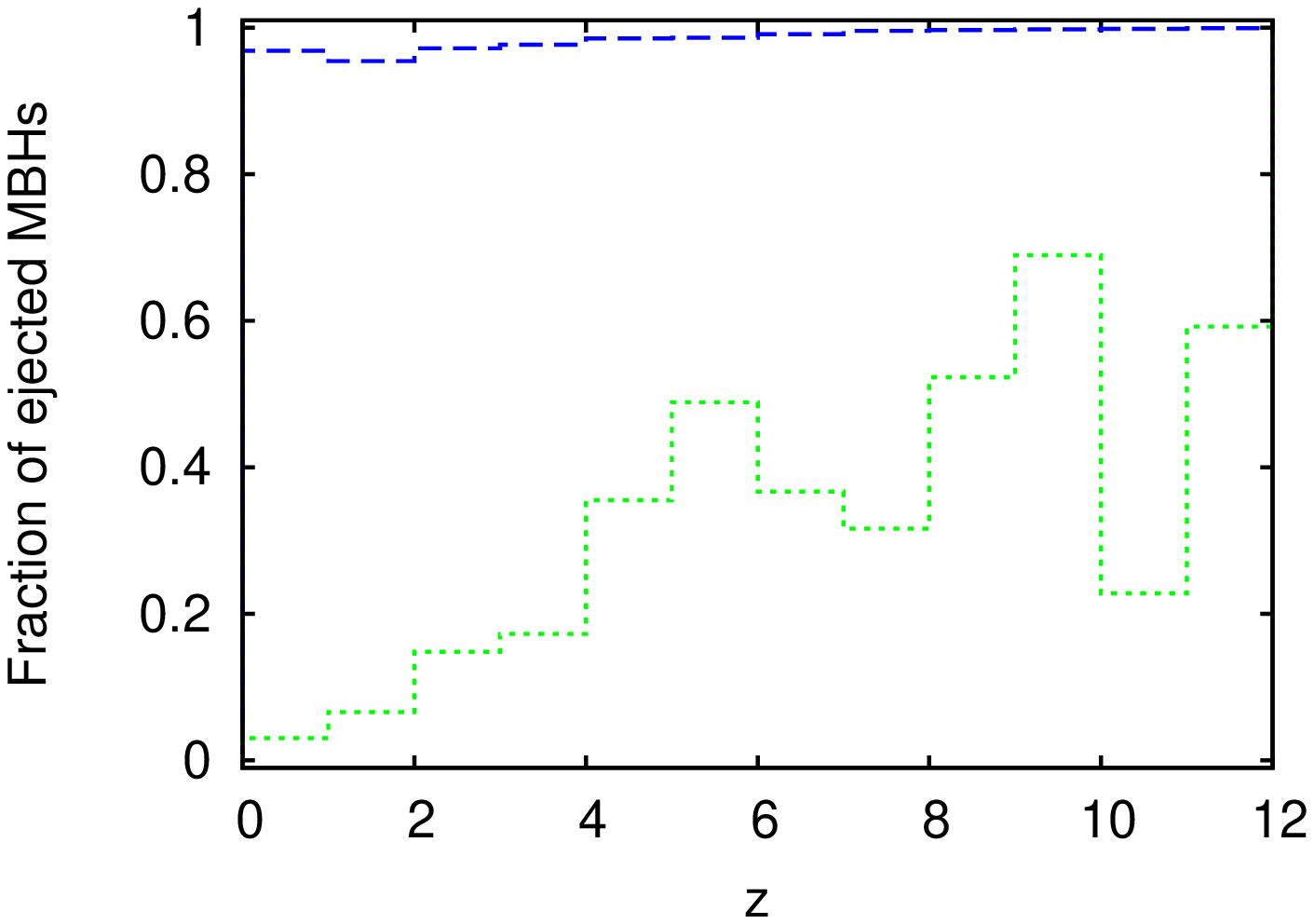}&
                \includegraphics[height=0.33\textwidth,angle=-90]{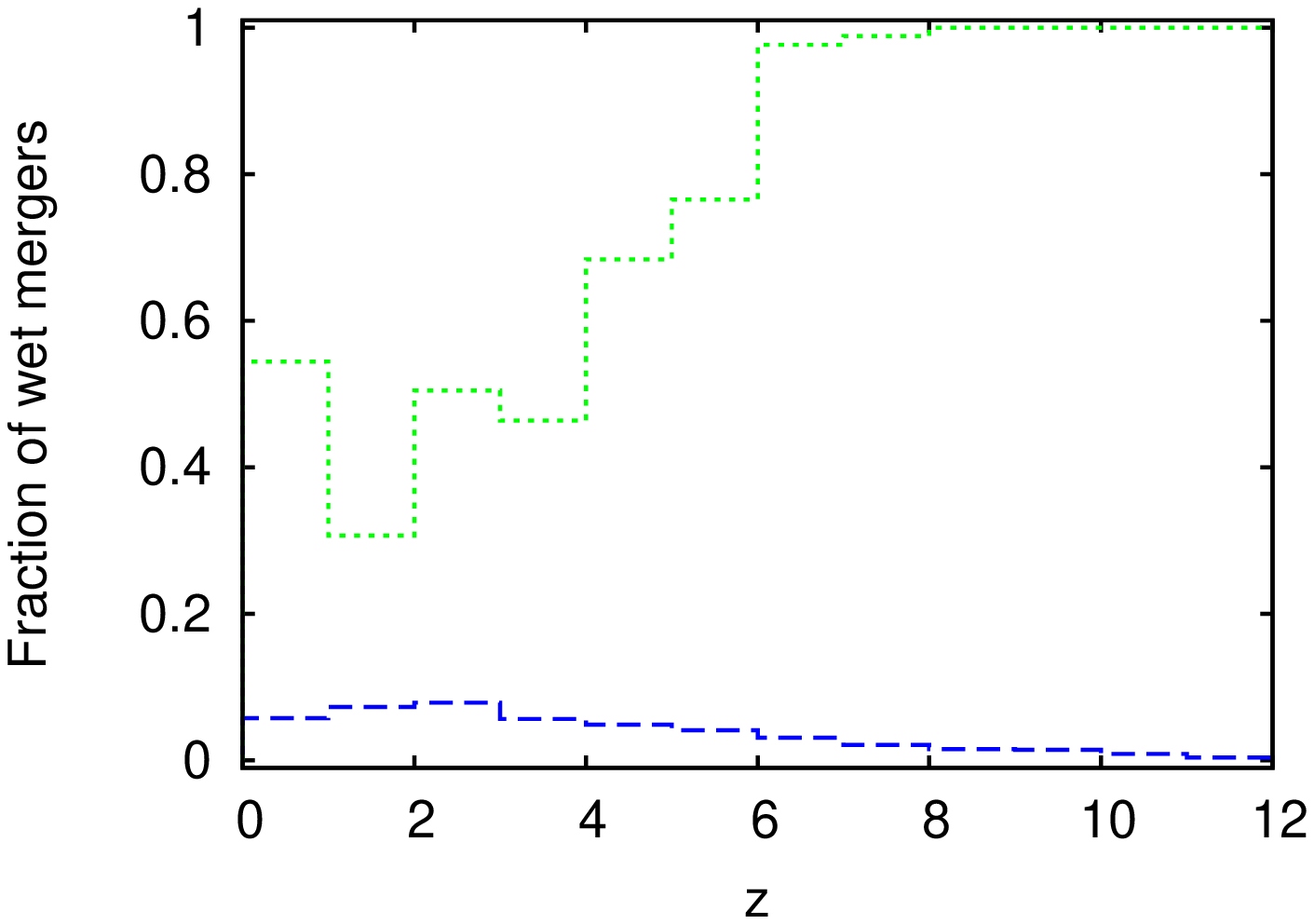} 
  \\
  \end{tabular}
\caption{The same as in Fig.~\ref{mergers_light}, but in the heavy-seed scenario. 
We notice that no mergers with  $M_{\rm bin}<10^4 M_\odot$ are present, because of the seed model.
\label{mergers_heavy}}
\end{figure*}

In this section we examine the predictions of our model for the MBH mergers, focusing in particular 
on their character, i.e. whether they happen in gas-rich (``wet'') nuclear environments
(where the gravito-magnetic torques align the MBH spins prior to the merger)  or in gas-poor (``dry'') ones, where the
spins prior to the merger are randomly oriented.

In Figs.~\ref{mergers_light} (light-seed scenario) and~\ref{mergers_heavy} (heavy-seed scenario) we present
predictions for the number of mergers observed in $1$ yr at $z=0$, for the
fraction of MBH remnants that are ejected from galactic spheroids~\footnote{As mentioned in
Sec.~\ref{sec:BHmergers}, our results for the ejection rates are overly pessimistic
if one is instead interested in the fraction of MBHs that are ejected from the whole composite
system (dark-matter halo and baryonic structures).}
as a result of the gravitational recoil,
and for the fraction of wet mergers,
in different ranges of the MBH binary's mass $M_{\rm bin}=M_{\rm bh\,1}+M_{\rm bh\,2}$ and as a function of redshift.
The predictions for the number of mergers, however, might be regarded as lower limits to the actual rates, 
because of the prescriptions described in Sec.~\ref{sec:DM} and aimed at keeping 
the computational time needed to follow the merger trees
up to high redshifts to an acceptable level. 

In spite of this note of caution, the results for the
light-seed scenario seem to confirm that LISA/SGO or a similar European mission
such as eLISA/NGO should detect at least a few MBH
merger events during its lifetime, although a detailed analysis will be necessary when the details of the mission
(e.g. its duration and sensitivity curve) are finalized. 
In fact, focusing for the moment on the $10^4 M_\odot < M_{\rm bin} <10^6 M_\odot$ and $M_{\rm bin} >10^6 M_\odot$ mass ranges
(which are the most relevant for LISA/SGO and eLISA/NGO),
we notice that the merger rates shown in Fig.~\ref{mergers_light} are qualitatively similar
 to the predictions of~\citet{sesana_light_heavy_rates} for the  ``BVRhf'' model (cf. their Fig. 1),
which yields event rates of a few per year for the original LISA mission.
In the heavy-seed scenario (Fig.~\ref{mergers_heavy}),
the events in the mass range $10^4 M_\odot<M_{\rm bin}<10^6 M_\odot$
are instead much more numerous ($\sim 680$ per year for $z<10$). This is indeed in agreement
with the analysis of \citet{sesana_light_heavy_rates},
who found that models with heavy seeds and large initial seed occupation numbers (i.e. their ``KBD'' model)
should give event rates of hundreds per year
for LISA/SGO or eLISA/NGO.
As for the low-mass range $M_{\rm bin}<10^4 M_\odot$, our light-seed scenario
predicts about 63 mergers per year for $z<10$. (Clearly, mergers in this mass range are not present
in the heavy-seed scenario.) These mergers may give a significant event rate 
for future third generation gravitational-wave detectors 
in the 0.1-10 Hz frequency band, such as DECIGO
or the Einstein Telescope~\citep{3rd_gen0,3rd_gen1,3rd_gen2}, or might
even be marginally detectable with LISA/SGO or eLISA/NGO at high redshifts if they have
$M_{\rm bin}\lesssim 10^4 M_\odot$.

\begin{figure*}
 \begin{tabular}{m{0.3\textwidth}m{0.3\textwidth}m{0.3\textwidth}} 
  \hskip-1cm\includegraphics[height=4cm]{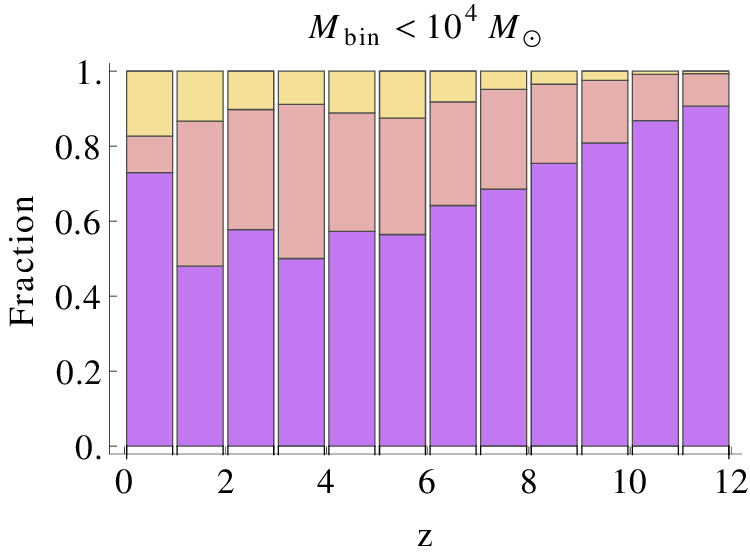}&
        \hskip-1cm     \includegraphics[height=4cm]{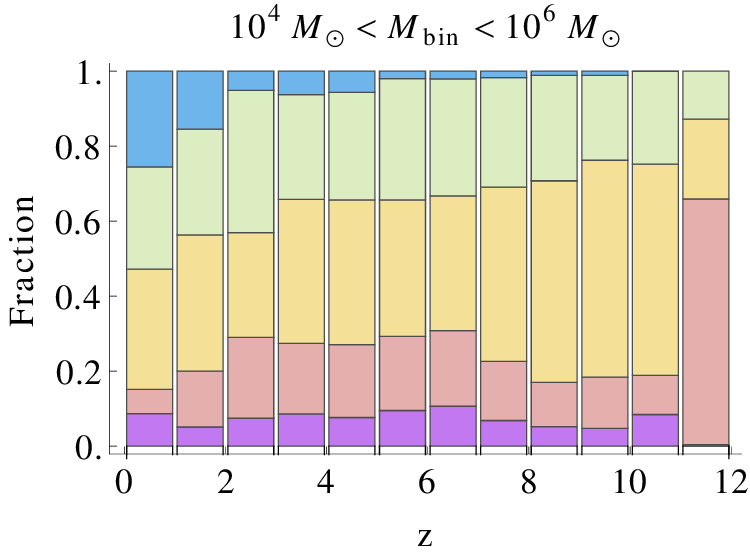}&
           \hskip-1cm     \includegraphics[height=4cm]{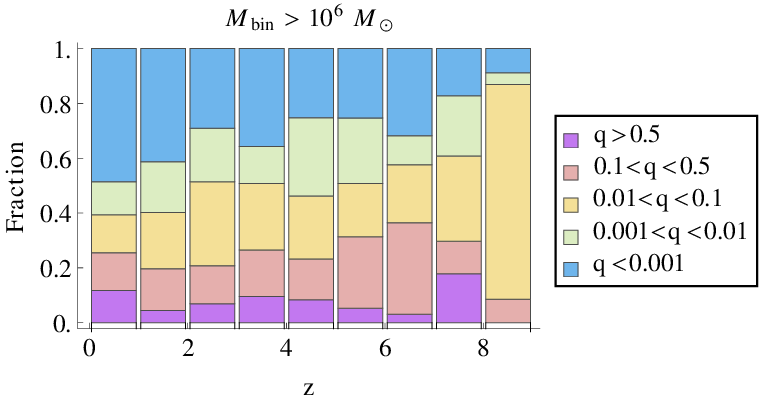} 
  \\
  \end{tabular}
\caption{The predictions of our model (in the light-seed scenario) for
the distribution of mass ratios $q=M_{\rm bh,2}/M_{\rm bh,1}$ (where $M_{\rm bh,2}\leq M_{\rm bh,1}$)
in MBH mergers, as a function of redshift and in different mass ranges.
\label{massratio_light}}
\end{figure*}

As for the fraction of MBHs ejected from galactic bulges and for the fraction of wet mergers, 
Figs.~\ref{mergers_light} and \ref{mergers_heavy}
confirm one's intuitive expectations. In the light-seed scenario, the mergers in the $M_{\rm bin}<10^4 M_\odot$
mass range almost always result in the MBH remnant being ejected. This is because the mass of 
MBHs remains small (i.e. $\sim M_{\rm seed}$) 
only if the host galaxies are disk-dominated, in which case little or no star formation happens in the bulge, and
no cold gas becomes available for the MBHs to grow. Naturally, if the bulge components are small they can hardly
retain the MBHs. Also, because the MBH seeds all have the same mass, mergers between seeds have mass ratio $q\sim 1$,
which gives larger recoil velocities (cf. the mass-ratio dependence of Eq.~\eqref{kick_v}).
Mergers in the  $10^4 M_\odot<M_{\rm bin}<10^6 M_\odot$ and $M_{\rm bin}>10^6 M_\odot$ mass ranges, instead, are less likely to result
in MBH ejections in the light-seed scenario. This is because if the MBHs have managed to grow 
beyond mere seeds, they have done so by accreting
the cold gas brought to galactic nuclei by star formation in the bulges (via radiation drag). As a result, the bulges are more
massive than in the case of mergers of MBH seeds, and they are more likely to retain the MBH remnants resulting
from mergers. Also, because the MBH have
grown to masses far from that of their seeds, the mergers are likely to involve mass ratios 
significantly different from $q=1$. This is indeed
shown in Fig.~\ref{massratio_light}, where we show the different 
mass ratios occurring in MBH mergers in the light-seed scenario, in different 
mass ranges. As can be seen, for $M_{\rm bin}<10^4 M_\odot$ the coalescing black holes 
often have comparable masses (essentially because many of them are still
seed black holes, or have grown little away from them), while the mass ratios become more varied at higher masses.

Similarly, as can be seen from Fig.~\ref{mergers_light}, the mergers in the mass 
range $M_{\rm bin}<10^4 M_\odot$ are wet only in $\sim30-40$\% of cases in the light-seed scenario, because if a significant amount of
gas were present in the galactic center, the black holes would have rapidly grown beyond $10^4 M_\odot$. In the   
$10^4 M_\odot<M_{\rm bin}<10^6 M_\odot$ and $M_{\rm bin}>10^6 M_\odot$ mass ranges, the mergers are mainly wet at high redshifts,
where a lot of cold gas is present in galactic nuclei,
but the fraction of wet mergers decreases as the cosmic evolution progresses, 
because the amount of gas shrinks as a result of both
accretion by the MBH and its quasar and radio-mode feedback on the galaxy (as well as 
a result of ram pressure and clumpy accretion, at large halo masses and low redshifts). 
Such a decrease in the fraction of gas-rich MBH mergers with redshift is a prediction that can be in principle
tested with gravitational-wave missions (e.g. LISA/SGO or eLISA/NGO). As already mentioned, the MBHs spins tend to be 
at least partially aligned in gas-rich environments
due to the gravito-magnetic torques exerted by the gas, while the spins are expected to
be randomly oriented and precess more strongly in
in gas-poor ones. 
Missions like LISA/SGO or eLISA/NGO should be able to tell a binary with nearly aligned spins from a binary with strongly
precessing ones by looking at the higher-order harmonics of the gravitational 
waveforms~\citep{harmonics,harmonics_err1,harmonics_err2,harmonics2}.

\begin{figure*}
 \begin{tabular}{m{0.12\textwidth}m{0.3\textwidth}m{0.01\textwidth}m{0.3\textwidth}m{0.3\textwidth}} 
&            
 \includegraphics[height=4cm]{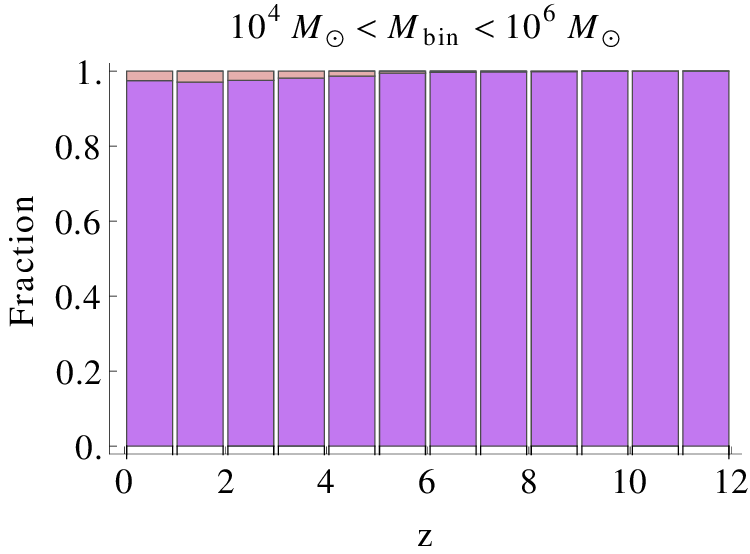}& &
             \includegraphics[height=4cm]{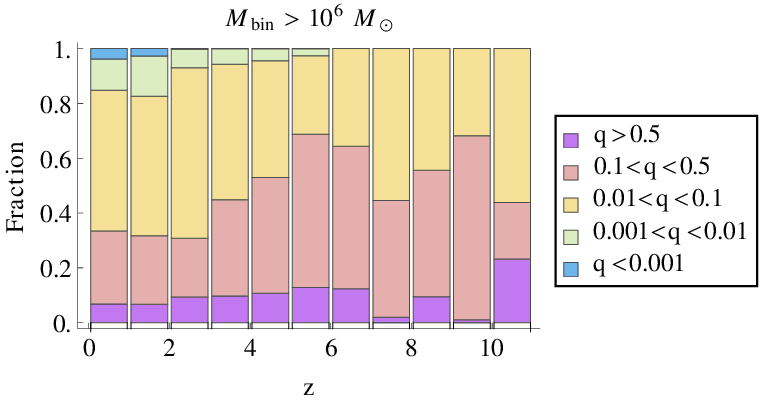} & 
  \\
  \end{tabular}
\caption{The same as in Fig.~\ref{massratio_light}, but in the heavy-seed scenario.
\label{massratio_heavy}}
\end{figure*}

The situation is similar in the heavy-seed scenario, as can be seen from Figs.~\ref{mergers_heavy} and~\ref{massratio_heavy}. 
The mergers
in the $10^4 M_\odot<M_{\rm bin}<10^6 M_\odot$ mass range happen between
MBHs that have barely grown away from their original  seeds of mass $M_{\rm seed}= 10^5 M_\odot$. Therefore, 
these mergers usually present comparable masses, happen in dry environments, and likely eject
the resulting MBH from the galactic spheroid.  
The mergers in the $M_{\rm bin}>10^6 M_\odot$ mass range, instead, consist of MBHs that have grown significantly larger than their 
seeds thanks to a gas rich environment and to significant star formation in the bulge, and therefore tend to be retained
in spheroids after mergers, and to merge in ``wet'' environments (at least at high redshifts, where circumbinary disks
have not yet been destroyed by accretion, AGN feedback, ram pressure and clumpy accretion).

We stress that in spite of the rather high ejection probability in the mergers between seeds
that one can observe in  Figs.~\ref{mergers_light} and~\ref{mergers_heavy}, our model correctly reproduces the
property of MBHs at low redshifts, as we showed in the previous section. In other words, the gravitational recoil
does not succeed at rooting out all the MBHs from their host galaxies. This is essentially
 because the occupation fraction of black-hole seeds
at high-redshifts is smaller than 1 (cf., in Sec.~\ref{sec:DM},  how we 
populate high-redshift halos with black-hole seeds), and this is enough to ensure that MBHs 
survive to low redshifts~\citep{lippai,zoltan_occ_frac,volonteri_kick,volonteri_kick2}. 
Also, as noted by \citet{schnittman_kick}, even if we populated
all halos with black-hole seeds at high redshift, 
the gravitational recoil in the first generation of
mergers would automatically decrease the MBH occupation fraction, 
and even in the case of very high ejection probabilities the occupation
fraction would settle to $\sim 50$ \% in a few more merger generations.

Another possibility would be to adopt the  galaxy-formation model of \citet{twophases,cook1,cook2}, in which
the baryonic evolution is driven by
the two-phase structural evolution of the dark-matter haloes~\citep{zhao1,zhao2,mo_mao,dieman}, and
presents two distinct phases: an
early ``fast collapse'' phase, where the dark-matter core structure is built 
through a series of violent merger events, corresponding
to an epoch where baryonic material collapses to directly form spheroidal structures (bulges);  
and a late ``slow collapse'' phase, where potentially large
amounts of material are added to the halo outskirts without affecting
the central regions, giving rise to the quiescent growth of disk structures
around the previously-formed bulges. This model can potentially 
reproduce the observed downsizing of baryonic structures more
naturally than standard semianalytical galaxy-formation models~\citep{twophases}, 
and may potentially have effects on the predictions for the MBH merger rates
and mass evolution. As already mentioned in Sec.~\ref{sec:calibration}, the standard
galaxy-formation model that we adopt here, in which disk galaxies form first and give rise to
spheroids by instabilities and major mergers, predicts the presence of a significant number of outliers
in the $M_{\rm bh}-\sigma$ and $M_{\rm bh}-M_{b}$ relations, while MBHs settle on the 
$M_{\rm bh}-\sigma$ and $M_{\rm bh}-M_{b}$ relations earlier in the case 
of this ``two-phase'' model (see \citet{cook2}).
This is because spheroidal structures form first, and as a result of radiation drag they feed the MBHs at higher redshifts
than in the ``standard'' model. Clearly, in such a scenario one does not have high-redshift mergers
between galaxies with little or no bulges, and therefore the ejection rate of black-hole seeds should be 
significantly reduced. Also, the earlier growth of the MBHs may boost the event rates for LISA/SGO or eLISA/NGO in the light-seed scenario.
We will explore in detail the effects of this alternative galaxy-formation model on 
the MBH mergers and evolution in a future paper.

Finally, as mentioned in the introduction, the results of this section depend on our assumption that
the MBH spins gets aligned by gravitomagnetic torques in gas-rich environments. However, due to 
processes such as star formation and feedback in the circumnuclear
disk~\citep{dotti,dotti2011} or clump formation in high-$z$ disk galaxies~\citep{bournaud,dubois}, this
alignment is likely to be only partial, which may result in higher ejection rates for the MBHs resulting from mergers.

\begin{figure*}
 \begin{tabular}{m{0.1\textwidth}m{0.3\textwidth}m{0.3\textwidth}m{0.3\textwidth}} 
 \hskip-0.5cm \includegraphics[width=0.12\textwidth]{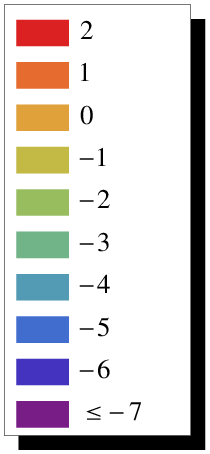}&
             \includegraphics[width=0.3\textwidth]{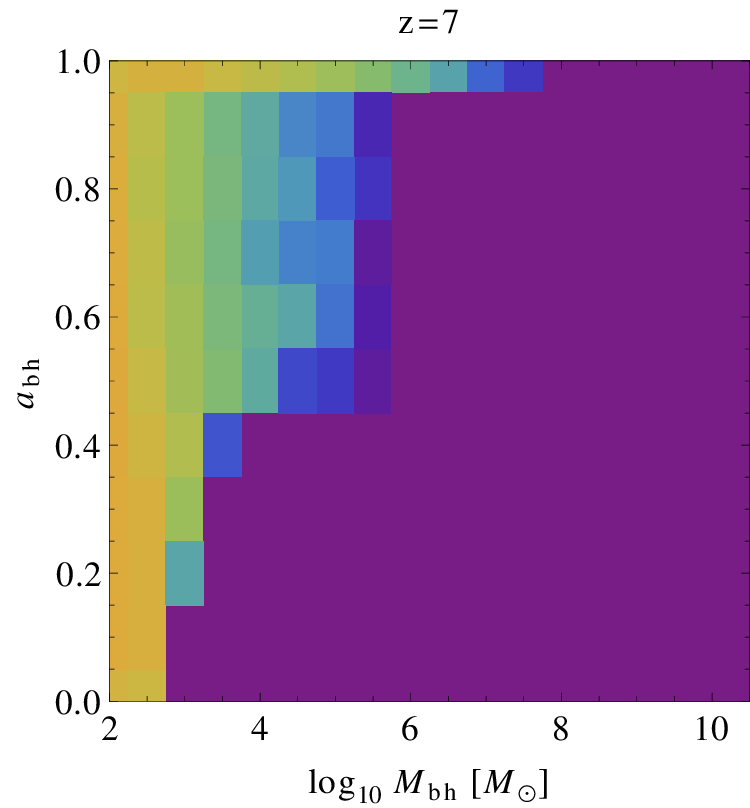}&
                \includegraphics[width=0.3\textwidth]{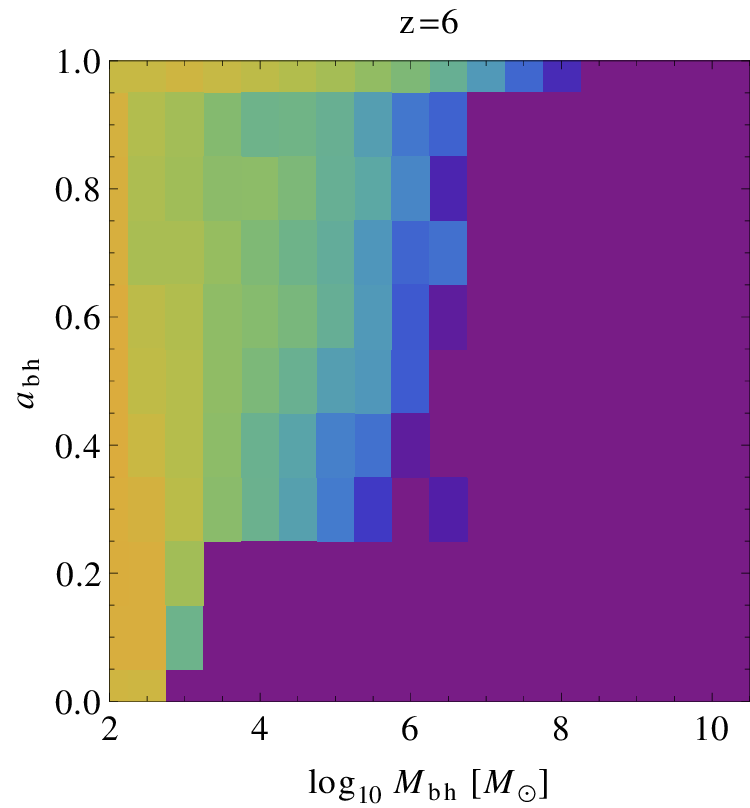} &
                \includegraphics[width=0.3\textwidth]{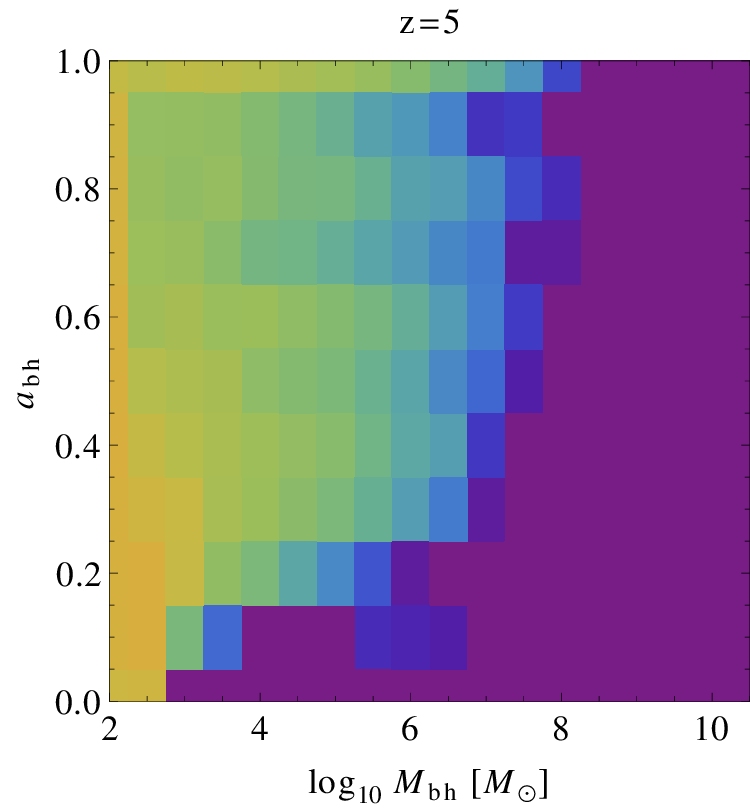} 
\\
 \vskip1cm
  \begin{tabular}{m{0.33\textwidth}m{0.33\textwidth}m{0.33\textwidth}}
\includegraphics[width=0.3\textwidth]{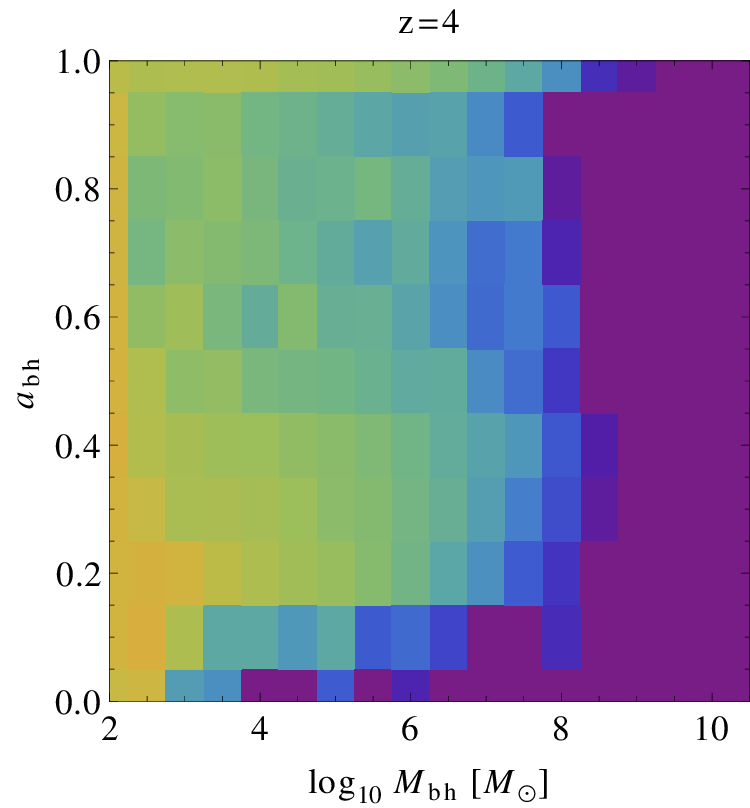} &
 \includegraphics[width=0.3\textwidth]{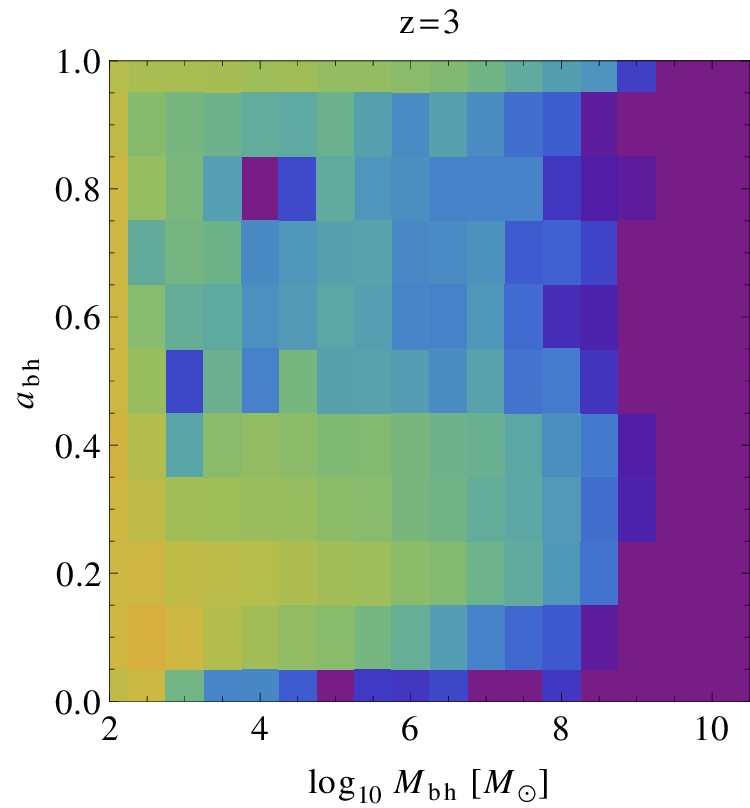} &
 \includegraphics[width=0.3\textwidth]{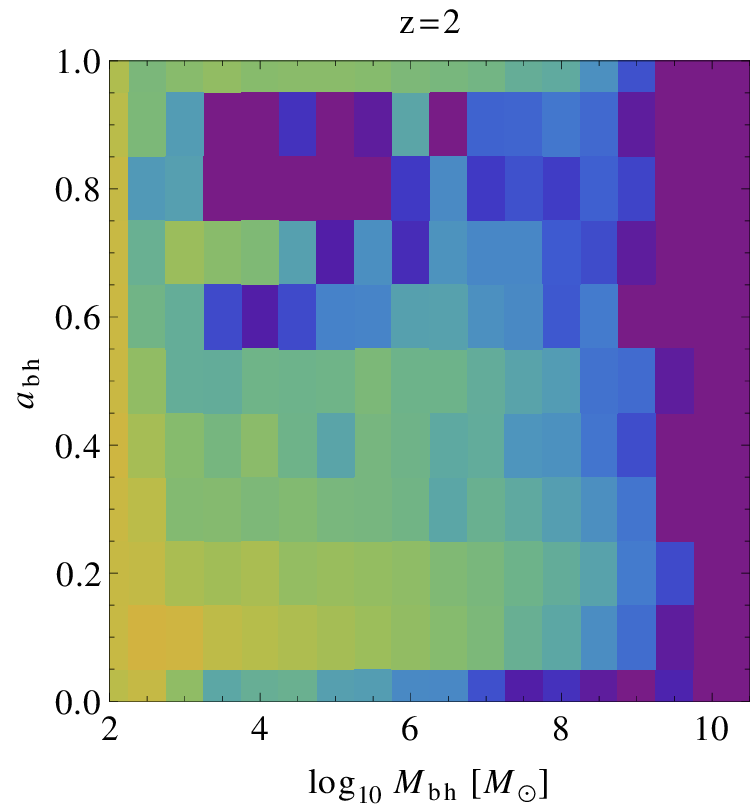} 
  \\
 \vskip 1cm
  \begin{tabular}{m{0.33\textwidth}m{0.33\textwidth}m{0.33\textwidth}}
   \includegraphics[width=0.3\textwidth]{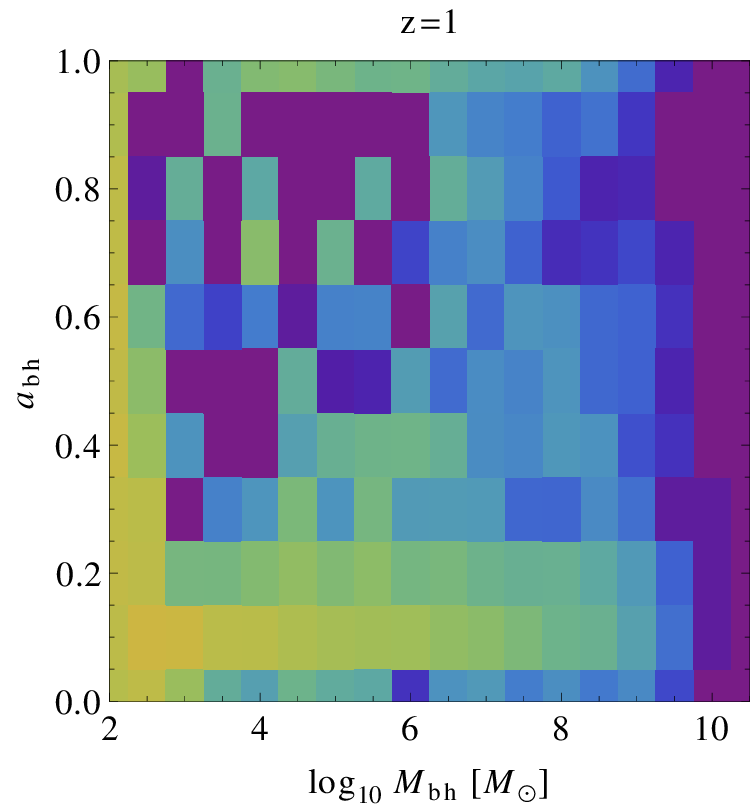} &
   \includegraphics[width=0.3\textwidth]{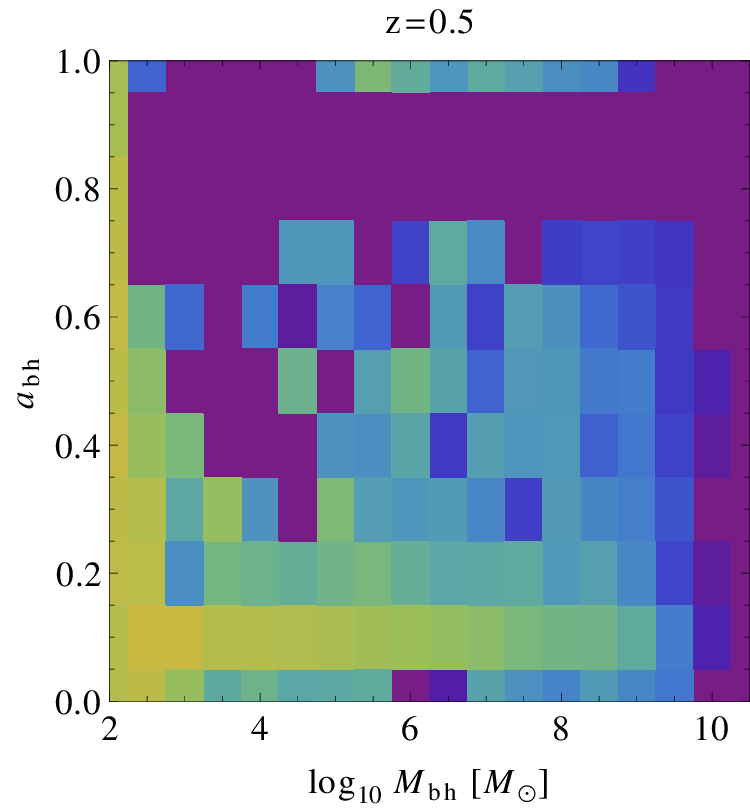} &
 \includegraphics[width=0.3\textwidth]{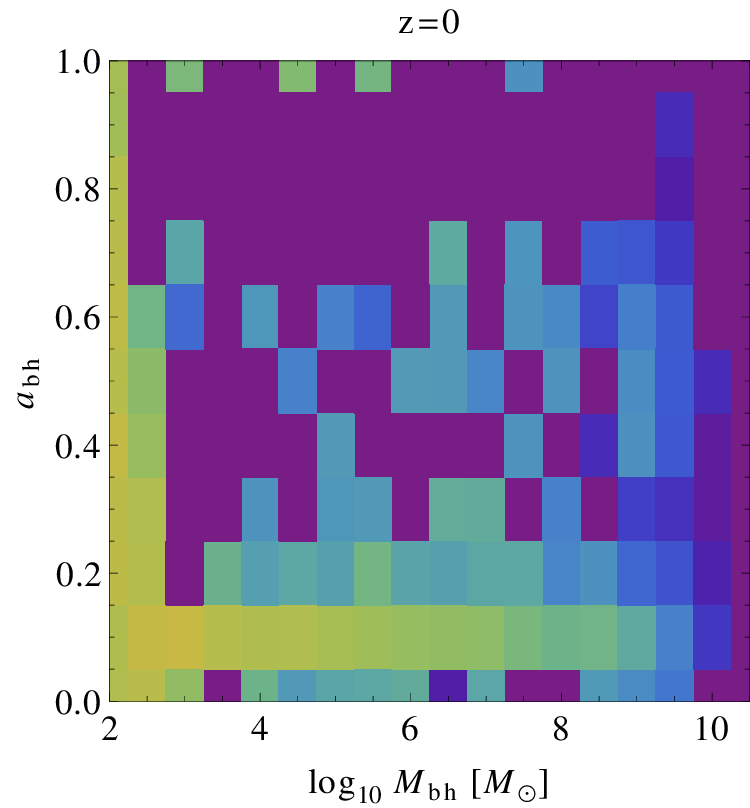} 
  \\
  \end{tabular}
  \\
  \end{tabular}
  \\
 \end{tabular}
\caption{The evolution of the MBH masses and spins with redshift, as predicted by our model
in the light-seed scenario. The color code represents the $\log_{10}$ of the density of
MBHs per unit (logarithmic) mass and unit spin, i.e.
$\log_{10}({\rm d}{\phi_{\rm bh}[{\rm Mpc}^{-3}]}/{{\rm d}a})=
\log_{10} ({{\rm d}^2 n_{\rm bh} [{\rm Mpc}^{-3}]}/({{\rm d}\log_{10} M_{\rm bh} 
 [M_\odot]\, {\rm d} a_{\rm})}))\,.$ 
\label{spin_evolution_light}}
\end{figure*}

\begin{figure*}
 \begin{tabular}{m{0.1\textwidth}m{0.3\textwidth}m{0.3\textwidth}m{0.3\textwidth}} 
  \hskip-0.5cm \includegraphics[width=0.12\textwidth]{legendColor.eps}&
             \includegraphics[width=0.3\textwidth]{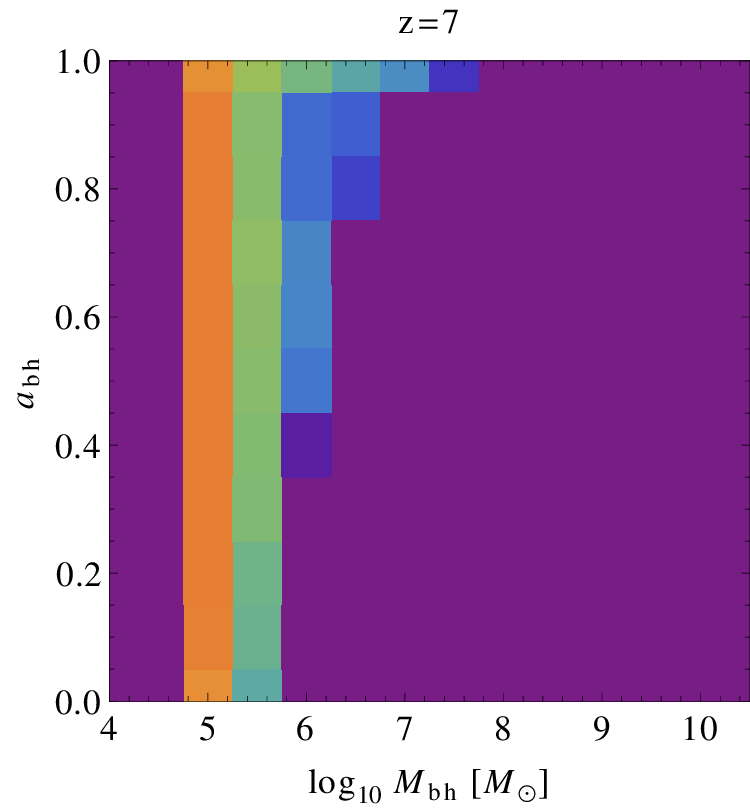}&
                \includegraphics[width=0.3\textwidth]{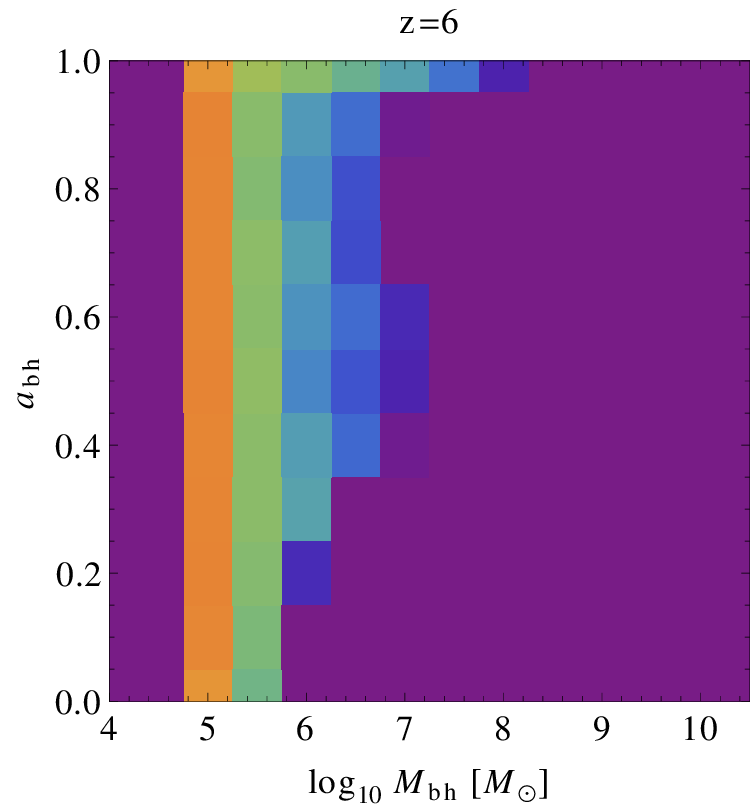} &
                \includegraphics[width=0.3\textwidth]{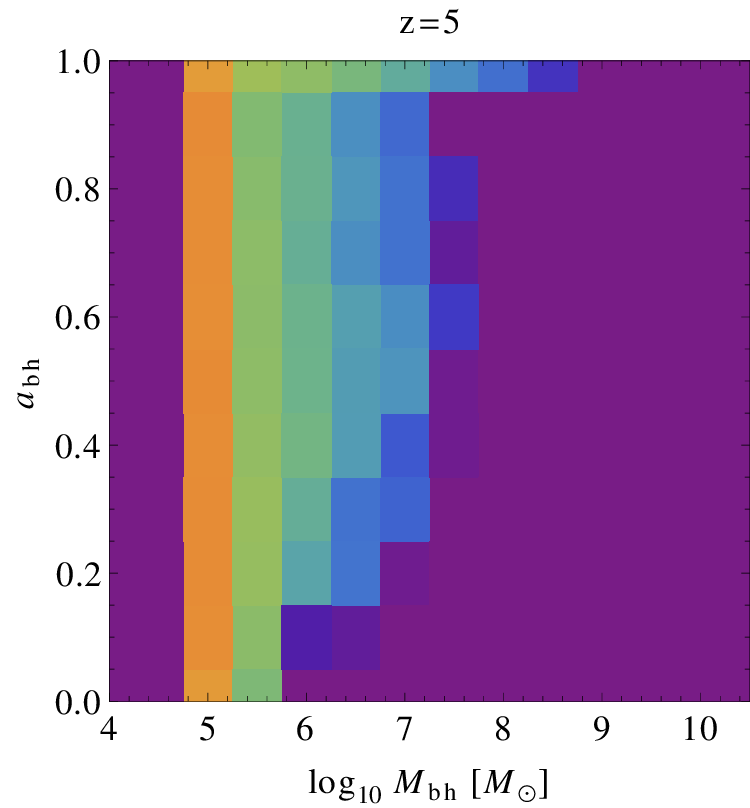} 
\\
 \vskip1cm
  \begin{tabular}{m{0.33\textwidth}m{0.33\textwidth}m{0.33\textwidth}}
\includegraphics[width=0.3\textwidth]{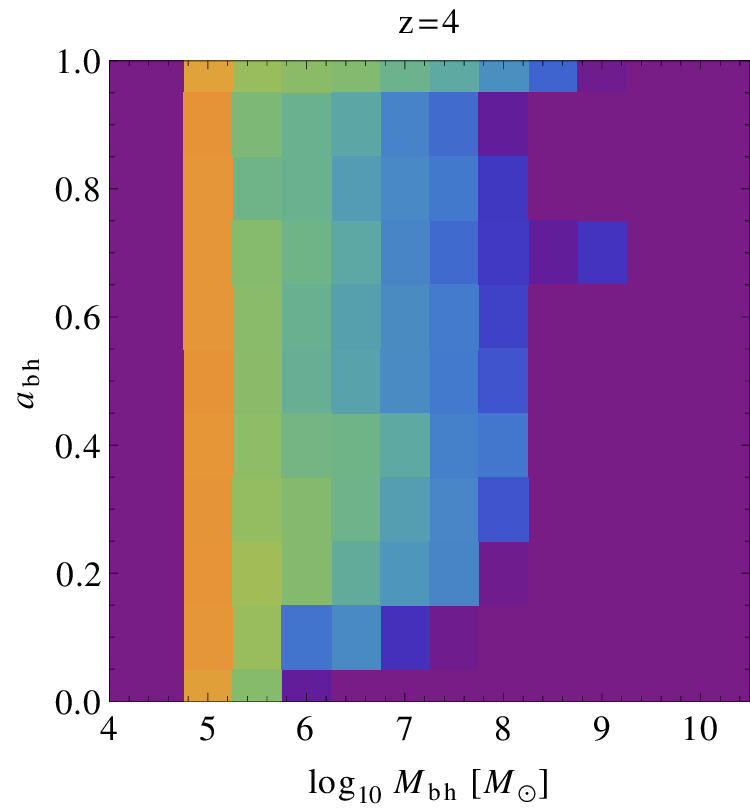} &
 \includegraphics[width=0.3\textwidth]{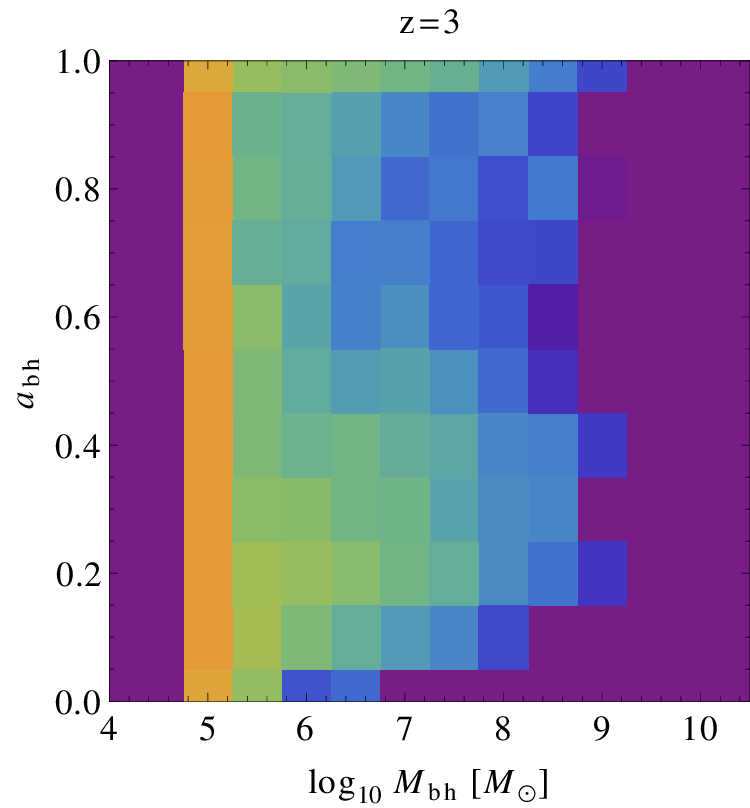} &
 \includegraphics[width=0.3\textwidth]{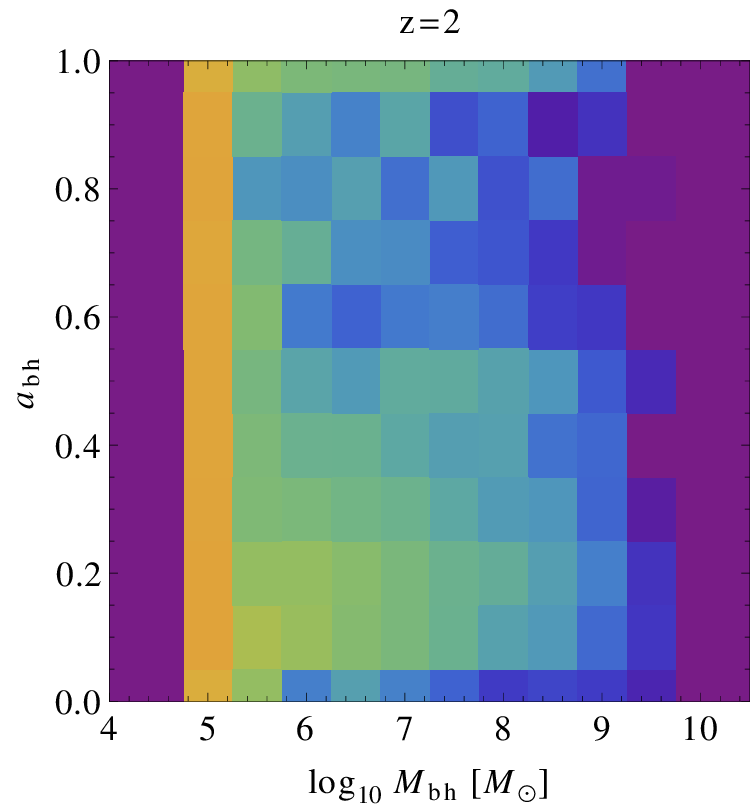} 
  \\
 \vskip 1cm
  \begin{tabular}{m{0.33\textwidth}m{0.33\textwidth}m{0.33\textwidth}}
   \includegraphics[width=0.3\textwidth]{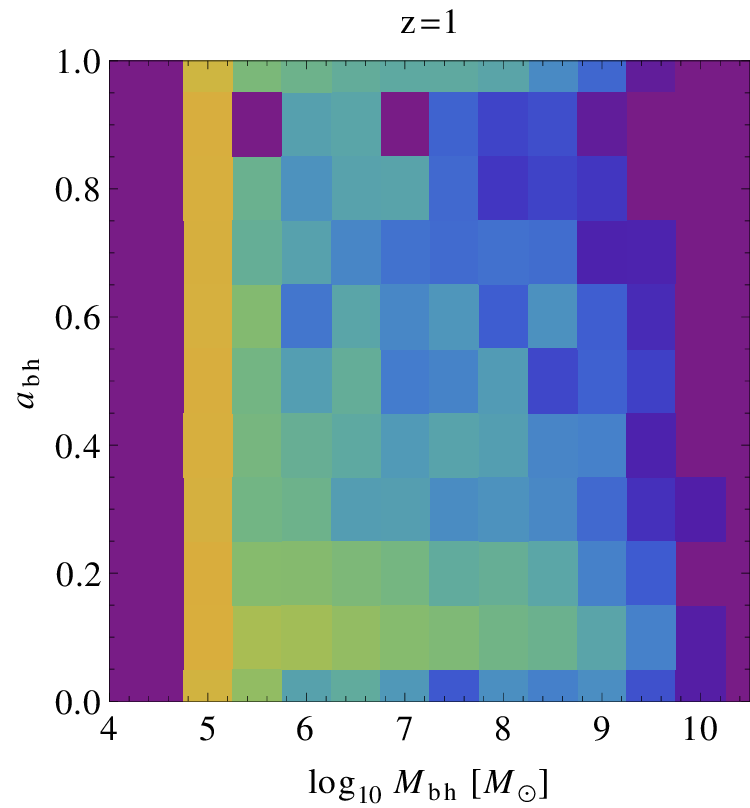} &
   \includegraphics[width=0.3\textwidth]{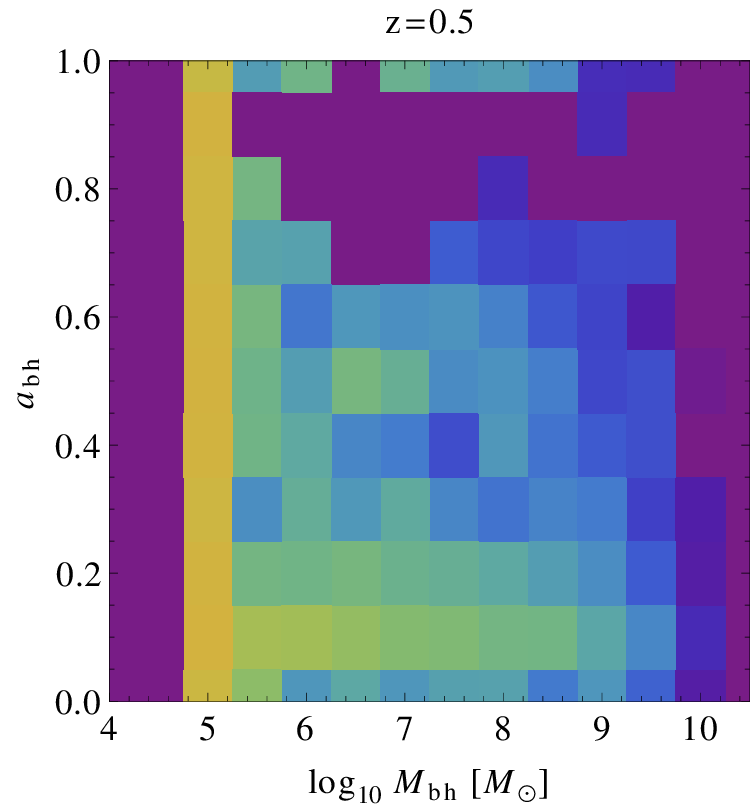} &
 \includegraphics[width=0.3\textwidth]{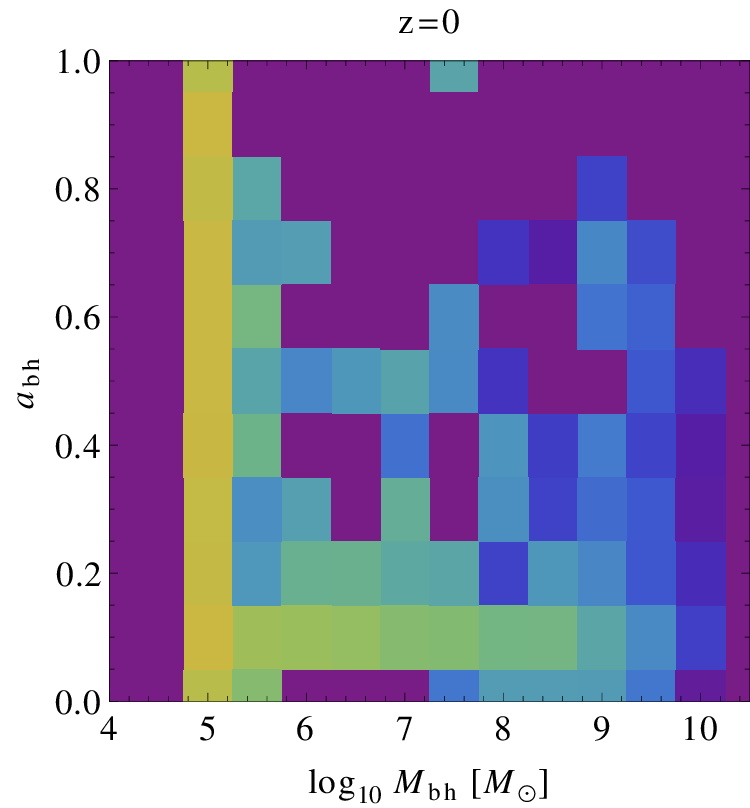} 
  \\
  \end{tabular}
  \\
  \end{tabular}
  \\
 \end{tabular}
\caption{The same as in Fig.~\ref{spin_evolution_light}, but in the heavy-seed scenario.
\label{spin_evolution_heavy}}
\end{figure*}

\section{The evolution of the MBH spins}

\label{sec:predictions2}

As a final application of our model, we study the redshift evolution of the MBH spins.  These
predictions will be readily testable by LISA/SGO or eLISA/NGO, which
will  measure the black-hole masses and spins with astonishing accuracy
($\sim 10^{-3}$ for the masses and $10^{-2}$ for 
the spins, see~\citet{berti_buonanno_will,harmonics,harmonics_err1,harmonics_err2,harmonics2})
and without the systematic uncertainties typically affecting electromagnetic (e.g. X-ray) determinations.

In Figs.~\ref{spin_evolution_light} (light-seed scenario) and~\ref{spin_evolution_heavy} (heavy-seed scenario), 
we present results for the distribution of masses and spins
of the MBHs residing in isolated galaxies or in the central galaxies of groups or clusters 
(i.e., we do not consider the MBHs residing in satellite galaxies), at redshifts ranging from $z=7$ to $z=0$. The color code
represents the $\log_{10}$ of the density of MBHs  per unit (logarithmic) mass and unit spin,
$\log_{10}({\rm d}{\phi_{\rm bh}[{\rm Mpc}^{-3}]}/{{\rm d}a})=
\log_{10} ({{\rm d}^2 n_{\rm bh} [{\rm Mpc}^{-3}]}/({{\rm d}\log_{10} M_{\rm bh} 
 [M_\odot]\, {\rm d} a_{\rm})}))\,.$
As can be seen, in the light-seed scenario, already at $z=7$ the MBH distribution has been skewed towards large spins from 
the initial uniform spin distribution of the seeds (still visible at $M_{\rm seed}=150 M_\odot$). 
This is because at high redshifts, where the AGN feedback is still ineffective and the MBHs small,  
large amounts of gas are present in galactic nuclei,
and the MBH spins grow  as a result of wet, spin-aligned mergers (cf. Sec.~\ref{sec:predictions1})
and most importantly because 
 accretion onto the MBHs is coherent and spins them up (cf. Sec.~\ref{smbh_evolution}).
We stress, as already mentioned in the introduction, that effects such as star formation and feedback in the circumnuclear disk, or
the formation of clumps in high-redshift disk galaxies fed by cold streams, may at least partially randomize
the accretion flows in gas-rich environments. As a result, the large spins $a_{\rm bh}\sim1$ shown in Figs.~\ref{spin_evolution_light} 
and~\ref{spin_evolution_heavy} at high redshift may be substantially reduced (e.g. \citet{dotti2011} show
that for quasars in merger remnants, sustained accretion results asymptotically in spins parameters $a_{\rm bh}\sim 0.7-0.9$).

At smaller redshifts this trends gets modified because the cold gas in the nuclear regions of galaxies becomes scarcer,
hence mergers tend to happen in dry environments (cf. Sec.~\ref{sec:predictions1})
and accretion turns chaotic. Chaotic accretion, in particular, appears to be the main
driving force behind the spin evolution in this phase, as can be seen from the appearance of 
a large number of MBHs with spin parameter $a_{\rm bh}\sim 0.1$. This is indeed what would be expected in
a chaotic-accretion scenario, where the black-hole spin oscillates around a small non-zero value~\citep{chaotic}.
In our model, the value $a_{\rm bh}\sim 0.1$ is easily explained. As mentioned in Sec.~\ref{smbh_evolution},
accretion turns chaotic when the mass of the gaseous reservoir drops below the black hole's mass (``dry'' environment).
Assuming that the MBH is almost maximally spinning as a result of the previous phase of coherent accretion, we can calculate
the spin of the MBH when the reservoir has been completed accreted by integrating Eq.~\eqref{adot_chaotic} 
from an initial
spin $a_{\rm bh}=1$, and assuming that the MBH accretes a mass of gas $M_{\rm res}=M_{\rm bh}^{\rm in}$. Doing so, one gets a final
spin $a_{\rm bh}\approx0.14$, which explains the large number of MBH with spin $a_{\rm bh}\sim 0.1$ at low redshifts.
The evolution of the spins is qualitatively similar in the heavy-seed scenario, with the difference that
no MBHs with $M_{\rm bh}< M_{\rm seed}=10^5 M_\odot$ are present, because 
of the seed-model described in Sec.~\ref{sec:DM}.

We stress that our results, and in particular the dichotomy between almost maximal spins at high redshifts and
small spins at $z\approx 0$, are qualitatively independent of our assumption that the seeds
are initially assigned spin parameters drawn from a uniform distribution $-1\leq a_{\rm bh}\leq1$,
at least for MBH masses $M_{\rm bh}\gtrsim 3 M_{\rm seed}$.
While it is unclear whether such a spin-parameter distribution makes sense physically, 
because little is known about the spins of the seeds,
at high redshifts the MBHs accrete coherently, and they lose memory
of their initial spin after accreting a mass comparable to their own 
(i.e., if a black hole of mass $M_{\rm bh}$ accretes
coherently, its spin becomes maximal after accreting a
mass $\lesssim 2 M_{\rm bh}$~\citep{bardeen70}).

In both the light and heavy-seed scenario, it would seem that the paucity of MBHs with large spins might be in contrast 
with the iron K$\alpha$ measurements of the  MBH spins in MCG-6-30-15 and in NGC3783, 
which were claimed to be 
respectively 
$a_{\rm bh} > 0.987$~\citep{reynolds} and   $a_{\rm bh} > 0.9$~\citep{suzaku,brenneman} at 90\% confidence level.
It should be noted, however, that both these measurements are still controversial, with
\citet{patrick} finding $a_{\rm bh}=0.49^{+0.20}_{-0.12}$ for MCG-6-30-15 and $a_{\rm bh}<-0.04$ for NGC3783,
and with other iron K$\alpha$ measurements of MBH spins giving 
smaller values, e.g. a spin between 0.3 and 0.77, according to the measurement, 
for Fairall 9~\citep{patrick,suzaku}.
Even more importantly, these spin measurements are necessarily biased by selection effects,
because large spins correspond to higher emission efficiencies $\eta(a_{\rm bh})$ and therefore higher AGN luminosities~\citep{suzaku,brenneman}.
Moreover, iron K$\alpha$ measurements are of course only possible in systems with accretion disks in the first place (i.e in AGNs),
and those systems are expected to host MBHs with high spins because coherent accretion spins them up to 
the maximal limit (cf. Sec.~\ref{smbh_evolution}).

\section{Conclusions}
\label{sec:conclusion}

We have utilized a semianalytical galaxy-formation model to track the evolution and mergers
of the dark-matter halos, the IGM, the baryonic structures (galactic disks and spheroids, in both their gaseous
and stellar components), and the MBHs that are thought to reside in the center of galaxies. The evolution
of the MBHs is deeply entangled with that of their host galaxies, because it is the star formation
in the galactic spheroid that funnels gas to the galactic center via e.g. radiation drag. 
This creates a reservoir that feeds the MBH, but the character of the accretion 
process (coherent or chaotic) depends on the amount of gas present in this reservoir. Similarly, when
two MBHs merge after their host galaxies have coalesced, their spins get aligned 
prior to the merger due to gravito-magnetic
torques if enough gas is present in the galactic nucleus, 
while the orientation of the spins remains essentially isotropic in gas-poor environments.
A further complication to this picture is that the MBHs also backreact on 
the larger-scale galactic evolution, through the
so-called AGN feedback, i.e. they are thought to quench star formation in high-mass systems by injecting energy
into the IGM via strong jets or accretion-disk winds. Indeed, the AGN feedback is 
a crucial ingredient of modern galaxy-formation models, and is 
needed to explain the ``anti-hierarchical'' evolution (or ``downsizing'')
of baryonic structures, i.e. the fact that the most massive galaxies are dominated by old stellar populations, 
while low-mass galaxies generally present young stellar populations and longer-lasting star-formation activity.

In this paper, we have made an attempt to study
how this complicated interdependence between
MBHs and their host galaxies affects the
black-hole mass and spin evolution,
considering both a scenario where
MBHs form from ``light''  $150 M_\odot$ seeds at $z\sim 15-20$ and one where
they form from ``heavy'' $\sim 10^5 M_\odot$ seeds at $z\sim 10-15$. 
Besides confirming that these two scenarios may be observationally distinguishable with LISA/SGO or a similar
European gravitational-wave mission (eLISA/NGO) by simply looking at the observed event rate for MBH mergers,
 we have studied the MBH mass and spin
evolution in detail. In particular, we have determined that accretion is mostly coherent and MBH
mergers happen in gas-rich environments at high redshifts, while at low redshifts, 
when AGN feedback, ram pressure and clumpy accretion have ``sterilized'' the galaxy,
accretion becomes mainly chaotic and mergers happen in gas-poor environments. 
This results in a spin distribution that is
skewed towards large spins at high redshifts, and towards spins $a_{\rm bh}\sim 0.1$ at low redshifts, 
a prediction that 
will be readily testable by LISA or a similar mission. 

In principle, LISA/SGO or eLISA/NGO will also be capable of 
testing our predictions for the character of
MBH mergers as function of redshift directly,
because 
gas-rich mergers tend to present aligned spins, while gas-poor mergers tend to present randomly oriented spins,
and these different orientations will produce an effect on the higher-order harmonics of the 
gravitational waveforms.
We will examine this more in detail in future work
aiming at calculating the eLISA/NGO signal-to-noise ratios and the effect
of higher-order harmonics. In particular, to accurately model
the gravitational waveforms for spinning black-hole binaries we will employ the 
effective-one-body model of \citet{BB_eob,BB_NNLO}, which 
successfully reproduces the exact (i.e. numerical-relativity) waveforms 
both in the extreme-mass ratio limit~\citep{TPL1,TPL2} and in the 
comparable-mass case~\citep{pan_etal}.

Finally we have briefly hinted at how the mass evolution of MBHs and the event rates for LISA/SGO or eLISA/NGO may be different 
in alternative galaxy-formation models such as the
``two-phase'' model of \citet{twophases,cook1,cook2}.
This model
reproduces the 
downsizing of cosmic structures more naturally
than the standard galaxy-formation model that we consider in this paper,
and predicts that galactic spheroids and MBHs should grow earlier 
than in the ``standard'' model, which might result in larger event rates for LISA/SGO  or eLISA/NGO
(at least in the case of light MBH seeds). We will 
study the predictions of this alternative galaxy-formation model in detail in a future paper.

\section*{Acknowledgments}
I thank Michael Cook, Gian Luigi Granato and Andrea Lapi 
for numerous discussions on semianalytical galaxy formation models,
and Emanuele Berti, Gian Luigi Granato, Andrea Lapi, Alberto Sesana and 
Marta Volonteri for critically reading a preliminary version of this
manuscript and providing useful and knowledgeable comments that significantly improved it.
Special thanks to Luciano Rezzolla for first bringing the problem of
the spin evolution of massive black holes to my attention, and for
facilitating access to several workstations at the Max Planck Institute for Gravitational Physics (Albert Einstein Institute).
I am also indebted to Manuel Tiglio for granting access
to the CSCAMM Cluster at the University of Maryland, and to Marcelo Ponce for making
available further computational resources at the University of Guelph.
Part of the simulations described in this paper were also run on the
SciNet clusters at the University of Toronto.
Finally, I acknowledge support from 
a CITA National Fellowship at the University of Guelph, and from 
NSF Grant PHY-0903631 while at the University of Maryland.

\label{lastpage}

\end{document}